\documentclass[english,11pt]{article}
\pdfoutput=1
\usepackage{epsf,amsmath,amssymb,graphicx,dcolumn}
\usepackage{scalefnt,ulem}
\usepackage{subcaption}
\usepackage{cite,hyperref}
\usepackage{cleveref}
\usepackage{todonotes}
\usepackage{color}
\usepackage{rotating}
\usepackage{microtype}
\usepackage{listings}
\usepackage{etoolbox}
\usepackage{booktabs}
\usepackage{titlesec}
\usepackage{enumitem}
\usepackage{multirow}
\usepackage{changepage}

\usepackage{longtable}
\usepackage{tabularx}

\usepackage{tikz}
\usetikzlibrary{shadows}

\newcommand*\keystroke[1]{%
  \tikz[baseline=(key.base)]
    \node[%
      draw,
      fill=white,
      drop shadow={shadow xshift=0.25ex,shadow yshift=-0.25ex,fill=black,opacity=0.75},
      rectangle,
      rounded corners=2pt,
      inner sep=1pt,
      line width=0.5pt,
      font=\scriptsize\sffamily
    ](key) {#1\strut}
  ;
}

\titleformat{\paragraph}
{\normalfont\normalsize\bfseries}{\theparagraph}{1em}{}
\titlespacing*{\paragraph}
{0pt}{3.25ex plus 1ex minus .2ex}{1.5ex plus .2ex}

\apptocmd{\thebibliography}{\setlength{\itemsep}{0.05cm}}{}{}
\definecolor{lightblue}{rgb}{.7,.8,1}
\Crefname{figure}{Fig.}{Figs.}
\newcommand{\Matrixversion}{\matrixparam{MATRIX\_v1.0.0}}

\def\bal#1\eal{\begin{align}#1\end{align}}
\newcommand{\gridrun}{{\it grid-run}}
\newcommand{\premainrun}{{\it pre-run}}
\newcommand{\mainrun}{{\it main-run}}
\newcommand{\resultrun}{{\it result-collection}}

\newcommand{\abbrev}{\scalefont{1}}

\newcommand{\mH}{m_{H}}

\newcommand{\slha}{{\abbrev SLHA}}
\newcommand{\susy}{{\abbrev SUSY}}

\newcommand{\block}[1]{{\tt Block #1}}
\newcommand{\blockentry}[2]{{\tt Block #1[#2]}}

\newcommand{\sm}{{\abbrev SM}}

\newcommand{\mb}{m_{\rm b}}

\newcommand{\citere}[1]{Ref.\cite{#1}}
\newcommand{\citeres}[1]{Refs.\cite{#1}}

\renewcommand{\arraystretch}{1.2}
\newcommand\Matrix{{\sc Matrix}}
\newcommand\Munich{{\sc Munich}}
\newcommand\OpenLoops{{\sc OpenLoops}}
\newcommand\Collier{{\sc Collier}}
\newcommand{\CutTools}{{\sc CutTools}}
\newcommand{\OneLOop}{{\sc OneLOop}}
\newcommand{\LHAPDF}{{\sc LHAPDF}}
\newcommand\FEWZ{{\sc FEWZ}}
\newcommand\NNLOjet{{\sc NNLOjet}}

\newcommand{\reffi}[1]{Fig.~\ref{#1}}

\newcommand{\refta}[1]{Table~\ref{#1}}
\newcommand{\refse}[1]{Section~\ref{#1}}

\newcommand{\ppppbar}{pp/p\bar{p}}

\def\qT{q_T}
\def\pT{p_T}

\newcommand{\matrixparam}[1]{\lstinline[basicstyle=\ttfamily]|#1|}
\newcommand{\ttt}{\matrixparam}

\newcommand{\listvariable}[2]{\hyperref[#1]{\ttt{#1:  #2}}}
\newcommand{\listswitch}[2]{\hyperref[#1]{\ttt{#1}}}
\newcommand{\eqn}[1]{Eq.\,(\ref{#1})}

\newcommand{\fig}[1]{{\reffi{#1}}}

\newcommand{\tab}[1]{{\refta{#1}}}
\newcommand{\sct}[1]{{\refse{#1}}}
\newcommand{\app}[1]{{\refapp{#1}}}
\newcommand{\scts}[1]{Sections~\ref{#1}}
\newcommand{\dd}{{\rm d}}

\newcommand{\pdf}{{\abbrev PDF}}

\newcounter{notecount}
\tikzstyle{boxblue} = [draw=blue, fill=blue!10, very thick,
    rectangle, rounded corners, inner sep=5pt, inner ysep=5pt]

\def\refeq#1{\mbox{Eq.~\eqref{#1}}}

\def\reffi#1{\mbox{Figure~\ref{#1}}}

\def\refta#1{\mbox{Table~\ref{#1}}}

\def\refse#1{\mbox{Section~\ref{#1}}}

\def\refapp#1{\mbox{\ref{#1}}}
\def\citere#1{\mbox{Ref.~\cite{#1}}}
\def\citeres#1{\mbox{Refs.~\cite{#1}}}

\newcommand{\rcut}{\ensuremath{r_{\mathrm{cut}}}}
\newcommand{\rcutmin}{\ensuremath{r_{\mathrm{cut}}^\mathrm{min}}}
\newcommand{\zz}{\ensuremath{ZZ}}
\newcommand{\ww}{\ensuremath{W^+W^-}}
\newcommand{\wgamma}{\ensuremath{W^\pm \gamma}}
\newcommand{\wz}{\ensuremath{W^\pm Z}}

\newcommand{\pt}{\ensuremath{p_T}}
\newcommand{\qt}{\ensuremath{q_T}}

\newcommand{\D}{\mathrm{d}}
\newcommand\as{\alpha_{\mathrm{S}}}

\makeatletter
\def\appendixname{Appendix}
\appto\appendix{%
  \addtocontents{toc}{\patch@l@section}
  \appto\appendixname{ }
}
\protected\def\patch@l@section{%
  \patchcmd{\l@section}{1.5em}{\widthof{\appendixname\space}+2.5em}{}{}%
}
\makeatother

\begin{document}
\begin{titlepage}

\begin{flushright}
CERN-TH-2017-232\\
ZU-TH 30/17
\end{flushright}
\vspace*{2cm}

\begin{center}
{\Large \bf Fully differential NNLO computations with {MATRIX}}
\end{center}
\par \vspace{2mm}
\begin{center}
{\bf Massimiliano Grazzini$^{(a)}$}, {\bf Stefan Kallweit$^{(b)}$} and {\bf Marius Wiesemann$^{(b)}$}\\[0.2cm]
\vspace{5mm}

$^{(a)}$Physik-Institut, Universit\"at Z\"urich, CH-8057 Z\"urich, Switzerland

$^{(b)}$TH Division, Physics Department, CERN, CH-1211 Geneva 23, Switzerland

\vskip 0.5cm
{\small\tt grazzini@physik.uzh.ch}\\[-0.1em]
{\small\tt stefan.kallweit@cern.ch}\\[-0.1em]
{\small\tt marius.wiesemann@cern.ch}\\[0.7em]

\vspace{5mm}

\end{center}

\par \vspace{2mm}
\begin{center} {\large \bf Abstract} \end{center}
\begin{quote}
\pretolerance 10000

We present the computational framework \Matrix{} \cite{MATRIX} which allows us to evaluate
fully differential cross sections for a wide class of processes at hadron colliders in
next-to-next-to-leading order (NNLO) QCD. The processes we consider are $2\to 1$
and $2\to 2$ hadronic reactions involving Higgs and vector bosons in the final state.
All possible leptonic decay channels of the vector bosons are included for the first time in the calculations,
by consistently accounting for all resonant and non-resonant diagrams, off-shell effects and spin correlations.
We briefly introduce the theoretical framework \Matrix{} is based on, discuss its relevant
features and provide a detailed description of how to use \Matrix{} to obtain NNLO accurate results for the various processes.
We report reference predictions for inclusive and fiducial cross sections of all the
physics processes considered here and discuss their corresponding uncertainties.
\Matrix{} features an automatic extrapolation procedure that allows us, for the first time,
to control the systematic uncertainties inherent to the applied
NNLO subtraction procedure down to the few permille level (or better).

\end{quote}

\vspace*{\fill}
\begin{flushleft}
November 2017

\end{flushleft}
\end{titlepage}

\tableofcontents

\section{Introduction}

Precision computations for Standard Model (SM) processes are vital for the rich physics programme at the LHC. The increasing 
amount of collected data pushes the experimental uncertainties down to the percent level, thereby demanding accurate predictions for
many relevant physics processes. This holds not only for SM measurements. Also new-physics  searches rely on a precise modelling 
of the SM backgrounds. In particular, the sensitivity to small deviations from the SM predictions directly depends on the size of
theoretical uncertainties. Besides single vector-boson and Higgs boson production processes, vector-boson pair production 
is particularly important in that respect since anomalous 
triple gauge couplings would be first uncovered
in cross sections and distributions of the diboson processes.

Precise SM computations require, in particular, the inclusion of QCD radiative corrections at the next-to-leading order (NLO), and if possible at the next-to-next-to-leading order (NNLO). NNLO QCD predictions for the simplest hadronic reactions have been available for quite some time. The pioneering computation of the inclusive cross section for vector-boson production was carried out in the '90s \cite{Hamberg:1990np}. The corresponding computation for Higgs boson production was performed about ten years later \cite{Harlander:2002wh,Anastasiou:2002yz,Ravindran:2003um}. They were followed by the calculation of the rapidity distribution of vector bosons \cite{Anastasiou:2003ds}.
Shortly after, fully differential calculations for Higgs and vector-boson production started to appear \cite{Anastasiou:2005qj,Melnikov:2006kv,Catani:2007vq,Anastasiou:2007mz,Grazzini:2008tf,Catani:2009sm}.
This further step was essential to obtain realistic predictions since fully differential computations allow us to apply selection cuts on the produced boson and on its decay products, and to directly address all the relevant kinematic distributions. 
The last decade has seen a revolution in the field of NNLO computations:
The calculations for associated production of a Higgs boson with a vector boson \cite{Ferrera:2011bk,Ferrera:2014lca,Ferrera:2017zex,Campbell:2016jau}, Higgs boson production in bottom-quark annihilation \cite{Harlander:2003ai,Harlander:2010cz,Harlander:2011fx,Buehler:2012cu}, top-mass effects in Higgs boson production \cite{Marzani:2008az,Harlander:2009mq,Harlander:2009my,Pak:2009dg,Neumann:2014nha}, $HH$ \cite{deFlorian:2013jea,deFlorian:2016uhr}, $\gamma\gamma$ \cite{Catani:2011qz,Campbell:2016yrh}, $Z\gamma$ \cite{Grazzini:2013bna,Grazzini:2015nwa,Campbell:2017aul}, $W\gamma$ \cite{Grazzini:2015nwa}, $ZZ$ \cite{Cascioli:2014yka,Grazzini:2015hta,Heinrich:2017bvg}, \ww{} \cite{Gehrmann:2014fva,Grazzini:2016ctr} and \wz{} \cite{Grazzini:2016swo,Grazzini:2017ckn} production have been completed.
NNLO results have been achieved also for further important processes like top-quark pair \cite{Czakon:2013goa,Czakon:2015owf} and single top \cite{Brucherseifer:2014ama} production, dijet production \cite{Currie:2017eqf}, Higgs production through vector-boson fusion \cite{Cacciari:2015jma}, $H$+jet \cite{Boughezal:2015aha,Caola:2015wna,Chen:2016zka}, $\gamma$+jet \cite{Campbell:2016lzl}, $Z$+jet \cite{Boughezal:2015ded,Gehrmann-DeRidder:2016jns} and $W$+jet \cite{Boughezal:2015dva}. Despite this tremendous progress, at present, publicly available NNLO programs typically carry out fully differential NNLO computations for a limited set of specific processes. Examples are \FEWZ{} \cite{Gavin:2012sy} and {\sc DYNNLO} \cite{Catani:2009sm} for vector-boson production, {\sc FehiPro} \cite{Anastasiou:2005qj,Anastasiou:2009kn} and {\sc HNNLO} \cite{Catani:2007vq,Grazzini:2008tf} for Higgs boson production, and {\sc 2$\gamma$NNLO} \cite{Catani:2011qz} for diphoton production. A notable exception is {\sc MCFM} \cite{Boughezal:2016wmq}, which in its current release features an NNLO implementation of single vector-boson and Higgs boson production, associated production of a Higgs boson with a vector-boson, and diphoton production.

In this paper, we present the computational framework \Matrix{}\footnote{\Matrix{} is the abbreviation of 
``\Munich{} Automates qT subtraction and Resummation to Integrate X-sections''.}, which features a parton-level Monte Carlo generator
capable of computing fiducial cross sections and distributions for Higgs boson, 
vector-boson and vector-boson pair production processes up to NNLO in QCD.
For the first time, we consider all possible leptonic decay channels of 
the vector bosons, and we include spin correlations and off-shell effects by 
accounting for all resonant and non-resonant diagrams, thereby allowing 
the user to apply
realistic fiducial cuts directly on the phase-space of the respective leptonic final state.
\Matrix{} achieves NNLO accuracy by using a process-independent 
implementation of the $\qT$-subtraction formalism~\cite{Catani:2007vq} in combination with
a fully automated implementation of the Catani--Seymour dipole subtraction method \cite{Catani:1996jh,Catani:1996vz}
within the Monte Carlo program \Munich{}\footnote{\Munich{} is the 
abbreviation of ``MUlti-chaNnel Integrator at Swiss~(CH) precision'' --- an automated parton-level NLO 
generator by S.~Kallweit.}.
All (spin- and colour-correlated) tree-level and one-loop amplitudes are obtained from \OpenLoops{}~\cite{Cascioli:2011va,Buccioni:2017yxi}. 
Early versions of \Matrix{} have
been used, in combination with the two-loop scattering amplitudes of~\citeres{Gehrmann:2011ab,Gehrmann:2015ora,deFlorian:2013uza}, 
for the NNLO calculations of $Z\gamma$~\cite{Grazzini:2013bna,Grazzini:2015nwa}, $W^\pm\gamma$~\cite{Grazzini:2015nwa} $ZZ$~\cite{Cascioli:2014yka,Grazzini:2015hta}, 
\ww{}~\cite{Gehrmann:2014fva,Grazzini:2016ctr}, \wz{}~\cite{Grazzini:2016swo,Grazzini:2017ckn} and $HH$ \cite{deFlorian:2016uhr}
production\footnote{A first application of the code to the resummed transverse-momentum spectra of $ZZ$ and \ww{} pairs 
has been presented in \citere{Grazzini:2015wpa} at NNLL+NNLO.} 
and the importance of including NNLO corrections for these processes is evident for both total rates and differential distributions. 
\Matrix{} provides a fully automated extrapolation procedure 
that allows us, for the first time, 
to control the systematic uncertainties inherent to the \qt{}-subtraction
procedure down to the few permille level (or better) for all NNLO predictions.
The \Matrix{} framework offers a simple interface to a powerful code to carry out such computations
in a relatively straightforward way, and its first public version is now available for download \cite{MATRIX}.

The manuscript is organized as follows: In \sct{sec:general} we give a general introduction into the \Matrix{} framework, 
where we review the \qt{}-subtraction formalism 
and describe the organization of the computations.  
We then provide detailed instructions on how to use the code:
This involves the generation, compilation and running of a process to compute LO, NLO and NNLO 
cross sections in \sct{sec:use}, and a detailed description of the relevant input files and parameters 
in \sct{sec:inputfiles}.
In \sct{sec:physics} we provide benchmark predictions for total and fiducial rates, respectively, for all processes, including the results of our novel 
extrapolation procedure, and 
we discuss the relevant physics features of each process. 
A discussion of the systematic uncertainties of NNLO cross sections
computed with \qt{} subtraction for a representative set of processes
and details on the extrapolation procedure 
are presented in \sct{sec:qtsubtraction}.
In \sct{sec:summary} we summarize our results. 
All predefined phase-space cuts are listed in \app{app:cuts}.
How to extend the predefined set of cuts, distributions and dynamic scales by modifying the underlying \textsc{C++} code is sketched in \app{app:advancedstuff}.
Finally, \app{app:Troubleshooting} provides a loose selection of solutions on compilation and running issues, 
which have been encountered in the testing phase of \Matrix{} and are expected to be potentially helpful for the user.

\section[NNLO computations in the M{\scriptsize ATRIX} framework]{NNLO computations in the M{\small ATRIX} framework}
\label{sec:general}

The computation of a QCD cross section at NNLO requires 
the evaluation of tree-level contributions with up to two additional unresolved partons,
of one-loop contributions with one unresolved parton and of purely virtual contributions.
The implementation of the corresponding scattering amplitudes in a complete NNLO calculation at the fully differential (exclusive) level is a highly non-trivial task because of the presence of infrared (IR) divergences at intermediate stages of the calculation. In particular, since the divergences affect real and virtual contributions in a different way, a straightforward combination of these components is not possible. Various methods have been proposed and used to
overcome these issues at NNLO~\cite{Kosower:1997zr, GehrmannDeRidder:2005cm,Daleo:2006xa,Currie:2013vh,Catani:2007vq,Somogyi:2005xz,DelDuca:2015zqa,DelDuca:2016csb,Czakon:2010td,Czakon:2011ve,Czakon:2014oma,Boughezal:2015dva,Boughezal:2015eha,Gaunt:2015pea,Caola:2017dug}. The method applied by \Matrix{} is transverse-momentum~(\qt{}) subtraction \cite{Catani:2007vq}, and it is briefly described below.

\subsection[The $\qT$-subtraction formalism]{The $\boldsymbol{\qT}$-subtraction formalism}
\label{subsec:formalism}

The \qt{}-subtraction formalism \cite{Catani:2007vq} is a method to handle and cancel IR divergences at NLO and NNLO.
The method exploits the fact that for the production of a colourless final-state system 
(i.e.\ a system composed of particles without QCD interactions) 
the behaviour of the $\qT$
distribution\footnote{Here and in the following, $\qT$ always refers
to the transverse momentum of the colourless final-state system under consideration.} at small $\qT$
has a universal (process-independent) structure that is explicitly known up to NNLO
through the formalism of transverse-momentum resummation \cite{Collins:1984kg,Bozzi:2005wk}.
This knowledge is sufficient to fully determine the $\qT$ dependence of the cross section at small $\qT$ and to
construct a non-local, but process-independent IR subtraction counterterm
 for this entire class of processes.\footnote{The extension to heavy-quark production has been discussed in \citere{Bonciani:2015sha}.}

In the \qt{}-subtraction method, the cross section for a generic process $pp\to F+X$, where $F$ is a colourless system as specified above, 
can be written up to (N)NLO as
\begin{equation}
\label{eq:main}
\dd{\sigma}^{\mathrm{F}}_{\mathrm{(N)NLO}}={\cal H}^{\mathrm{F}}_{\mathrm{(N)NLO}}\otimes \dd{\sigma}^{\mathrm{F}}_{\mathrm{LO}}
+\left[ \dd{\sigma}^{\mathrm{F + jet}}_{\mathrm{(N)LO}}-
\dd{\sigma}^{\mathrm{CT}}_{\mathrm{(N)NLO}}\right].
\end{equation}
The term $d{\sigma}^{\mathrm{F + jet}}_{\mathrm{(N)LO}}$ represents the cross
section for the production of the system $F+$jet at (N)LO accuracy.
If \refeq{eq:main} is applied at NLO, the LO cross section $d{\sigma}^{\mathrm{F + jet}}_{\mathrm{LO}}$ can be obtained by direct integration of the corresponding tree-level amplitudes. If \refeq{eq:main} is applied at NNLO, the NLO cross section $d{\sigma}^{\mathrm{F + jet}}_{\mathrm{NLO}}$ can be evaluated by using any available NLO subtraction method \cite{Frixione:1995ms,Frixione:1997np,Catani:1996jh,Catani:1996vz} to handle and cancel the corresponding IR divergencies.
Therefore, $d{\sigma}^{\mathrm{F + jet}}_{\mathrm{NLO}}$ is finite provided that $\qT\neq 0$, but it diverges in the limit $\qT\to 0$.
The process-independent counterterm 
$\D{\sigma}^{\mathrm{CT}}_{\mathrm{(N)NLO}}$ guarantees the cancellation of this divergence of
the $F+$jet cross section, and its general expression is provided in \citere{Bozzi:2005wk}.
The numerical implementation of the contribution in the square bracket in~\eqn{eq:main}, which is 
by construction finite in the limit $\qT\to0$, is discussed in detail in \refse{subsec:matrix}.
The computation is completed by evaluating the first term on the right-hand side of \refeq{eq:main}, which
depends on the hard-collinear coefficients ${\cal H}^{\mathrm{F}}_{\mathrm{NLO}}$ and ${\cal H}^{\mathrm{F}}_{\mathrm{NNLO}}$, respectively, at NLO and NNLO.
The structure of the NLO coefficient ${\cal H}^{\mathrm{F}}_{\mathrm{NLO}}$ has been obtained in a universal way 
from the one-loop corrections to the respective Born subprocess~\cite{deFlorian:2001zd}.
The general form of ${\cal H}^{\mathrm{F}}_{\mathrm{NNLO}}$ is also known \cite{Catani:2013tia}: It has been derived from
the explicit results for Higgs~\cite{Catani:2011kr} and vector-boson~\cite{Catani:2012qa} production
in terms of the suitably subtracted two-loop corrections to the respective Born subprocesses.
Thus, if the $q{\bar q}\to F$ (or $gg\to F$) two-loop amplitude is available, the coefficient ${\cal H}^{\mathrm{F}}_{\mathrm{NNLO}}$ can be straightforwardly extracted.

\subsection[Implementation within the M{\scriptsize ATRIX} framework]{Implementation within the M{\small ATRIX} framework}
\label{subsec:matrix}

\Matrix{} provides a process library for the computation of colour-singlet processes at NNLO QCD. 
The core of the \Matrix{} framework is the Monte Carlo program \Munich{}, which is 
capable of computing both QCD and EW~\cite{Kallweit:2014xda,Kallweit:2015dum} corrections to any SM process at NLO accuracy.
\Munich{} employs an automated implementation of the Catani--Seymour dipole-subtraction method for massless~\cite{Catani:1996jh,Catani:1996vz}
and massive~\cite{Catani:2002hc} partons, and contains a general implementation of a very efficient, multi-channel based  phase-space integration.
All amplitudes up to one-loop level are supplied by \OpenLoops{}\footnote{\OpenLoops{} relies on the fast and stable tensor reduction
 of \Collier{}~\cite{Denner:2014gla,Denner:2016kdg}, supported by a rescue system based on 
quad-precision \CutTools\cite{Ossola:2007ax} with \OneLOop\cite{vanHameren:2010cp} to deal with exceptional phase-space points.}~\cite{Cascioli:2011va} through an automated interface. 
With this functionality inherited from \Munich{}, \Matrix{} is immediately able to perform in principle any SM calculation up to NLO accuracy.
To promote \Munich{} to a Monte Carlo integrator at NNLO QCD, the $F+$jet cross section at NLO ($d{\sigma}^{\mathrm{F + jet}}_{\mathrm{NLO}}$) is combined
with a process-independent implementation of the \qt{}-subtraction formalism for both $gg$- and $q{\bar q}$-initiated processes
within the \Matrix{} framework. The universal nature of the counterterm
$\D{\sigma}^{\mathrm{CT}}_{\mathrm{NNLO}}$ and the hard-collinear coefficients  
${\cal H}^{\mathrm{F}}_{\mathrm{NNLO}}$ in \refeq{eq:main} allows us to perform NNLO QCD 
computations\footnote{On the same basis \Matrix{} automates also the small-\qt{} 
resummation of logarithmically enhanced terms at NNLL
accuracy (see \citere{Grazzini:2015wpa}, and \citere{Wiesemann:2016tae} for more details), which, however, is not yet included in the first release.} 
for the hadroproduction of an arbitrary set of colourless final-state particles,
provided that the corresponding two-loop virtual amplitudes to the Born-level subprocesses are available.

To this end, \Matrix{} includes the hard-collinear coefficients of \citere{Catani:2011kr,Catani:2012qa}, relevant for single Higgs and vector-boson production, and employs own implementations of the two-loop amplitudes
for the associated production of a $W/Z$ boson with a photon~\cite{Gehrmann:2011ab} and $\gamma\gamma$~\cite{Anastasiou:2002zn}  production, whereas external codes are used for on-shell 
$ZZ$~\cite{Cascioli:2014yka} and \ww{}~\cite{Gehrmann:2014fva} production\footnote{Private code provided by T.~Gehrmann, A.~von~Manteuffel and L.~Tancredi.}. 
The two-loop amplitudes for off-shell production of massive vector-boson pairs\cite{Gehrmann:2015ora} are taken from the publicly 
available code \textsc{VVamp}~\cite{hepforge:VVamp}.
Any new production process of colour singlets can be supplemented to the \Matrix{} library upon implementation 
of the corresponding two-loop amplitudes, since all remaining process-dependent ingredients are available
in \Munich{}+\OpenLoops{} and the implementation of the \qt{}-subtraction formalism is fully general.

While the idea behind the \qt{}-subtraction formalism has been outlined in the
previous Section, one point deserves some additional discussion.
The contribution in the square bracket in \refeq{eq:main} is formally finite in the limit $\qT\to 0$,
but both $\D{\sigma}^{\mathrm{F + jet}}_{\mathrm{(N)LO}}$ and $\D{\sigma}^{\mathrm{CT}}_{\mathrm{(N)NLO}}$
are separately divergent. Since the subtraction is non-local,
we introduce a technical cut-off $r_{\mathrm{cut}}$ on the dimensionless quantity $r=\qt/M$ ($M$ being the invariant mass of the colourless system) which renders both terms separately finite. Below this cut-off, $\D{\sigma}^{\mathrm{F + jet}}_{\mathrm{(N)LO}}$ 
and $\D{\sigma}^{\mathrm{CT}}_{\mathrm{(N)NLO}}$ are assumed to be identical, which is correct up to power-suppressed contributions.
The latter vanish in the limit $r_{\mathrm{cut}}\to0$ and can be controlled by monitoring the $r_{\mathrm{cut}}$ dependence of the cross section.
The absence of any residual logarithmic dependence on $r_{\mathrm{cut}}$ thus provides strong evidence of the correctness of the 
computation since any mismatch between the contributions would result in a divergence of the cross section for $r_{\mathrm{cut}}\to0$.
The cut-off on $r$ acts as a slicing parameter, and, correspondingly, the \qt{}-subtraction method as implemented in \Matrix{} works 
very similar to a phase-space slicing method. 

To monitor the $r_{\mathrm{cut}}$ dependence without the need of repeated CPU-intensive runs,
\Matrix{} simultaneously computes the cross section at several $r_{\mathrm{cut}}$ values.
The numerical information on the $r_{\mathrm{cut}}$ dependence is used to address the limit $r_{\mathrm{cut}}\to0$ by using a fit based on the results at finite \rcut{} values. 
The extrapolated result, including an estimate of the uncertainty of the extrapolation 
procedure, is provided at the end of each run.
Details on the $\rcut{}\rightarrow 0$ extrapolation and its uncertainty estimate
are presented in \sct{sec:qtsubtraction}, where we also discuss the 
$r_{\mathrm{cut}}$ dependence of a representative set of the processes available in the 
first release of \Matrix{}.

\section[How to use M{\scriptsize ATRIX}]{How to use M{\small ATRIX}}\label{sec:use}

The code is engineered in a way that guides the user from the very first execution of 
\Matrix{} to the very end of a run of a specific process, obtaining all relevant results.
In-between there are certain steps/decisions to make (such as choosing the process, 
inputs, parameters, \ldots), which will be described in more detail throughout this and the next Section.

The only thing we require the user of \Matrix{} to provide on the machine where the code 
is executed is a working installation of {\sc LHAPDF}, which is a well-known standard code 
by now, such that \matrixparam{lhapdf-config} is recognized as a terminal command, or that the path to the \matrixparam{lhapdf-config}
executable is specified in the file \matrixparam{MATRIX\_configuration} (see \sct{sec:config} for more details on the configuration 
of \Matrix{}).\footnote{\Matrix{} has been tested to work with {\sc LHAPDF} versions 5 and 6.}

\renewcommand{\baselinestretch}{1.0}
\begin{table}[!htbp]
\begin{center}
\begin{tabular}{lclcl}
\toprule
\matrixparam{\$\{process\_id\}}  &&   {\bf process}  &&  {\bf description} \\
\midrule
\matrixparam{pph21}         &&   $\ppppbar \to H$                       &&   on-shell Higgs-boson production\\
\matrixparam{ppz01}         &&   $\ppppbar \to Z$                       &&   on-shell $Z$ production\\
\matrixparam{ppw01}         &&   $\ppppbar \to W^-$                     &&   on-shell $W^-$ production with CKM\\
\matrixparam{ppwx01}        &&   $\ppppbar \to W^+$                     &&   on-shell $W^+$ production with CKM\\
\matrixparam{ppeex02}       &&   $\ppppbar \to e^- e^+$                 &&   $Z$ production with decay\\
\matrixparam{ppnenex02}     &&   $\ppppbar \to \nu_e \bar\nu_e$             &&   $Z$ production with decay\\
\matrixparam{ppenex02}      &&   $\ppppbar \to e^- \bar\nu_e$               &&   $W^-$ production with decay and CKM\\
\matrixparam{ppexne02}      &&   $\ppppbar \to e^+ \nu_e$               &&   $W^+$ production with decay and CKM\\
\matrixparam{ppaa02}        &&   $\ppppbar \to \gamma\gamma$             &&   $\gamma\gamma$ production\\
\matrixparam{ppeexa03}      &&   $\ppppbar \to e^- e^+ \gamma$          &&   $Z \gamma$ production with decay\\
\matrixparam{ppnenexa03}    &&   $\ppppbar \to \nu_e \bar\nu_e \gamma$       &&   $Z \gamma$ production with decay\\
\matrixparam{ppenexa03}     &&   $\ppppbar \to e^- \bar\nu_e \gamma$         &&   $W^- \gamma$ with decay\\
\matrixparam{ppexnea03}     &&   $\ppppbar \to e^+ \nu_e \gamma$         &&   $W^+ \gamma$ with decay\\
\matrixparam{ppzz02}        &&   $\ppppbar \to Z Z$                     &&   on-shell $ZZ$ production\\
\matrixparam{ppwxw02}       &&   $\ppppbar \to W^+ W^-$                     &&   on-shell $W^+W^-$ production\\
\matrixparam{ppemexmx04}    &&   $\ppppbar \to e^- \mu^- e^+ \mu^+$       &&   $ZZ$ production with decay\\
\matrixparam{ppeeexex04}    &&   $\ppppbar \to e^- e^- e^+ e^+$         &&   $ZZ$ production with decay\\
\matrixparam{ppeexnmnmx04}  &&   $\ppppbar \to e^- e^+ \nu_\mu \bar\nu_\mu$       &&   $ZZ$ production with decay\\
\matrixparam{ppemxnmnex04}  &&   $\ppppbar \to e^- \mu^+ \nu_\mu \bar\nu_e $   &&   $W^+W^-$ production with decay\\
\matrixparam{ppeexnenex04}  &&   $\ppppbar \to e^- e^+ \nu_e \bar\nu_e$         &&   $ZZ$ / $W^+W^-$ production with decay\\
\matrixparam{ppemexnmx04}   &&   $\ppppbar \to e^- \mu^- e^+ \bar\nu_\mu $     &&   $W^-Z$ production with decay\\
\matrixparam{ppeeexnex04}   &&   $\ppppbar \to e^- e^- e^+ \bar\nu_e $     &&   $W^-Z$ production with decay\\
\matrixparam{ppeexmxnm04}   &&   $\ppppbar \to e^- e^+ \mu^+ \nu_\mu$     &&   $W^+Z$ production with decay\\
\matrixparam{ppeexexne04}   &&   $\ppppbar \to e^- e^+ e^+ \nu_e$     &&   $W^+Z$ production with decay\\
\bottomrule
\end{tabular}
\end{center}
\renewcommand{\baselinestretch}{1.0}
\caption{\label{tab:processes} Available processes in {\sc \Matrix}.}
\end{table}
\renewcommand\arraystretch{1.1}

\subsection{Compilation and setup of a process}\label{sec:setup}
Assuming that the \Matrixversion{}\matrixparam{.tar.gz} package is extracted and \LHAPDF{} is installed, the simple command\footnote{Note that global compilation settings (if necessary) must be set before starting the code; for options 
see \sct{sec:config}.}
\lstset{basicstyle=\small}
\lstset{keepspaces=true}
\lstset{columns=flexible}
\lstset{showstringspaces=false}
{\tt
\begin{lstlisting}[language=bash]
 $ ./matrix
\end{lstlisting}
}
executed from the folder \Matrixversion{} opens
the \Matrix{} shell, an interactive steering interface for the compilation and the 
setup of a certain process. In principle, one can always follow the on-screen instructions; auto-completion 
of commands should work in all the \Matrix{}-related shells. The first thing to do is to 
choose the desired process that should be created and compiled, by typing the respective \matrixparam{\$\{process\_id\}}, e.g.\
{\tt
\begin{lstlisting}[language=bash]
 |===>> ppz01
\end{lstlisting}
}
for on-shell $Z$-Boson production. To find a certain \matrixparam{\$\{process\_id\}}, the command
{\tt
\begin{lstlisting}[language=bash]
 |===>> list
\end{lstlisting}
}
will print a list of all available processes on screen, in the same format as given in \tab{tab:processes}.
After entering the process, you will be asked to agree with the terms to use \Matrix{}. 
They require you to acknowledge the work of various groups that went into the computation of the present \Matrix{} process by citing the references provided in the file \matrixparam{CITATION.bib}. This file is provided with the results in every \Matrix{} run. 
In particular, a separate dialog appears for
external computations if the implementation of a process is based on them. 
Simply type 
{\tt
\begin{lstlisting}[language=bash]
 |===>> y
\end{lstlisting}
}
for each of these dialogs. Once agreed to the usage terms of \Matrix{}, the compilation script will automatically pursue the following steps:
\begin{itemize}
\item linking to {\sc LHAPDF} \cite{Buckley:2014ana};
\item download and installation of \OpenLoops{} \cite{Cascioli:2011va,hepforge:OpenLoops} (skipped if already installed);
\item installation of {\sc Cln} \cite{cln} (skipped if already installed);
\item installation of {\sc Ginac} \cite{Bauer:2000cp} (skipped if already installed);
\item download of the relevant tree-level and one-loop amplitudes through \OpenLoops{} (skipped if they already exist);
\item compilation of \Matrix{} process (asked for recompilation if executable exists);
\item setting up of the \Matrix{} process folder under the path \matrixparam{run/\$\{process\_id\}\_MATRIX}\,.
\end{itemize}
Thereafter, the \Matrix{} shell exits and the process is ready to be run from the created process folder. As instructed 
on screen, enter that folder,
{\tt
\begin{lstlisting}[language=bash]
 $ cd run/${process_id}_MATRIX
\end{lstlisting}
}
and start a run for this process by continuing with the instructions given in \sct{sec:runprocess}.

We note that a process folder created by \Matrix{} may 
be moved to and used from essentially any location on the present machine. 
Moreover, a \Matrix{} process folder can be shipped to another system
that contains a working installation of the respective process in \Matrix{}.
This requires, however, to change the soft links for 
\matrixparam{bin/run_process} and \matrixparam{input/MATRIX_configuration} 
inside the process folder
to the correct files of the \Matrix{} installation on the new system.

\subsection{Compilation with arguments}

The \Matrix{} script also features compilation directly via arguments: Type
\lstset{basicstyle=\small, frame=none}
{\tt
\begin{lstlisting}[language=bash]
 $ ./matrix --help
\end{lstlisting}
}
in order to see the available options.

We summarize a few useful examples in the following:\\[-0.8cm]
\begin{enumerate}[label={\arabic*.)}]
\item To directly compile some specific process with ID \matrixparam{\$\{process\_id\}}, simply use the following command:
\lstset{basicstyle=\small, frame=none}
{\tt
\begin{lstlisting}[language=bash]
 $ ./matrix ${process_id}
\end{lstlisting}
}
\item To clean the process before compiling (remove object files and executable), add the following option:
\lstset{basicstyle=\small, frame=none}
{\tt
\begin{lstlisting}[language=bash]
 $ ./matrix ${process_id} --clean_process
\end{lstlisting}
}
\item One can force the code to download the latest \OpenLoops{} version even if there is an \OpenLoops{} version found on the system, by using
\lstset{basicstyle=\small, frame=none}
{\tt
\begin{lstlisting}[language=bash]
 $ ./matrix ${process_id} --install_openloops
\end{lstlisting}
}
\item The command
\lstset{basicstyle=\small, frame=none}
{\tt
\begin{lstlisting}[language=bash]
 $ ./matrix ${process_id} --folder_name_extension _my_process_extension
\end{lstlisting}
}
will add an extension to the created process folder such that the default name will be changed to 
\matrixparam{run/\$\{process\_id\}\_MATRIX_my_process_extension}\,.
\end{enumerate}

\subsection{General structure of a process folder}\label{sec:folder}

\begin{figure}[t]
\begin{center}
\hspace{0.15cm}
\includegraphics[trim = 7mm -7mm 0mm 0mm, width=0.97\textwidth]{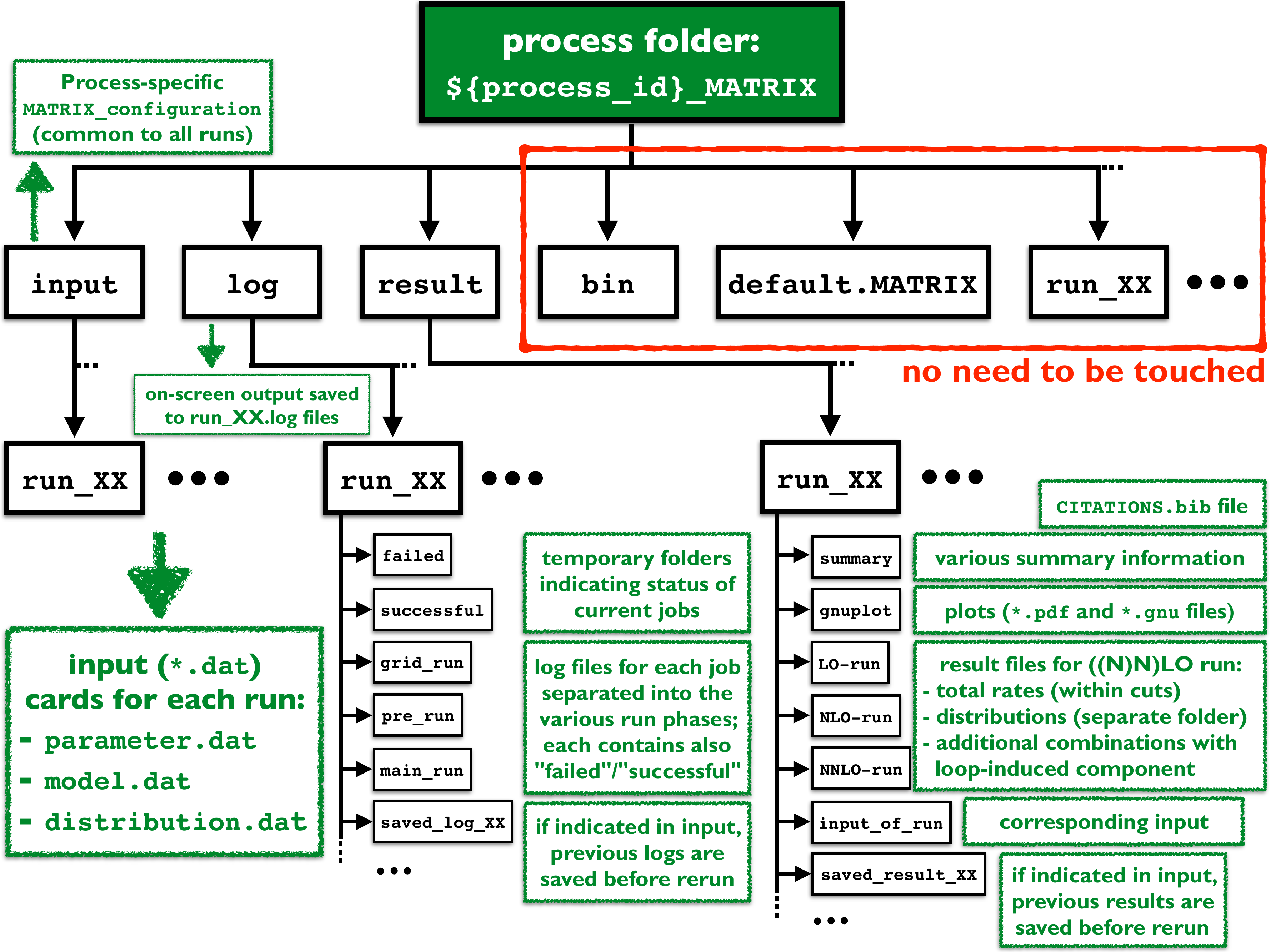}
\caption[]{\label{fig:folderstructure}{Overview of the folder structure inside a \Matrix{} process folder.
}}
\end{center}
\end{figure}

Before providing details on how to actually start the run in a \Matrix{} process folder, it is useful 
to understand the essential parts of the general folder structure the code uses and produces 
while running. This will significantly simplify the comprehension of the code behaviour in the 
upcoming Section. \fig{fig:folderstructure} visualizes the general structure: 
The folders relevant to a user are \matrixparam{input}, \matrixparam{log} and \matrixparam{result}. They will be discussed in detail 
below, while the others should not be touched/are not of interest (especially for an unexperienced user); 
the folder \matrixparam{bin} contains the executable and will only be used to start the \Matrix{} run shell; the folder 
\matrixparam{default.MATRIX} is the default folder for a run of this process, which is copied upon creation 
of each new run; the run folders denoted by \matrixparam{run\_XX} contain the actual runs started by the user, where \matrixparam{XX} stands for 
the name given by the user, or an increasing number starting with  \matrixparam{01} in case no name is given 
(see \sct{sec:runprocess} for more details).

The folders \matrixparam{input}, \matrixparam{log} and \matrixparam{result} all follow the same structure: They contain subfolders of the form 
\matrixparam{run\_XX} that correspond to each run started by the user, so that the relevant information is kept 
strictly separated between those different runs. The organization of these subfolders is identical for each run up 
to differences controlled by the inputs. We note that parts of the folder structure 
are created in the course of running. \fig{fig:folderstructure} shows the folder structure at the very end of a 
complete run of the most complex type (i.e.\ including LO, NLO and NNLO with separate PDF choices). 
In the following we discuss the purpose and the organization of the relevant folders for such a run:
\begin{itemize}
\item \matrixparam{input}:\\[-0.7cm]
\begin{itemize}[leftmargin=*]
\item Three different cards can be modified in order to adjust all the run settings (of physics-related and technical kind), model parameters and distributions to be generated in the run. The respective files can be accessed directly or through the interface of the \Matrix{} run shell; see \sct{sec:inputfiles} for details on the input cards.
\begin{itemize}[leftmargin=*]
\item The file \matrixparam{parameter.dat} controls the physics-related run settings, such as collider type, machine energy, PDFs, etc., but also
technical parameters, such as which orders in perturbation theory should be computed, which precision is to be achieved in the run, if distributions are computed, if the loop-induced $gg$ contribution is included, etc.
\item The file \matrixparam{model.dat} sets all relevant model parameters, such as masses, widths, etc.
\item The file \matrixparam{distribution.dat} gives the possibility to define distributions from the final-state particles with certain ranges, bin sizes, etc. (only effective if distributions are turned on in the file \matrixparam{parameter.dat}).
\end{itemize}
\item The process-specific file \matrixparam{MATRIX\_configuration} for general \Matrix{} configurations inside the folder \matrixparam{input} is the same for all runs of this process and can be modified to use an individual configuration for this process (by default it is a soft link to the global file \matrixparam{MATRIX\_configuration} inside the folder \Matrixversion{}\matrixparam{/config}, but may be replaced by a copy of this file, see \sct{sec:config})
\end{itemize}
\item \matrixparam{log}:\\[-0.7cm]
\begin{itemize}[leftmargin=*]
\item This folder is for debugging purposes only. Log files (\matrixparam{*.log} files) are saved for every single job that is started during a run. Once a job has finished successfully, this is indicated by a file created in the \matrixparam{successful} folder. If a job fails (even after a certain number of retries) a corresponding file will be added to the \matrixparam{failed} folder.
\item At the end of each running phase (\gridrun{}, \premainrun{}, \mainrun{}; see \sct{sec:runinteractive}) the respective log files (including the \matrixparam{successful} and \matrixparam{failed} folders) are moved into the folders \matrixparam{grid\_run}, \matrixparam{pre\_run} and \matrixparam{main\_run}, respectively.
\item If an existing run, which has already created log files 
in the respective log folder, is picked up and started again, those log files are saved into a subfolder \matrixparam{saved\_log\_XX}, where \matrixparam{XX} is an increasing number starting at \matrixparam{01} (only working if the respective switch in the file \matrixparam{parameter.dat} is turned on; default: turned off).
\item The on-screen output of the \Matrix{} run script is saved for each run to a file \matrixparam{run\_XX.log}.
\end{itemize}
\item \matrixparam{result}:\\[-0.7cm]
\begin{itemize}[leftmargin=*]
\item This folder contains all relevant results that are generated during and collected at the end of a run.
\item A file \matrixparam{CITATIONS.bib} is created with every run, which contains the citation keys for all publications that were relevant for the specific run.
Please cite these papers if you use the results of \Matrix{} to acknowledge the efforts that have been made to obtain these results with \Matrix{}.
\item The folder \matrixparam{summary} contains information on the respective run. In particular, 
the summary of all total rates (possibly within cuts), which are also printed on screen at the end of each run, is 
saved to the file \matrixparam{result\_summary.dat} (currently the only file there).
\item In the folder \matrixparam{gnuplot} one finds (automatically generated) \matrixparam{*.gnu} and \matrixparam{*.pdf} files for every distribution created during the run. Its \matrixparam{histogram} subfolder contains the data prepared and used for these plots. Additionally, all pdf files are combined into a single file \matrixparam{all\_plots.pdf} using the 
\matrixparam{pdfunite} binary. If either \matrixparam{gnuplot} or \matrixparam{pdfunite} do not exist on the system, 
the corresponding \matrixparam{*.pdf} files are not created.
\item The total rates and distributions are saved to plain text files in the folders \matrixparam{LO-run}, \matrixparam{NLO-run} and \matrixparam{NNLO-run}. This separation reflects the different PDF sets that can be chosen for each of the three runs (in the file \matrixparam{parameter.dat}; see \sct{sec:paramOrder}). Total rates (possibly within cuts) are saved to the files \matrixparam{rate\_XX.dat} (including scale variations if turned on in the file \matrixparam{parameter.dat}; see \sct{sec:scale}), where, depending on the considered order, \matrixparam{XX} can be \matrixparam{LO}, \matrixparam{NLO\_QCD}, \matrixparam{NNLO\_QCD} and \matrixparam{loop-induced\_QCD}. Additional files \matrixparam{rate\_extrapolated\_XX} are created for total rates, which are computed with the \qt{}-subtraction method (NNLO, and possibly NLO): They provide extrapolated results for $\rcut{}\rightarrow 0$ as the final results, see \sct{sec:qtsubtraction}, while the original rate files contain only the cross sections calculated at a finite \rcut{} value. Inside each of the \matrixparam{*-run} folders the distributions are saved to a folder \matrixparam{distributions} (including minimum and maximum results of the scale variations). There is an extra distribution folder 
\matrixparam{distributions\_NLO\_plus\_loop-induced} 
inside the folder \matrixparam{NLO-run}, which contains the results of the NLO 
distributions combined with the loop-induced contribution (if turned on). Besides, there are 
folders \matrixparam{distributions\_NLO\_prime\_plus\_loop-induced} and \matrixparam{distributions\_only\_loop-induced} inside the folder \matrixparam{NNLO-run}
which contain the combined NLO$'$+$gg$ contribution (both evaluated with NNLO PDFs) and the pure $gg$ contribution, respectively.
\item The folder \matrixparam{input\_of\_run} contains the three input cards (\matrixparam{parameter.dat}, \matrixparam{model.dat}, \matrixparam{distribution.dat}), which were copied at the beginning of the respective run.
\item If an existing run, which has already created results in the respective result folder, 
is picked up and started again, those results are saved into a subfolder \matrixparam{saved\_result\_XX}, where \matrixparam{XX} is an increasing number, starting with \matrixparam{01} (only working if the respective switch in the file \matrixparam{parameter.dat} is turned on; default: turned on).
\end{itemize}
\end{itemize}

\subsection{Running a process}\label{sec:runprocess}

\subsubsection{Running with interactive shell}
\label{sec:runinteractive}

From the \Matrix{} process folder (default:
\matrixparam{run/\$\{process\_id\}\_MATRIX}) the command\footnote{Note that the global configuration (if necessary) must be set in the  file \matrixparam{input/MATRIX\_configuration}  before starting the run; 
see \sct{sec:config} for a description of the general options.}
\lstset{basicstyle=\small}
{\tt
\begin{lstlisting}[language=bash]
 $ ./bin/run_process
\end{lstlisting}
}
opens the \Matrix{} run shell, an interactive steering interface for handling all run-related 
settings, inputs and options.\footnote{The script can also be started with certain arguments, see \sct{sec:arg}.}
From here on one can simply follow the on-screen instructions of the \Matrix{} run shell; we thus only summarize the most 
relevant steps in the following.

First, one must choose a name,
{\tt
\begin{lstlisting}[language=bash]
 |===>> ${run_name}
\end{lstlisting}
}
for the run, which has to begin with \matrixparam{run\_}, to generate a new run. 
Alternatively, one can also 
list and choose one of the runs which already exist (have been created before). As in all \Matrix{} shells, auto-complete 
should work here. The general idea is that each run is separate, i.e.\
each of these runs will create its own run folder (\matrixparam{\$\{run_name\}}) and the corresponding subfolders inside \matrixparam{input}, \matrixparam{log} and \matrixparam{result}. An old run can only be picked up when the previous 
one is not running any more. One should, however, be careful with this option since all data
of the old run will be overwritten (except for possibly the
results and the log files, see \sct{sec:folder}).

Next, we can choose from a list of several commands printed on screen. These commands are divided into three 
groups: general commands, input to modify, run modes. Information on each individual command (\matrixparam{\$\{command\}}) can be received through 
the help menu by typing
{\tt
\begin{lstlisting}[language=bash]
 |===>> help ${command}
\end{lstlisting}
}
In order to directly modify the input cards from the shell (opened in the default editor\footnote{The default editor can be set through the \matrixparam{default\_editor} variable of the file \matrixparam{MATRIX\_configuration}, or by exporting directly the \matrixparam{EDITOR} environment variable on the system, e.g.\ \matrixparam{export EDITOR=emacs}, where the respective editor (here: \matrixparam{emacs}) must be installed and recognized as a terminal command.}), one can simply type the name of the input file
{\tt
\begin{lstlisting}[language=bash]
 |===>> ${name_input_file}
\end{lstlisting}
}
where \matrixparam{\$\{name\_input\_file\}} can be either \matrixparam{parameter}, \matrixparam{model} or \matrixparam{distribution}. Changes will be done directly 
to the respective files \matrixparam{parameter.dat}, \matrixparam{model.dat} or \matrixparam{distribution.dat} inside the folder \matrixparam{input/\$\{run_name\}} (see \sct{sec:folder}). Details on the impact of the various parameters, which can be accessed through the input files, are described in \sct{sec:inputfiles}.\footnote{By default the inputs are already set to use reasonable 
cuts and parameters for each process; the default run (without changing the cards) 
computes a simple LO cross section with 1\% precision, which we recommend to use 
when running for the first time in order to test whether everything is working 
properly. As this run should be very quick (a few minutes), this test can be done 
in local mode (see \sct{sec:config} for the settings in the file \matrixparam{input/MATRIX\_configuration}).}

After adjusting the input cards to obtain the desired results, we can start the run by typing
{\tt
\begin{lstlisting}[language=bash]
 |===>> run
\end{lstlisting}
}
This will start a complete run, no human intervention is needed from now on. Once the run is finished the 
results from the run are collected in the respective folder \matrixparam{result/\$\{run_name\}} as printed on screen 
(see \sct{sec:folder} for details on the result-folder structure), the most relevant results, which are the total rates, are also printed on 
screen at the very end of the run.\footnote{Note that \Matrix{} provides the extrapolated cross section for $\rcut{}\rightarrow 0$ as a final result at NNLO (and at NLO if the \qt{}-subtraction procedure is used also at NLO) including an extrapolation uncertainty, see \sct{sec:qtsubtraction}, which is printed on screen after the cross section with a fixed \rcut{} value.} We emphasize that for every run a \matrixparam{CITATION.bib} file is created and provided inside the folder \matrixparam{result/\$\{run_name\}}. Please cite these papers if you use the 
results of \Matrix{} to acknowledge the efforts that have been made to obtain these results with \Matrix{}.

When performing a time-extensive (NNLO) run, we recommend to start \Matrix{} from 
a window manager (e.g.\ \matrixparam{screen} or \matrixparam{tmux}) in order 
to be able to logout from the present machine during the run. An alternative
is to start \Matrix{} with \matrixparam{nohup} as explained in the second example
of \sct{sec:arg}.

\subparagraph{\bf Running phases of a complete run}\quad

A complete run is divided into various stages ({\it running phases}), each of which may be started directly from the run shell by typing 
the name of the respective run mode (\matrixparam{\$\{run_mode\}}). One must bear in mind, however, that every run stage depends on all 
previous run stages and will fail in case one of the previous ones has not finished successfully. The order of the run stages is as follows: 
\begin{itemize}
\item \gridrun{}:\quad First, the integration grids are created in the warm-up phase (\matrixparam{run\_grid}).
\item \premainrun{}:\quad Next, the expected runtimes for the \mainrun{} are extrapolated from a quick pre-run 
phase (\matrixparam{run\_pre}); some preliminary results are already printed on screen. 
\item \mainrun{}:\quad Then, the main run is started, computing all results to the desired precision (\matrixparam{run\_main}).
\item \resultrun{}:\quad Finally, the results are collected, and all distributions are automatically plotted if gnuplot is installed (\matrixparam{run\_results}).
\end{itemize}
Note that the \resultrun{} will always be started automatically after a successful \mainrun{}. 
Furthermore, if the run mode \matrixparam{run\_pre\_and\_main} is used, the code will start 
from the \premainrun{} (assuming a successful \gridrun{}) and automatically 
continue with the \mainrun{} and \resultrun{}.

Starting from one of these intermediate stages can be useful in many respects. One example is 
the continuation of a run after some unwanted behaviour, if some stages have already 
passed successfully and one would like to restart from one of the later stages. 
Note that all jobs in the requested run stage are removed and started from
scratch. To continue a run while keeping already successfully finished jobs 
of the requested run stage, or to run with increased precision, the 
\matrixparam{--continue} command can be used, see example seven of \sct{sec:arg}.
Another example is running again with a modified set of inputs.
In the latter case it is sufficient to only restart the \mainrun{} as follows:
\lstset{basicstyle=\small}
{\tt
\begin{lstlisting}[language=bash]
 $ ./bin/run_process
\end{lstlisting}
}
to start the script,
{\tt
\begin{lstlisting}[language=bash]
 |===>> ${run_name}
\end{lstlisting}
}
to pick up the old run with name \matrixparam{\$\{run_name\}}, and
{\tt
\begin{lstlisting}[language=bash]
 |===>> parameter
\end{lstlisting}
}
to change, e.g., the \pdf{} set in the file 
\matrixparam{parameter.dat} (if not done by hand before).
It is essential to also uncomment \matrixparam{include_pre_in_results = 0} in the same file
in order to avoid mixing of the different settings in \premainrun{} and \mainrun{} in the result 
combination.
After that, the \mainrun{} is started by
{\tt
\begin{lstlisting}[language=bash]
 |===>> run_main
\end{lstlisting}
}

Other run modes to be selected involve different behaviour of the code, such as only setting up the folder \matrixparam{\$\{run_name\}} and the corresponding subfolders inside \matrixparam{input}, \matrixparam{log} and \matrixparam{result} (\matrixparam{setup\_run})
without starting the run (this is helpful if one wants to change the inputs by hand, but not through the interface, e.g.\ by 
copying the input files from somewhere else); deleting a given run including its respective 
subfolders inside \matrixparam{input}, \matrixparam{log} and \matrixparam{result} (\matrixparam{delete\_run}); etc.

\subsubsection{Running with arguments}\label{sec:arg}

The run script allows some of the various settings, which are typically controlled interactively, 
to be controlled directly by arguments in its shell command. This enables, e.g., the 
possibility to directly start a certain run without having to interact with the interface. Type

\lstset{basicstyle=\small, frame=none}
{\tt
\begin{lstlisting}[language=bash]
 $ ./bin/run_process --help
\end{lstlisting}
}
in order to see all available options.

We summarize a few useful examples in the following:\\[-0.8cm]
\begin{enumerate}[label={\arabic*.)}]
\item The command
\lstset{basicstyle=\small, frame=none}
{\tt
\begin{lstlisting}[language=bash]
 $ ./bin/run_process ${run_name} --run_mode run
\end{lstlisting}
}
will create (pick up, if \matrixparam{\$\{run_name\}} exists) the run with name \matrixparam{\$\{run_name\}}, and
directly start a complete run (due to \matrixparam{--run\_mode run}), with the default inputs
(or the ones already set in \matrixparam{\$\{run_name\}}).\footnote{Note that in the default inputs only a simple LO run is enabled.} 
The \matrixparam{\$\{run_mode\}} may be chosen as any of the various commands outlined at the end of the 
previous Section, e.g.\ \matrixparam{--run\_mode run\_pre\_and\_main} to start the run directly from the \premainrun{} (assuming a successful 
\gridrun{} has already been done).
\item The same command can be used in combination with \matrixparam{nohup}
\lstset{basicstyle=\small, frame=none}
{\tt
\begin{lstlisting}[language=bash]
 $ nohup ./bin/run_process ${run_name} --run_mode run > run.out &
\end{lstlisting}
}
to run \Matrix{} in the background while one is still able to logout from the 
present machine. The on-screen output of \Matrix{} in this 
example is written to the file \matrixparam{run.out}.
\item The command
\lstset{basicstyle=\small, frame=none}
{\tt
\begin{lstlisting}[language=bash]
 $ ./bin/run_process ${run_name} --delete_run
\end{lstlisting}
}
will delete the run with name \matrixparam{\$\{run_name\}}, including its respective subfolders inside \matrixparam{input}, \matrixparam{log} and \matrixparam{result}.
\item The command
\lstset{basicstyle=\small, frame=none}
{\tt
\begin{lstlisting}[language=bash]
 $ ./bin/run_process ${run_name} --setup_run
\end{lstlisting}
}
will create a run with name \matrixparam{\$\{run_name\}} including its respective 
subfolders inside \matrixparam{input}, \matrixparam{log} and \matrixparam{result}
without starting the run. One may 
then modify the input files directly by hand (without using the \Matrix{} shell) and continue with starting the run as described under 1.).
\item One may want to copy, e.g.\ as a backup, some existing (possibly finished) run. The command
\lstset{basicstyle=\small, frame=none}
{\tt
\begin{lstlisting}[language=bash]
 $ ./bin/run_process ${run_name} --copy_run_from ${run_another_name}
\end{lstlisting}
}
allows to make a complete copy of an existing run with name \matrixparam{\$\{run\_another\_name\}} to a new run with name \matrixparam{\$\{run_name\}}. This 
may take quite a while in case a finished run is copied, as the run folder could have a rather large size.
\item In certain situations it may be helpful to use inputs other than the default inputs when creating a new run. The command
\lstset{basicstyle=\small, frame=none}
{\tt
\begin{lstlisting}[language=bash]
 $ ./bin/run_process ${run_name} --input_dir ${any_folder_inside_input}
\end{lstlisting}
}
will create a run with name \matrixparam{\$\{run_name\}}, and the three input files will be copied from the folder \matrixparam{\$\{any\_folder\_inside\_input\}} inside 
the folder \matrixparam{input}. This may, of course, also be the name of another run whose inputs should be used. The only requirement is that 
a folder with the given name exists inside the folder \matrixparam{input} and contains the files \matrixparam{parameter.dat}, \matrixparam{model.dat} and \matrixparam{distribution.dat}.
\item \Matrix{} provides the possibility to continue a run, while deleting only the content of later run stages, but not of the current run stage. This is very useful in two situations: First, a run has crashed in the middle or at the end of a run stage, but several jobs have already finished successfully. Second, the precision of a run should be improved by adding more statistics to a previous run. In both cases the command
\lstset{basicstyle=\small, frame=none}
{\tt
\begin{lstlisting}[language=bash]
 $ ./bin/run_process ${run_name} --run_mode run_main --continue
\end{lstlisting}
}
will continue the run with name \matrixparam{\$\{run_name\}} and not delete any job that has already finished successfully. Note that it is absolutely required not to change any of the inputs, except for a possibly increased precision, with respect to the previous run if the flag \matrixparam{--continue} is used. Any other \matrixparam{\$\{run_mode\}} may be chosen to be continued in this way.
\end{enumerate}

\subsection{Configuration file}\label{sec:config}
\begin{table}
\begin{center}
\resizebox{\columnwidth}{!}{%
\begin{tabular*}{\textwidth}{llplpl}
\toprule
{\bf variable}  & \multicolumn{2}{p{10.2cm}}{\raggedright \bf description}    \\
\midrule
\matrixparam{default\_editor}  & \multicolumn{2}{p{10.2cm}}{\raggedright Sets the editor to be used for interactive access to input files. Alternatively, the default editor may be configured directly by exporting the \matrixparam{EDITOR} environment variable on the system.}\\
\matrixparam{mode}  & \multicolumn{2}{p{10.2cm}}{\raggedright Switch to choose local (multicore) run mode or cluster mode.}    \\
\matrixparam{cluster\_name} & \multicolumn{2}{p{10.2cm}}{\raggedright Name of the cluster; currently supported: {Slurm},  {LSF} (e.g.\ lxplus),  {Condor},  {HTCondor} (e.g.\ lxplus),  {PBS},  {Torque},  {SGE}.}\\
\matrixparam{cluster\_queue} & \multicolumn{2}{p{10.2cm}}{\raggedright Queue/Partition of the cluster to be used for cluster submit; not required  in most cases.}\\
\matrixparam{cluster\_runtime} & \multicolumn{2}{p{10.2cm}}{\raggedright Runtime of jobs in cluster submit; not required  in most cases.}\\
\matrixparam{cluster\_submit\_line[1-99]} & \multicolumn{2}{p{10.2cm}}{\raggedright Lines in cluster submit file to add cluster-specific options.}\\
\matrixparam{max\_nr\_parallel\_jobs} & \multicolumn{2}{p{10.2cm}}{\raggedright Number of cores to be used in multicore mode; maximal number of available cores on cluster.}\\
\matrixparam{parallel\_job\_limit} & \multicolumn{2}{p{10.2cm}}{\raggedright Upper threshold for number of parallel jobs; if exceeded, user intervention required to continue.}\\
\matrixparam{max\_jobs\_in\_cluster\_queue} & \multicolumn{2}{p{10.2cm}}{\raggedright If cluster queue contains more jobs than this value, \Matrix{} will wait until jobs finish before submitting further jobs.}\\
\matrixparam{path\_to\_executable} & \multicolumn{2}{p{10.2cm}}{\raggedright This path can be set to the folder that contains the executables of the processes (usually \matrixparam{bin} in the \Matrix{} main folder), 
and provides the possibility to use an executable from a different \Matrix{} 
installation; 
not required in most cases.}\\
\matrixparam{max\_restarts} & \multicolumn{2}{p{10.2cm}}{\raggedright If there are still jobs left that failed after all jobs finished, \Matrix{} will restart all failed jobs $n$ times 
when this parameter is set to $n$.}\\
\matrixparam{nr\_cores} & \multicolumn{2}{p{10.2cm}}{\raggedright Number of cores to be used for the compilation; determined automatically by the number of available cores on the machine if not set.}\\
\matrixparam{path\_to\_lhapdf} & \multicolumn{2}{p{10.2cm}}{\raggedright Path to \matrixparam{lhapdf-config}; not required in most cases.}\\
\matrixparam{path\_to\_openloops} & \multicolumn{2}{p{10.2cm}}{\raggedright Path to the \matrixparam{openloops} executable; not required in most cases.}\\
\matrixparam{path\_to\_ginac} & \multicolumn{2}{p{10.2cm}}{\raggedright Path to the \matrixparam{ginac} installation; not required in most cases.}\\
\matrixparam{path\_to\_cln} & \multicolumn{2}{p{10.2cm}}{\raggedright Path to the \matrixparam{cln} installation; not required  in most cases.}\\
\matrixparam{path\_to\_libgfortran} & \multicolumn{2}{p{10.2cm}}{\raggedright Path to the \matrixparam{libgfortran} library; not required  in most cases. This path can also be used if the \matrixparam{libquadmath} library is not found, to be set to the respective \matrixparam{lib} folder.}\\
\matrixparam{path\_to\_gsl} & \multicolumn{2}{p{10.2cm}}{\raggedright Path to \matrixparam{gsl-config}; not required in most cases.}\\
\bottomrule
\end{tabular*}
}
\end{center}
\renewcommand{\baselinestretch}{1.0}
\caption{Parameters to be set in the file \matrixparam{MATRIX\_configuration}.} 
 \label{tab:config}
\end{table}
\renewcommand\arraystretch{1.1}

Before turning to physics-related and technical settings relevant for a specific \Matrix{} run in \sct{sec:inputfiles}, we discuss the global 
parameters that steer the general behaviour of the code.
The file \matrixparam{MATRIX\_configuration} controls various global settings for both the compilation and the running of the code. 
The general idea is that these configurations can be made once and for all, depending on the respective environment 
one is working on: One can, e.g., set the relevant paths for the compilation (if not found automatically), choose 
local running or specify the cluster scheduler available on the present machine, etc. The global settings that affect the running of a process 
may still be altered at a later stage (before starting the respective run) and 
can be chosen different for different process folders. The main file \matrixparam{MATRIX\_configuration} can be found in the folder \matrixparam{config} 
inside the \Matrix{} main folder. This file is linked during each setup of a process (see \sct{sec:setup}) into the folder
\matrixparam{input} of the respective process folder. This soft link may be replaced by the actual file such that each process 
can have its own configuration file. This allows for process-specific run settings, and
one can, e.g., change from cluster 
to local run mode for a specific process (or even only for a dedicated run and change it back after having started the run).\footnote{Since the file
\matrixparam{MATRIX\_configuration} is read only at the beginning of a run, any change done after that has no effect.}

The options controlled by the file \matrixparam{MATRIX\_configuration} are listed in \tab{tab:config}.

\section[Settings of a M{\scriptsize ATRIX} run]{Settings of a M{\small ATRIX} run}\label{sec:inputfiles}

In this Section all relevant input settings are discussed. Most of them are directly physics-related, but there are also a few more technical parameters. 

\subsection{Process-independent settings}

Every run of a process contains three input files in its respective subfolder inside \matrixparam{input}, which can be modified by the user. The generic inputs in the files
\matrixparam{parameter.dat}, \matrixparam{model.dat} and \matrixparam{distribution.dat} of each \Matrix{} run are described in the following.

\subsubsection{Settings in \matrixparam{parameter.dat}}\label{sec:parameter}

All main parameters, related to the run itself or the behaviour of the code, 
are specified in the file \matrixparam{parameter.dat}. Most of them should be completely 
self-explanatory, and we will focus our discussion on the essential ones. The 
settings can be organized into certain groups and are discussed in the order 
they appear in the file \matrixparam{parameter.dat} for the sample case of $Z\gamma$ production (where applicable).

\paragraph{General run settings}\label{sec:gensettings}
\lstset{basicstyle=\scriptsize, frame=single}
{\tt
\begin{lstlisting}
process_class   =  pp-emepa+X  #  process id
E               =  6500.       #  energy per beam
coll_choice     =  1           #  (1) PP collider; (2) PPbar collider
\end{lstlisting}
}

\matrixparam{process_class}\quad A unique identifier for the process under consideration; 
it should never be touched by the user; in particular, no other process can be 
chosen at this stage. Its sole purpose is to identify which process the respective 
parameter file belongs to.

\matrixparam{E}\quad Value of the energy per beam; assumed to be identical for both initial hadrons, i.e.\ equal to half of the collider energy. Here and in what follows, all input parameters with energy dimension 
are understood in units of GeV.

\paragraph{Scale settings}\label{sec:scale}
\lstset{basicstyle=\scriptsize, frame=single}
{\tt
\begin{lstlisting}
scale_ren       =  91.1876     #  renormalization (muR) scale
scale_fact      =  91.1876     #  factorization (muF) scale
dynamic_scale   =  6           #  dynamic ren./fac. scale
                               #  0: fixed scale above
                               #  1: invariant mass (Q) of system (of the colourless final states)
                               #  2: transverse mass (mT^2=Q^2+pT^2) of the colourless system
                               #  3: transverse mass of photon (note: mT_photon=pT_photon)
                               #  4: transverse mass of Z boson (lepton system, mT_lep1+lep2)
                               #  5: geometric avarage of mT of photon and mT of Z boson
                               #  6: quadratic sum of Z mass and mT of the photon (mu^2=m_Z^2+mT_photon^2)
factor_central_scale = 1.      #  relative factor for central scale
scale_variation  = 1           #  switch for muR/muF variation (0) off; (1) 7-point; (2) 9-point
variation_factor = 2           #  symmetric scale variation factor up and down
\end{lstlisting}
}

\matrixparam{dynamic\_scale}\quad This parameter allows the user to choose between the specified fixed renormalization and factorization 
scales (\matrixparam{scale\_ren}/\matrixparam{scale\_fact}) and dynamic ones. A dynamic scale must be implemented individually for 
the process under consideration. For all processes two dynamic scales are provided by default: the invariant 
mass (\matrixparam{dynamic\_scale = 1}) and the transverse mass (\matrixparam{dynamic\_scale = 2}) of the colourless final-state system. 
The relevant file of the C++ code is \matrixparam{prc/\$\{process\_id\}/user/specify.scales.cxx} in the \Matrix{} main folder 
(recompilation needed if modified!). All additional dynamic scale choices for each process 
are discussed in \sct{sec:processes}.
A user interested in setting a specific dynamic scale which has not been implemented yet for this process
is advised to contact the authors.\footnote{A short description on how to
add user-specified scales, cuts and distributions to the C++ code is given in \app{app:advancedstuff} for the advanced user.}

\matrixparam{factor\_central\_scale}\quad A relative factor that multiplies the central scale; particularly useful for dynamic scales.

\matrixparam{variation\_factor}\quad This (integer) value determines by which factor 
with respect to the central scale the scale variation is performed.

\paragraph{Order-dependent run settings}\label{sec:paramOrder}
\lstset{basicstyle=\scriptsize, frame=single}
{\tt
\begin{lstlisting}
# LO
run_LO          =  1           #  switch for LO cross section (1) on; (0) off 
LHAPDF_LO       =  NNPDF30_lo_as_0118 #  LO LHAPDF set
PDFsubset_LO    =  0           #  member of LO PDF set
precision_LO    =  1.e-2       #  precision of LO cross section

# NLO
run_NLO         =  0           #  switch for NLO cross section (1) on; (0) off 
LHAPDF_NLO      =  NNPDF30_nlo_as_0118 #  NLO LHAPDF set
PDFsubset_NLO   =  0           #  member of NLO PDF set
precision_NLO   =  1.e-2       #  precision of NLO cross section
NLO_subtraction_method = 1     #  switch for (1) Catani-Seymour (2) qT subtraction at NLO

# NNLO
run_NNLO        =  0           #  switch for NNLO cross section (1) on; (0) off 
LHAPDF_NNLO     =  NNPDF30_nnlo_as_0118 #  NNLO LHAPDF set
PDFsubset_NNLO  =  0           #  member of NNLO PDF set
precision_NNLO  =  1.e-2       #  precision of NNLO cross section
loop_induced    =  1           #  switch to turn on (1) and off (0) loop-induced gg channel

switch_qT_accuracy = 0         #  switch to improve qT-subtraction accuracy (slower numerical convergence) 

\end{lstlisting}
}

A single run of a process in \Matrix{} involves up to three different orders (\matrixparam{\$\{order\}}), namely LO, NLO and NNLO. 
For each of these orders we may choose the following inputs:

\matrixparam{run\_\$\{order\}}\quad Switch to turn on and off the order \matrixparam{\$\{order\}} in the run.

\matrixparam{LHAPDF\_\$\{order\}}\quad LHAPDF string that determines the PDF set used at this order with the respective member \matrixparam{PDFsubset\_\$\{order\}}. 

\matrixparam{precision\_\$\{order\}}\quad Desired numerical precision of the cross section (within cuts) of this run.

\matrixparam{NLO\_subtraction\_method}\quad Switch to choose the NLO subtraction scheme: For the NLO part of the computation two 
different subtraction schemes are available. The default is the Catani--Seymour 
dipole subtraction, which comes with the advantage of being fully local and thus does not lead to any $\rcut{}$ 
dependence. The NLO computation can also be performed by means of the \qt{}-subtraction method.
The option to use both subtraction schemes in the same run is currently not supported.

\matrixparam{loop_induced}\quad For certain processes (such as \zz{}, \ww{}, \ldots{}) a loop-induced $gg$ contribution enters at the NNLO; this contribution is separately finite and 
can be included or excluded by this switch; if a process has no loop-induced $gg$ 
component, the switch is absent.

\matrixparam{switch_qT_accuracy}\quad Switch specific to processes with large $\rcut$ dependence (in particular processes with final-state photons). The lowest calculated value of \rcut{} is
changed from $\rcut=0.15$\% (\matrixparam{switch_qT_accuracy = 0}) to $\rcut=0.05$\% (\matrixparam{switch_qT_accuracy = 1}) in order to improve
the accuracy of the \qt{}-subtracted NNLO cross section, at the cost of numerical convergence. 
We refer to \sct{sec:processes} for further information.

\paragraph{Settings for fiducial cuts}
\label{sec:fiducial_cuts}

We first note that certain settings, such as photon isolation, naturally 
only affect dedicated processes. The default input files are adapted such that they only
contain options that are of relevance for the respective process.
It is not recommended to add any new blocks to the input files in order to avoid unwanted
behaviour, although such additional settings would usually just not have any impact
on the run.

\subparagraph{Jet algorithm}\quad
{\tt
\begin{lstlisting}
jet_algorithm =	3              #  (1) Cambridge-Aachen (2) kT (3) anti-kT
jet_R_definition = 0           #  (0) pseudo-rapidity (1) rapidity
jet_R = 0.4                    #  DeltaR
\end{lstlisting}
}

\matrixparam{jet\_algorithm}\quad Switch to choose between three predefined jet-clustering algorithms: 
Cambridge-Aachen \cite{Dokshitzer:1997in,Wobisch:1998wt}, $k_T$ \cite{Catani:1993hr} or anti-$k_T$ \cite{Cacciari:2008gp}.\footnote{We note that, for the processes considered in the first release of \Matrix{}, the three algorithms are actually equivalent, since the final state contains at most two partons. Also parameter \matrixparam{jet_R_definition} has no impact for final states with at 
most two partons, as the pseudo-rapidity and rapidity of massless partons 
is identical.}

\matrixparam{jet\_R\_definition}\quad According to the setting of this switch, the distance $\Delta R$ of jets is defined either via pseudo-rapidity or rapidity,
\begin{align}
\Delta R_{jj}=\sqrt{\Delta\eta_{jj}^2+\Delta\phi_{jj}^2}\qquad\textnormal{or}\qquad \Delta R_{jj}=\sqrt{\Delta y_{jj}^2+\Delta\phi_{jj}^2}\, .
\label{eq:dRjjdefinition}
\end{align}

\matrixparam{jet\_R}\quad Value of the jet radius used for the jet definition.

This sets the relevant parameters for the jet algorithm. Selection cuts on jets, including the setting for their minimal transverse momenta and maximal rapidity, are described below under the paragraph {\it Particle definition and generic cuts}.

\subparagraph{Photon isolation}\quad

For all processes involving identified final-state photons, \Matrix{} relies on the smooth-cone photon isolation procedure from \citere{Frixione:1998jh}, which works as follows:
For every cone of radius \mbox{$\delta=\sqrt{(\Delta \eta)^2+(\Delta \phi)^2}<\delta_0$} around a final-state photon, the total amount of hadronic (partonic) transverse energy $E_T$
inside the cone has to be smaller than $E_{T}^{\rm max}(\delta)$,
\begin{align}
\sum_{i=\mathrm{hadrons(partons)}}p_{T,i}\;\Theta(\delta-\delta_{i\gamma})\;\leq\;E_{T}^{\rm max}(\delta)\;=\; E_T^{\mathrm{ref}} \left(\frac{1-\cos \delta}{1- \cos \delta_0}\right)^n \qquad \forall\; \delta\;\leq\;\delta_0\, ,
\label{eq:frixione}
\end{align}
where $E_T^{\mathrm{ref}}$ is a reference transverse-momentum scale that can be chosen to be either a fraction $\epsilon_\gamma$ of the transverse momentum of the respective photon ($p_{T,\gamma}$) or a fixed value ($\pT^0$),
\begin{align}
E_T^{\mathrm{ref}}\;=\;\epsilon_\gamma \,p_{T,\gamma} \qquad\textnormal{or}\qquad E_T^{\mathrm{ref}}\;=\;\pT^0 \, .
\label{eq:frixionepTscale}
\end{align}

{\tt
\begin{lstlisting}
frixione_isolation = 1         #  switch for Frixione isolation (0) off;
                               #  (1) with frixione_epsilon, used by ATLAS;
                               #  (2) with frixione_fixed_ET_max, used by CMS
frixione_n = 1                 #  exponent of delta-term
frixione_delta_0 = 0.4         #  maximal cone size
frixione_epsilon = 0.5         #  photon momentum fraction
#frixione_fixed_ET_max = 5      #  fixed maximal pT inside cone
\end{lstlisting}
}

\matrixparam{frixione_isolation}\quad  Switch for smooth-cone photon isolation with three possible settings: turned off; using one or the other alternative in \eqn{eq:frixionepTscale}.

\matrixparam{frixione_n}\quad Value of $n$ in \eqn{eq:frixione}.

\matrixparam{frixione_delta_0}\quad Value of $\delta_0$ in \eqn{eq:frixione}.

\matrixparam{frixione_epsilon}\quad Value of $\varepsilon_\gamma$ in \eqn{eq:frixionepTscale}. Only used for \matrixparam{frixione_isolation = 1}, 
and must be commented if \matrixparam{frixione_isolation = 2}.

\matrixparam{frixione_fixed_ET_max}\quad Value of $\pT^0$ in \eqn{eq:frixionepTscale}. Only used for \matrixparam{frixione_isolation = 2}, 
and must be commented if \matrixparam{frixione_isolation = 1}.

Selection cuts on photons, including the setting for their minimal transverse momenta and maximal rapidity, are described in the following paragraph.

\renewcommand{\baselinestretch}{1.0}
\begin{table}
\begin{center}
\begin{tabular}{lcl}
\toprule
{\bf identifier}  &&   {\bf description}    \\
\midrule
\matrixparam{jet} && parton-level jets, 5 light quarks+gluons, clustered according to jet algorithm\\ [0.5ex]
\matrixparam{ljet} && light jets: same as \matrixparam{jet}, but without bottom jets\\ [0.5ex]
\matrixparam{bjet} && bottom jets: jets with a bottom charge (see main text)\\ [0.5ex]
\matrixparam{photon} && photons, isolated according to chosen smooth-cone isolation\\ [0.5ex]
\matrixparam{lep} && charged leptons, i.e.\ electrons and muons, including particles and anti-particles\\ [0.5ex]
\matrixparam{lm} && negatively charged leptons, i.e.\ electrons and muons\\ [0.5ex]
\matrixparam{lp} && positively charged leptons, i.e.\ positrons and anti-muons\\ [0.5ex]
\matrixparam{e} && electrons and positrons\\ [0.5ex]
\matrixparam{em} && electrons\\ [0.5ex]
\matrixparam{ep} && positrons\\ [0.5ex]
\matrixparam{mu} && muons and anti-muons\\ [0.5ex]
\matrixparam{mum} && muons\\ [0.5ex]
\matrixparam{mup} && anti-muons\\ [0.5ex]
\matrixparam{z} && Z bosons\\ [0.5ex]
\matrixparam{w} && $W^+$ and $W^-$ bosons\\ [0.5ex]
\matrixparam{wp} && $W^+$ bosons\\ [0.5ex]
\matrixparam{wm} && $W^-$ bosons\\ [0.5ex]
\matrixparam{h} && Higgs bosons\\ [0.5ex]
\matrixparam{nua} && neutrinos and anti-neutrinos\\ [0.5ex]
\matrixparam{nu} && neutrinos\\ [0.5ex]
\matrixparam{nux} && anti-neutrinos\\ [ 0.5ex]
\matrixparam{nea} && electron-neutrinos and anti-electron-neutrinos\\ [0.5ex]
\matrixparam{ne} && electron-neutrinos\\ [0.5ex]
\matrixparam{nex} && anti-electron-neutrinos\\ [0.5ex]
\matrixparam{nma} && muon-neutrinos and anti-muon-neutrinos\\ [0.5ex]
\matrixparam{nm} && muon-neutrinos\\ [0.5ex]
\matrixparam{nmx} && anti-muon-neutrinos\\ [0.5ex]
\matrixparam{missing} && sum of all neutrino momenta, containing only one entry (special group)\\ [0.5ex]

\bottomrule
\end{tabular}
\end{center}
\renewcommand{\baselinestretch}{1.0}
  \caption{All relevant particle groups predefined in \Matrix{}. Each group is ordered by the transverse momenta of the respective particles, 
  starting with the hardest one. These groups are most important to recognize by the user in two situations: 
  when using the predefined blocks for fiducial cuts and when defining distributions (see \sct{sec:distributions}). Furthermore, they can be accessed directly in 
  the C++ code which is essential to the advanced user when defining user-specified scales, cuts and distributions, see
  \app{app:advancedstuff}.} 
  \label{tab:predefinedparticles}
\end{table}
\renewcommand\arraystretch{1.1}

\subparagraph{Particle definition and generic cuts}\quad

Some fiducial cuts are defined in a general, i.e.\ process-independent, way by requiring a minimum and maximum 
multiplicity of a certain (group of) particle(s) with given requirements (such as minimal transverse momentum or maximal rapidity). 
For that purpose, the user can define which requirements (clustered) parton-level objects 
need to fulfil in order to be considered particles that can 
be accessed in scale definitions, phase-space cuts and distributions.
\refta{tab:predefinedparticles} summarizes the content of all relevant predefined particle groups. All objects entering these groups will be ordered by
their transverse momenta, starting with the hardest one.

The parameters \matrixparam{define_y $\{particle_group\}} and \matrixparam{define_eta $\{particle_group\}} set the geometric range for the acceptance of
particles in \matrixparam{$\{particle_group\}}, 
in terms of upper limits on the absolute value of rapidity and/or pseudo-rapidity, respectively, in the hadronic frame.
Objects that do not fulfil these requirements are discarded in the respective particle group. For example, \matrixparam{define_eta lepton = 2.5} defines all leptons in the 
respective group with a maximal absolute pseudo-rapidity of $2.5$.

The parameter \matrixparam{define_pT $\{particle_group\}} sets a threshold on the transverse momentum of particles in \matrixparam{$\{particle_group\}}. Objects below that threshold are not discarded, but they do not increase
the multiplicity counter of accepted particles in the respective \matrixparam{$\{particle_group\}}. 
They enter the respective ($\pT{}$-ordered) particle groups at the very end of the group.

Setting only the above parameters does not result in selection cuts yet.
To define requirements on the multiplicity counter of accepted 
particles of that \matrixparam{$\{particle_group\}}, 
the parameters \matrixparam{n_observed_min $\{particle_group\}}, 
and \matrixparam{n_observed_max $\{particle_group\}} 
are used:
They define how many particles of that group must be observed at least and at most, respectively, in the final state for an event to be accepted. 
No cut is applied here if the minimum and maximum requirements do not impose an actual restriction.%

These parameters are organized in blocks for each \matrixparam{\$\{particle\_group\}} in the file \matrixparam{parameter.dat} with the following general structure:

\matrixparam{define_eta $\{particle_group\}}\\
\matrixparam{define_y $\{particle_group\}}\\
\matrixparam{define_pT $\{particle_group\}}\\
\matrixparam{n_observed_min $\{particle_group\}}\\
\matrixparam{n_observed_max $\{particle_group\}}
\vspace{0.1cm}

Such blocks are predefined for the relevant particle groups of each process 
in the respective file \matrixparam{parameter.dat}. They should be sufficient 
for most practical purposes, and it is generally recommended to stick to the 
predefined blocks. Nevertheless, it is possible to add additional blocks 
also for the other particle groups using the structure above. In this case, care 
has to be taken to avoid unwanted behaviour. 
In particular, requiring a certain number of particles which actually do not exist in the final state of a given process must be avoided.
Below, we provide examples of the respective 
blocks available in various processes.

\subparagraph{Jet cuts}\quad
\lstset{basicstyle=\scriptsize, frame=single}
{\tt
\begin{lstlisting}
define_pT jet =	30.            #  requirement on jet transverse momentum (lower cut)
define_eta jet = 4.4           #  requirement on jet pseudo-rapidity (upper cut)
define_y jet = 1.e99           #  requirement on jet rapidity (upper cut)
n_observed_min jet = 0         #  minimal number of observed jets (with cuts above)
n_observed_max jet = 99	       #  maximal number of observed jets (with cuts above)
\end{lstlisting}
}

This defines the particle group \matrixparam{jet} with a minimal transverse momentum of $30$\,GeV and a maximal absolute rapidity of $4.4$, 
using the jet-clustering algorithm specified above. No phase-space cut is effective, since this process has a maximum of two jets 
in the final state at NNLO, and neither a minimal ($>0$) nor a maximal number ($<2$) of observed jets is required.\footnote{Note that setting \matrixparam{n_observed_min jet = 1} would effectively reduce any (N)NLO calculation for the production of a final state $F$ in \Matrix{} to be only a (N)LO
accurate calculation for the production of $F$+jet. On the other hand, setting \matrixparam{n_observed_max jet = 0} would impose a veto against events that contain any jets that fulfil the defined requirements on jets.}
 However, the particle group \matrixparam{jet} with 
the requirements defined here can be accessed in the definition of distributions, see \sct{sec:distributions}.\footnote{Accordingly, the defined particle group 
is also accessible within the C++ code as discussed for the definition of new dynamic scales, cuts and observables for distributions by the advanced user in \app{app:advancedstuff}.}

Analogous blocks can be processed by \Matrix{} for the particle groups \matrixparam{bjet} and \matrixparam{ljet}, 
which denote bottom jets and light jets (i.e.\ all jets, but the bottom jets), 
respectively. Note that a computation with bottom quarks treated as massless requires a jet 
involving a $b\bar{b}$ pair from a $g\to b\bar{b}$ splitting to be considered a light jet, 
but not a bottom-jet, to guarantee observables to be IR safe.

\subparagraph{Lepton cuts}\quad
{\tt
\begin{lstlisting}
define_pT lep = 25.            #  requirement on lepton transverse momentum (lower cut)
define_eta lep = 2.47          #  requirement on lepton pseudo-rapidity (upper cut)
define_y lep = 1.e99           #  requirement on lepton rapidity (upper cut)
n_observed_min lep = 2         #  minimal number of observed leptons (with cuts above)
n_observed_max lep = 99        #  maximal number of observed leptons (with cuts above)
\end{lstlisting}
}
This block defines each lepton in the particle group \matrixparam{lep} to 
have a minimal transverse momentum of 25\,GeV and a maximal absolute pseudo-rapidity 
of 2.47. It further requires the presence of at least two such leptons. All 
events not passing this criterion are discarded from the fiducial 
phase space.\footnote{We stress again that any lepton in the 
particle group \matrixparam{lep} fulfils the defined (rapidity) requirements, irrespective 
of whether \matrixparam{n_observed_min lep} or \matrixparam{n_observed_max lep} 
require the presence of a minimal or maximal number of such leptons in the event. 
This is important to 
bear in mind when using \matrixparam{lep} to define distributions 
in \sct{sec:distributions}. Even in a fully inclusive phase space without 
fiducial cuts, any distribution using \matrixparam{lep} will be affected 
by the defined (rapidity) requirements in the file \matrixparam{parameter.dat} on the leptons.
Of course, the equivalent is true for any of the other particle groups.}

Analogous blocks are available for other particle groups of charged leptons,
namely \matrixparam{lm}, \matrixparam{lp}, \matrixparam{e}, \matrixparam{mu}, \matrixparam{em}, \matrixparam{ep}, \matrixparam{mum} and \matrixparam{mup}.

\subparagraph{Photon cuts}\quad
{\tt
\begin{lstlisting}
define_pT photon = 15.         #  requirement on photon transverse momentum (lower cut)
define_eta photon = 2.37       #  requirement on photon pseudo-rapidity (upper cut)
define_y photon = 1.e99        #  requirement on photon rapidity (upper cut)
n_observed_min photon = 1      #  minimal number of observed photons (with cuts above)
n_observed_max photon = 99     #  maximal number of observed photons (with cuts above)
\end{lstlisting}
}

Similarly, due to this block the photons in the particle group \matrixparam{photon}, 
which have passed the isolation criterion defined above, have a transverse momentum greater 
than $15$\,GeV and absolute pseudo-rapidity smaller than $2.37$, and the presence of 
at least one such isolated photon is required. Note that for the cross section to 
be IR finite, the number of identified photons in the final state must be equal to 
the total number of photons in the final state of a process.

\subparagraph{Heavy-boson cuts}\quad
{\tt
\begin{lstlisting}
define_pT w = 0.               #  requirement on W-boson transverse momentum (lower cut)
define_eta w = 1.e99           #  requirement on W-boson pseudo-rapidity (upper cut)
define_y w = 1.e99             #  requirement on W-boson rapidity (upper cut)
n_observed_min w = 0           #  minimal number of observed W-bosons (with cuts above)
n_observed_max w = 99          #  maximal number of observed W-bosons (with cuts above)
\end{lstlisting}
}
Equivalent blocks are available for the particle groups of 
heavy bosons, namely \matrixparam{w}, \matrixparam{wm}, \matrixparam{wp}, \matrixparam{z} and \matrixparam{h}. The above example does not impose any requirements on $W$ bosons, 
as needed for a fully inclusive cross section.

\subparagraph{Neutrino cuts}\quad
{\tt
\begin{lstlisting}
define_pT missing = 30.        #  requirement on pT of sum of all neutrinos (lower cut)
\end{lstlisting}
}
The particle group \matrixparam{missing} contains only the missing energy vector, given by the sum of all neutrino momenta.
In processes with neutrinos this particle group can be used to impose
a minimum requirement on the total missing transverse momentum in the event. The example above sets $p_T^{\rm miss}>30$\,GeV.

In particular for technical checks it might be useful to access neutrinos also as 
individual particles. To do so, \Matrix{} can process blocks for the particle groups 
\matrixparam{nua},
\matrixparam{nu}, \matrixparam{nux}, 
\matrixparam{nea}, \matrixparam{nma}, 
\matrixparam{ne}, \matrixparam{nex}, \matrixparam{nm} and \matrixparam{nmx}.

\subparagraph{Process-specific cuts}\quad

A number of cuts are defined individually for each process. They enable a realistic definition of fiducial phase spaces as
used in experimental measurements. For every process-specific cut there is usually one integer-valued switch
(\matrixparam{user\_switch}) to either turn on and off a certain cut or to choose between different options. Moreover, each switch typically comes with one or more 
real-valued parameters (\matrixparam{user\_cut}) which are only active if the respective switch is turned on. There are a number of predefined 
process-specific cuts for each process, all of which are defined directly inside the C++ code in the file 
\Matrixversion{}\matrixparam{/prc/\$\{process\_id\}/user/specify.cuts.cxx}; the list of predefined (process-specific) cuts for each 
process is documented in \refse{sec:processes}. A user interested in setting a specific cut which has not been implemented yet for a 
certain process is advised to contact the authors.\footnote{A short description on how to
add user-specified scales, cuts and distributions to the C++ code is given in \app{app:advancedstuff} for the advanced user.}

For $Z\gamma$ production, e.g., the following predefined cuts are accessible in the file \matrixparam{parameter.dat}.
{\tt
\begin{lstlisting}
user_switch M_leplep = 1       #  switch to turn on (1) and off (0) cuts on lepton-lepton invariant mass
user_cut min_M_leplep = 40.    #  requirement on lepton-lepton invariant mass (lower cut)
user_cut max_M_leplep = 1.e99  #  requirement on lepton-lepton invariant mass (upper cut)

user_switch M_lepgam = 0       #  switch to turn on (1) and off (0) cuts on lepton-photon invariant mass
user_cut min_M_lepgam = 40.    #  requirement on lepton-photon invariant mass (lower cut)

user_switch R_leplep = 0       #  switch to turn on (1) and off (0) cuts on lepton-lepton separation
user_cut min_R_leplep = 0.5    #  requirement on lepton-photon separation in y-phi-plane (lower cut)

user_switch R_lepgam = 1       #  switch to turn on (1) and off (0) cuts on lepton-photon separation
user_cut min_R_lepgam = 0.7    #  requirement on lepton-photon separation in y-phi-plane (lower cut)

user_switch R_lepjet = 1       #  switch to turn on (1) and off (0) cuts on lepton-jet separation
user_cut min_R_lepjet = 0.3    #  requirement on lepton-jet separation in y-phi-plane (lower cut)

user_switch R_gamjet = 1       #  switch to turn on (1) and off (0) cuts on photon-jet separation
user_cut min_R_gamjet = 0.3    #  requirement on photon-jet separation in y-phi-plane (lower cut)

user_switch pT_lep_1st = 0     #  switch to turn on (1) and off (0) cuts on pT of hardest lepton
user_cut min_pT_lep_1st = 25   #  requirement on pT of hardest lepton (lower cut)
\end{lstlisting}
}
They should be rather self-explanatory and enable standard invariant-mass and $R=\sqrt{y^2+\phi^2}$-separation cuts 
on the final-state leptons, photons and jets, as well as a lower transverse-momentum cut on the hardest lepton.

\paragraph{M{\footnotesize ATRIX} behaviour}

\lstset{basicstyle=\scriptsize, frame=single}
{\tt
\begin{lstlisting}
max_time_per_job = 12          #  very rough time (in hours) one main-run job shall take
switch_distribution = 1        #  switch to turn on (1) and off (0) distributions
save_previous_result = 1       #  switch to save previous result if rerun
save_previous_log = 0          #  switch to save previous log if rerun
#include_pre_in_results = 0     #  switch to include (0) main-run (1) main+pre-run in results
reduce_workload = 0            #  switch to keep full output (0) or reduce the workload (1)
random_seed = 0                #  specify integer value between 0-100 (grid-/pre-run reproducible)
\end{lstlisting}
}

\matrixparam{max\_time\_per\_job}\quad Essential (real-valued) parameter to control the parallelization 
of the jobs in the \mainrun{} (the \gridrun{} and \premainrun{} are unaffected, i.e.\ they will always run the same number of jobs). The given value sets a very rough requirement on the time (in hours) a 
single job in the \mainrun{} may take. It should be regarded as a tuning parameter rather than an 
exact measure; the actual runtime of the jobs may deviate significantly 
(factor of $\sim [0.5,2]$) in certain cases. Together with the precision that can be 
set individually for each order (see \sct{sec:scale}) \matrixparam{max\_time\_per\_job} 
determines the level of parallelization; clearly, the higher the precision (with constant 
\matrixparam{max\_time\_per\_job}), the higher the level of parallelization.
One must bear in mind that too small values of \matrixparam{max\_time\_per\_job} (below $
\sim 1$ hour for a NNLO run) become unreliable, i.e.\ the jobs would take significantly longer
than specified in that case. For heavy NNLO runs 
($\lesssim 0.1\%$ precision for one of the most complicated processes) we recommend 
not to use values $\lesssim 5$\,hours, as too small values lead to a huge 
parallelization which may have a negative effect on the result combination.
Also note that this parameter becomes ineffective as soon as the number of jobs 
is larger than \matrixparam{max\_nr\_parallel\_jobs}, which can be set in the file \matrixparam{MATRIX\_configuration} (see \sct{sec:config}), or the number of cores 
in local mode. 

\matrixparam{switch\_distribution}\quad Switch to control whether distributions are 
generated during the run.

\matrixparam{save\_previous\_result}\quad This switch is effective when rerunning in a run folder which already contained a full run 
including results. If the switch is turned on, the previous results are saved into a subfolder \matrixparam{saved\_result\_XX} of the result folder for 
the respective run, where \matrixparam{XX} is an increasing number starting at \matrixparam{01} for each time an old result is saved; default: turned on.

\matrixparam{save\_previous\_log}\quad This switch is effective when rerunning in a run folder which already contained a run with written log files. If the switch is turned on, the previous log files are saved into a subfolder \matrixparam{saved\_log\_XX} of the log folder for the respective run, where \matrixparam{XX} is an increasing number starting at \matrixparam{01} for each time old log files are saved; default: turned off.

\matrixparam{include\_pre\_in\_results}\quad This switch affects the result combination. It allows the user to include/exclude the 
results of the \premainrun{} 
into/from the \resultrun{}, 
which always includes the \mainrun{}. 
If the switch is absent, i.e.\ commented (default), this decision is made 
internally in the \Matrix{} code independently for each contribution by a certain 
algorithm which is designed to optimize the total precision, while excluding 
irrelevant low-statistic runs of the \premainrun{} phase.
Excluding the \premainrun{} from the \resultrun{} is particularly useful if  
the \mainrun{} is restarted with a slightly modified setup, in order to avoid mixing of the two setups.

\matrixparam{reduce_workload}\quad Switch to reduce the output of the jobs to a minimum. May be used to improve the speed 
on clusters with slow access to the file system.

\matrixparam{random_seed}\quad Sets starting seed for run. {\it grid-} and \premainrun{} for same seed are reproducible.

\subsubsection{Settings in \matrixparam{model.dat}}\label{sec:model}
All model-related parameters are set in the file \matrixparam{model.dat}. We adopt 
the \susy{} Les Houches accord (\slha{}) format \cite{Skands:2003cj}. This standard 
format is used in many codes and thus simplifies the settings of common 
model parameters. In the \slha{} format inputs are organized in blocks which have different entries characterized by a number. For 
simplicity, we introduce the following short-hand notation: \blockentry{example}{i} corresponds to entry $i$ in \block{example}. 
For example, entry 25 of \block{mass} (\blockentry{mass}{25}) in the \slha{} format corresponds to the Higgs mass in the \sm{}, which 
is required as an input in the file \matrixparam{model.dat}. Only the format for decay widths is slightly different and not organized in a \matrixparam{Block}, but defined by the keyword \matrixparam{DECAY}, followed by a number which specifies the respective particle. A typical model file is shown below.

\lstset{basicstyle=\scriptsize, frame=single}
{\tt
\begin{lstlisting}
##########################
# MATRIX model parameter #
##########################

#--------\
# masses |
#--------/
Block MASS
  1 0.000000     # M_d 
  2 0.000000     # M_u
  3 0.000000     # M_s
  4 0.000000     # M_c
  5 0.000000     # M_b 
  6 1.732000e+02 # M_t
 11 0.000000     # M_e 
 12 0.000000     # M_ve
 13 0.000000     # M_mu
 14 0.000000     # M_vm
 15 1.777000e+00 # M_tau
 16 0.000000     # M_vt
 23 9.118760e+01 # M_Z 
 24 8.038500e+01 # M_W
 25 1.250000e+02 # M_H 

#-------------------\
# inputs for the SM |
#-------------------/
Block SMINPUTS
  2 1.166390e-05 # G_F 

#------------------\
# Yukawa couplings |
#------------------/
#Block YUKAWA 
#  5 4.750000e+00 # M_YB 
#  6 1.730000e+02 # M_YT 
# 15 1.777000e+00 # M_YTAU 

#---------------\
# decays widths |
#---------------/
DECAY  6 1.442620e+00 # WT 
DECAY 23 2.495200e+00 # WZ 
DECAY 24 2.085400e+00 # WW 
DECAY 25 4.070000e-03 # WH
\end{lstlisting}
}
The \matrixparam{Block} \matrixparam{Yukawa} is currently not used, which is why it is commented. 

In the first release of \Matrix{}, only on- and off-shell $W$-boson 
production allow for a non-trivial CKM matrix. This feature will be added also for other 
processes like $W\gamma$ and \wz{} production in a future update.
\label{sec:CKM}
The CKM parameters are controlled in the file \matrixparam{model.dat} of these processes through additional \matrixparam{Block}s. 
The user may choose between three different setups. The default is a complete CKM matrix, where each of the entries may be set 
individually using \block{CKM} as defined below.

\lstset{basicstyle=\scriptsize, frame=single}
{\tt
\begin{lstlisting}                                                                                                                                                                                                                                                                    
#------------\                                                                                                                                                                                                                                                                             
# CKM matrix |                                                                                                                                                                                                                                                                             
#------------/                                                                                                                                                                                                                                                                             
Block CKM                                                                                                                                                                                                                                                                                  
 11 0.974170e+00 # V_ud                                                                                                                                                                                                                                                                    
 12 0.224800e+00 # V_us                                                                                                                                                                                                                                                                    
 13 0.004090e+00 # V_ub                                                                                                                                                                                                                                                                    
 21 0.220000e+00 # V_cd                                                                                                                                                                                                                                                                    
 22 0.995000e+00 # V_cs                                                                                                                                                                                                                                                                    
 23 0.040500e+00 # V_cb                                                                                                                                                                                                                                                                    
 31 0.008200e+00 # V_td                                                                                                                                                                                                                                                                    
 32 0.040000e+00 # V_ts                                                                                                                                                                                                                                                                    
 33 1.009000e+00 # V_tb                                                                                                                                                                                                                                                                    
\end{lstlisting}
}

The default values are chosen according to the SM CKM matrix as reported by the PDG in \citere{Olive:2016xmw}. Note that any 
top-related CKM entry has no effect on the processes considered in \Matrix{}.

A second option to use a non-trivial CKM matrix is through the Cabibbo angle $\theta_c$, by adding the \block{VCKMIN} as follows:

\lstset{basicstyle=\scriptsize, frame=single}
{\tt
\begin{lstlisting}                                                                                                                                                                                                                                                                    
#---------------\                                                                                                                                                                                                                                                                          
# Cabibbo angle |                                                                                                                                                                                                                                                                          
#---------------/                                                                                                                                                                                                                                                                          
Block VCKMIN                                                                                                                                                                                                                                                                              
  1 0.227000e+00 # Cabibbo angle  
\end{lstlisting}
}

This enables mixing only between the first two generations, while turning off any mixing with the third generation, i.e.\ by setting internally 
$V_{ud}=\cos(\theta_c)$, $V_{us}=\sin(\theta_c)$, $V_{cd}=-\sin(\theta_c)$, $V_{cs}=\cos(\theta_c)$, $V_{tb}=1$, and $V_{ub}=V_{cb}=V_{td}=V_{ts}=0$.
Note that only \block{CKM} or \block{VKCMIN} may be present in the file \matrixparam{model.dat} at the same time.

Finally, if both blocks are absent, a trivial CKM matrix (no mixing) is used.

\subsubsection{Settings in \matrixparam{distribution.dat}}
\label{sec:distributions}

\renewcommand\arraystretch{1.5}
\begin{table}
\begin{center}
\resizebox{\columnwidth}{!}{%
\begin{tabular}{lclcp{.70\textwidth}}
\toprule
{\bf identifier} && {\bf binned variable}  && {\bf description}     \\
\midrule
\matrixparam{pT} && $\sum_{j=1}^m p^j_{T}$ && scalar sum of transverse momenta of \matrixparam{particle 1} to \matrixparam{particle m} \\
\matrixparam{m} && $m(p^1)$ && invariant mass of \matrixparam{particle 1}\\
\matrixparam{dm} && $m(p^1)-m(p^2)$ && invariant-mass difference between \matrixparam{particle 1} and \matrixparam{particle 2}\\
\matrixparam{absdm} && $\left|m(p^1)-m(p^2)\right|$ && absolute invariant-mass difference between \matrixparam{particle 1} and \matrixparam{particle 2}\\
\matrixparam{mmin} && $\min\left(m(p^1),m(p^2)\right)$ && minimal invariant-mass of \matrixparam{particle 1} and \matrixparam{particle 2}\\
\matrixparam{mmax} && $\max\left(m(p^1),m(p^2)\right)$ && maximal invariant-mass of \matrixparam{particle 1} and \matrixparam{particle 2}\\
\matrixparam{y} && $y(p^1)$ && rapidity of \matrixparam{particle 1}\\
\matrixparam{absy} && $\left|y(p^1)\right|$ && absolute rapidity of \matrixparam{particle 1}\\
\matrixparam{dy} && $y(p^1)-y(p^2)$ && rapidity difference between \matrixparam{particle 1} and \matrixparam{particle 2}\\
\matrixparam{absdy} && $\left|y(p^1)-y(p^2)\right|$ && absolute rapidity difference between \matrixparam{particle 1} and \matrixparam{particle 2}\\
\matrixparam{dabsy} && $\left|y(p^1)\right|-\left|y(p^2)\right|$ && difference between absolute rapidities of \matrixparam{particle 1} and \matrixparam{particle 2}\\
\matrixparam{absdabsy} && $\left|\left|y(p^1)\right|-\left|y(p^2)\right|\right|$ && absolute difference between absolute rapidities of \matrixparam{particle 1} and \matrixparam{particle 2}\\
\matrixparam{eta} && $\eta(p^1)$ && pseudo-rapidity of \matrixparam{particle 1}\\
\matrixparam{abseta} && $\left|\eta(p^1)\right|$ && absolute pseudo-rapidity of \matrixparam{particle 1}\\
\matrixparam{deta} && $\eta(p^1)-\eta(p^2)$ && pseudo-rapidity difference between \matrixparam{particle 1} and \matrixparam{particle 2}\\
\matrixparam{absdeta} && $\left|\eta(p^1)-\eta(p^2)\right|$ && absolute pseudo-rapidity difference between \matrixparam{particle 1} and \matrixparam{particle 2}\\
\matrixparam{dabseta} && $\left|\eta(p^1)\right|-\left|\eta(p^2)\right|$ && difference between absolute pseudo-rapidities of \matrixparam{particle 1} and \matrixparam{particle 2}\\
\matrixparam{absdabseta} && $\left|\left|\eta(p^1)\right|-\left|\eta(p^2)\right|\right|$ && absolute difference between absolute pseudo-rapidities of \matrixparam{particle 1} and \matrixparam{particle 2}\\
\matrixparam{phi} && $\phi(p^1)$  && azimuthal angle of \matrixparam{particle 1}\\
\matrixparam{phi} && $\Delta\phi(p^1,p^2)$  && difference in azimuthal angle between \matrixparam{particle 1} and \matrixparam{particle 2}\\
\matrixparam{dR} && $\sqrt{\left[\Delta y(p^1,p^2)\right]^2+\left[\Delta\phi(p^1,p^2)\right]^2}$  && distance in $y$-$\phi$-plane between \matrixparam{particle 1} and \matrixparam{particle 2}\\
\matrixparam{dReta} && $\sqrt{\left[\Delta \eta(p^1,p^2)\right]^2+\left[\Delta\phi(p^1,p^2)\right]^2}$  && distance in $\eta$-$\phi$-plane between \matrixparam{particle 1} and \matrixparam{particle 2}\\
\matrixparam{ET} && $\sum_{j=1}^m E_T(p^j) \equiv\sum_{j=1}^m \sqrt{\left[m(p^j)\right]^2+\big[p^j_T\big]^2}$  && scalar sum of transverse masses of \matrixparam{particle 1} to \matrixparam{particle m}\\
\matrixparam{mT} && $E_T(p^1)$ && transverse mass of \matrixparam{particle 1} \\
\matrixparam{mT} && $\sqrt{ \left[\sum_{j=1}^m E_T(p^j)\right]^2 - \left[p_T\left(\sum_{j=1}^m p^j\right)\right]^2}$ && transverse mass, defined with all neutrinos in \matrixparam{particle 1} and all other particles in \matrixparam{particle 2} to \matrixparam{particle m}\\
\matrixparam{pTveto} && $\sigma(p^1_T<p_{T,\rm veto})$ && cumulative cross section with a veto on \pt{} of \matrixparam{particle 1} as a function of $p_{T,\rm veto}$\\
\matrixparam{multiplicity} && $N$ && distribution in number of identified objects of type \matrixparam{particle 1}\\
\matrixparam{muR} && $\mu_R$ &&    distribution in renormalization scale (no \matrixparam{particle j} definition)\\
\matrixparam{muF} && $\mu_F$ &&    distribution in factorization scale (no \matrixparam{particle j} definition)\\
\bottomrule
\end{tabular}
}
\end{center}
\renewcommand{\baselinestretch}{1.0}
\caption{Predefined distributions available in \Matrix{}. 
These distributions can be used in a distribution block of the file \matrixparam{distribution.dat} and require to specify the parameter(s) \matrixparam{particle j}, $j=1,\ldots ,m$ ($m\ge 1$). 
Some observables behave differently for a different number of defined particles $m$: In these cases the respective options are given in separate rows.
We use the shorthand notation $p^j$ for the momentum of \matrixparam{particle j}.
Note that, if \mbox{\matrixparam{particle j}} has several entries of particles, $p^j$ is the sum of their momenta, which provides a straightforward way to access 
trivially reconstructed particles (e.g.\ a single $Z$ boson from its decay leptons).}
\label{tab:predefineddistributions}
\end{table}
\renewcommand\arraystretch{1.1}

\paragraph{General structure}

In the file \matrixparam{distribution.dat} the user can define histograms for distributions which are filled during the run. Each distribution is 
represented by one block containing the following parameters:

\matrixparam{distributionname}\quad Unique user-defined label (string) of the distribution for identification at the end of the run; every \matrixparam{distributionname} starts a new block. Code will stop if the same distribution identifier is used twice.

\matrixparam{distributiontype}\quad Type identifier (string) of the observable to be binned. \Matrix{} has a number of predefined 
observables, which are summarized in \tab{tab:predefineddistributions}.  A user interested in a specific distribution which 
has not been implemented yet is advised to contact the authors.\footnote{A short description on how to
add user-specified scales, cuts and distributions to the C++ code is given in \app{app:advancedstuff} for the advanced user.}

\matrixparam{particle j}\quad Specification of particles entering the definition of the observable to be binned. Several final-states particles may 
be grouped into one \matrixparam{particle}. The general form is as follows:
\lstset{basicstyle=\scriptsize, frame=single}
{\tt
\begin{lstlisting}
particle 1 = ${particle_group_1} ${position_in_pT_ordering_1}
particle 1 = ${particle_group_2} ${position_in_pT_ordering_2}
particle 1 = ${particle_group_3} ${position_in_pT_ordering_3}
...
particle 2 = ${particle_group_4} ${position_in_pT_ordering_4}
particle 2 = ${particle_group_5} ${position_in_pT_ordering_5}
particle 2 = ${particle_group_6} ${position_in_pT_ordering_6}
...
particle 3 = ${particle_group_7} ${position_in_pT_ordering_7}
particle 3 = ${particle_group_8} ${position_in_pT_ordering_8}
particle 3 = ${particle_group_9} ${position_in_pT_ordering_9}
...
\end{lstlisting}
}
Each \matrixparam{\$\{particle\_group\_i\}} is given by one of the particle groups defined in \tab{tab:predefinedparticles}, and 
\matrixparam{\$\{position\_in\_pT\_ordering\_i\}} is an integer which determines the desired position in the \pt{}-ordering of the respective group.
For instance, \matrixparam{lep 2} corresponds to the second-hardest lepton in the final state.
If \matrixparam{particle j} has several entries, the respective $4$-momenta are summed to define the momentum 
of \matrixparam{particle j}.\footnote{This provides a simple way to access distributions of combined particles, such as 
a $Z$ boson determined by its two decay leptons. We note that combined (reconstructed) particles are defined for certain processes (see, e.g., \sct{sec:ZZSF}) as additional 
particle groups via user-defined particles. This is particularly useful if the definition of such particle requires a certain pairing prescription, 
e.g.\ the reconstruction of a $Z$ boson in a same-flavour channel with more than two leptons. An advanced user may use this concept to define his own particle groups, see Appendix \app{app:particlegroup}.} How many \matrixparam{particle}s ($j=1,2,3,...$) are allowed or required depends on the observable under consideration. Many observables 
use only one \matrixparam{particle} entry, i.e.\ only \matrixparam{particle 1}, others that determine the distance or angle between two particles require 
two \matrixparam{particle}s, i.e.\ \matrixparam{particle 1} and \matrixparam{particle 2}. \tab{tab:predefineddistributions} specifies this behaviour for each 
of the predefined observables.

\matrixparam{binning\_type}\quad Defines how the binning is performed. It may be set to 
\matrixparam{linear}, \matrixparam{logarithmic} or \matrixparam{irregular} (if not specified, \matrixparam{linear} is used as default):
\vspace{-0.5cm}\begin{itemize}[leftmargin=*]
\setlength\itemsep{-0.3em}
\item[--] The setting \matrixparam{linear} requires the definition of three inputs out of \matrixparam{startpoint}, \matrixparam{endpoint}, 
\matrixparam{binnumber} and \matrixparam{binwidth}. The fourth one is uniquely defined then. Defining all four parameters 
results in a stop of the C++ code if they are inconsistent.
\item[--] The setting \matrixparam{logarithmic} requires the definition of \matrixparam{startpoint}, \matrixparam{endpoint} and \matrixparam{binnumber}.
The widths of the resulting bins are determined equidistantly on a logarithmic scale from this input.
\item[--] The setting \matrixparam{irregular} facilitates the definition of an arbitrary (not necessarily equidistant) binning, which is specified by the input parameter
\matrixparam{edges}.
\end{itemize}
\vspace{-0.32cm}
\matrixparam{startpoint}\quad Left endpoint of the first bin (real number).

\matrixparam{endpoint}\quad Right endpoint of the last bin (real number).

\matrixparam{binnumber}\quad Number of bins in the histogram (integer).

\matrixparam{binwidth}\quad Width of each bin in the histogram (real number).

\matrixparam{edges}\quad Edges (real numbers) of an irregular histogram, specified by $a_0:a_1:\dots:a_n$ for $n$ bins.

\paragraph{Examples}

We give a few examples on how proper distributions may be defined for the sample process of $Z\gamma$ production 
(examples can be found also in the file \matrixparam{distribution.dat} of each process).

\begin{itemize}
\item Transverse momentum of the hardest lepton, regularly binned in 200 bins from $0-1000$\,GeV (i.e.\ in $5$\,GeV steps):
\lstset{basicstyle=\scriptsize, frame=single}
{\tt
\begin{lstlisting}
distributionname  =  pT_lep1
distributiontype  =  pT
particle 1        =  lep 1
startpoint        =  0.
endpoint          =  1000.
binnumber         =  200 
\end{lstlisting}
}
\item Transverse momentum of the second-hardest lepton, regularly binned from $0-1000$\,GeV in $5$\,GeV steps (i.e.\ in 200 bins):
\lstset{basicstyle=\scriptsize, frame=single}
{\tt
\begin{lstlisting}
distributionname  =  pT_lep2
distributiontype  =  pT
particle 1        =  lep 2
startpoint        =  0.
endpoint          =  1000.
binwidth          =  5. 
\end{lstlisting}
}
\item Transverse momentum of the hardest photon with irregular edges (as used by ATLAS in the 7\,TeV analysis for $Z\gamma$ \cite{Aad:2013izg}):
\lstset{basicstyle=\scriptsize, frame=single}
{\tt
\begin{lstlisting}
distributionname  =  pT_gamma_ATLAS
distributiontype  =  pT
particle 1        =  photon  1
binningtype       =  irregular
edges             =  0.:15.:20.:30.:40.:60.:100.:1000.:3500.
\end{lstlisting}
}
\item Invariant mass of the pair formed by the hardest and the second-hardest lepton, binned from $0-1000$\,GeV in $10$\,GeV steps:
\lstset{basicstyle=\scriptsize, frame=single}
{\tt
\begin{lstlisting}
distributionname  =  m_lep1_lep2
distributiontype  =  m
particle 1        =  lep 1
particle 1        =  lep 2
startpoint        =  0.
endpoint          =  1000.
binwidth          =  10.
\end{lstlisting}
}
\item Distance in $y$--$\phi$ plane between the hardest electron and the hardest positron, binned from $0-10$ in $0.1$\,steps:
\lstset{basicstyle=\scriptsize, frame=single}
{\tt
\begin{lstlisting}
distributionname  =  dR_em1_ep1
distributiontype  =  dR
particle 1        =  em 1
particle 2        =  ep 1
startpoint        =  0.
endpoint          =  10.
binwidth          =  0.1
\end{lstlisting}
}

\end{itemize}

The default file \matrixparam{distribution.dat} contains further examples and information, as well as instructions on how 
to define distributions in this format.

\subsection{Process-specific settings}
\label{sec:processes}

In this Section we provide information specific to the individual processes. Below we list all processes available in \Matrix{} by their
respective \matrixparam{\$\{process\_id\}}, summarize the predefined process-specific cuts and dynamic 
scales, and, where applicable, we give additional process-specific information.

In addition to the standard cuts on particle groups, discussed in \sct{sec:fiducial_cuts}, process-specific fiducial cuts are 
predefined via an integer-valued parameter \ttt{user_switch} 
in combination with none, one or more real-valued parameters \ttt{user_cut}. For a \ttt{user_switch XXX} 
together with corresponding \ttt{user_cut XXX_A},  \ttt{user_cut XXX_B}, and so forth, we adopt the notation

\ttt{XXX: XXX_A, XXX_B, ...}

to list all available predefined cuts in the respective file \matrixparam{parameter.dat} of each process. A detailed explanation for each of these cuts is 
given in \app{app:cuts}.\footnote{The links embedded for each cut in this Section can be used to jump to the corresponding explanation in \app{app:cuts}, if supported by the PDF viewer in use.}

As outlined in \sct{sec:scale}, dynamic scales are set by the switch 
\matrixparam{dynamic_scale} in the file \matrixparam{parameter.dat}, and there are two default scales for all processes: the 
invariant and the transverse mass of the colourless system. Any additional predefined scale implemented for a process is stated below, 
and the adopted nomenclature is summarized in \refta{tab:symbols}.
\renewcommand\arraystretch{1.1}

\begin{table}
\resizebox{\columnwidth}{!}{%
\begin{tabular}{ll}
\toprule

$m_Z$: & mass of the $Z$ boson\\
$m_W$: & mass of the $W$ boson\\
$p_{T,e^-e^+}$: & transverse momentum of the reconstructed $Z$ boson (electron pair)\\
$p_{T,\mu^-\mu^+}$: & transverse momentum of the reconstructed $Z$ boson (muon pair)\\
$p_{T,\nu_\mu\bar\nu_\mu}$: & transverse momentum of the reconstructed $Z$ boson (neutrino pair)\\
$p_{T,Z_{\rm rec}}$: & transverse momentum of the reconstructed $Z$ boson (see main text)\\
$p_{T,Z_{i,\rm rec}}$: & transverse momentum of the respective reconstructed $Z$ boson (see main text)\\
$p_{T,e^\pm\nu_e}$: & transverse momentum of the reconstructed $W$ boson (electron--neutrino pair)\\
$p_{T,\mu^\pm\nu_\mu}$: & transverse momentum of the reconstructed $W$ boson (muon--neutrino pair)\\
$p_{T,W^\pm_{\rm rec}}$: & transverse momentum of the reconstructed $W$ boson (see main text)\\
$m_{T,e^-e^+}$: & transverse mass of the reconstructed $Z$ boson (electron pair)\\
$m_{T,\mu^-\mu^+}$: & transverse mass of the reconstructed $Z$ boson (muon pair)\\
$m_{T,\nu_e\bar\nu_e}$: & transverse mass of the reconstructed $Z$ boson (neutrino pair)\\
$m_{Z_{\rm rec}}$: & transverse mass of the reconstructed $Z$ boson (see main text)\\
$m_{T,Z_{i,\rm rec}}$: & transverse mass of the respective reconstructed $Z$ boson (see main text)\\
$m_{T,e^\pm\nu_e}$: & transverse mass of the reconstructed $W$ boson (electron--neutrino pair)\\
$m_{T,\mu^\pm\nu_\mu}$: & transverse mass of the reconstructed $W$ boson (muon--neutrino pair)\\
$m_{T,W^\pm_{\rm rec}}$: & transverse mass of the reconstructed $W$ boson (see main text)\\
$m_{T,\gamma}(p_{T,\gamma})$: & transverse mass (momentum) of the photon\\
\bottomrule
\end{tabular}}
\caption{\label{tab:symbols} Symbols used in the definition of dynamic scales throughout this Section.}
\end{table}

We note that all leptons are considered massless throughout all computations. This implies that, e.g., electrons may be considered as 
muons and vice versa in order to get results for other lepton flavours. Thus, a process like $\ppppbar\to e^-e^+$ is fully equivalent to $\ppppbar\to \mu^-\mu^+$, and only the former is provided in \Matrix{}. 
The same holds for more involved processes such as $\ppppbar\to e^- \mu^- e^+ \bar\nu_\mu$ and $\ppppbar\to \mu^- e^- \mu^+ \bar\nu_e$ if the 
cuts do not depend on the lepton flavour. Since we provide only the $\ppppbar\to e^- \mu^- e^+ \bar\nu_\mu$ channel, 
$\ppppbar\to \mu^- e^- \mu^+ \bar\nu_e$ for 
different muon and electron cuts can be simply computed by using $\ppppbar\to e^- \mu^- e^+ \bar\nu_\mu$ with muon cuts implemented 
for electrons and vice versa.

An alternative which will be supported in a future release is an exchange of electrons and muons 
by means of the parameter \matrixparam{process_class}. For every process where this is relevant, a 
separate file \matrixparam{parameter.dat} will be provided inside its folder \matrixparam{input}, which can 
be used instead of the original file \matrixparam{parameter.dat} of the process to run with exchanged 
electrons and muons. For example, for different-flavour \wz{} production (\matrixparam{\$\{process_id\} = ppemexnmx04}) an additional file with 
\matrixparam{process_class = ppmemxnex04} instead of \matrixparam{process_class = ppemexnmx04}  will be used to calculate the 
process $\ppppbar\to \mu^- e^- \mu^+ \bar\nu_e$ instead of $\ppppbar\to e^- \mu^- e^+ \bar\nu_\mu$, and all scales, cuts, distributions, 
etc.\ are to be formulated directly for the actual particles of this new process.

All processes available in \Matrix{} are discussed in the following, grouped into Higgs boson production (\refse{sec:Higgsbosonprocesses}), vector-boson production (\refse{sec:vectorbosonprocesses}), diphoton and vector-boson plus photon production (\refse{sec:photonprocesses}), and vector-boson pair production (\refse{sec:vectorbosonpairprocesses}).

\subsubsection{Higgs boson production}\label{sec:Higgsbosonprocesses}
\paragraph{\matrixparam{pph21} ($\ppppbar\to H$)}
On-shell Higgs boson production has no process-specific cuts or dynamic scales. The process is computed in the 
infinite-top-mass approximation by using an effective field theory where the top quark is integrated out.

\subsubsection{Vector-boson production}\label{sec:vectorbosonprocesses}

This group contains both the on-shell and the off-shell production of a single vector boson. 
Whereas the former processes feature cuts and distributions only with respect to the on-shell final state, 
the off-shell processes give access to, in principle, arbitrary phase-space selection cuts and distributions of the leptons. 
The phenomenologically irrelevant process of $\ppppbar\to\nu_e\bar\nu_e$ production has been added as it might be useful for technical checks.

\paragraph{\matrixparam{ppz01} ($\ppppbar\to Z$)}
On-shell $Z$-boson production has no process-specific cuts or dynamic scales.

\paragraph{\matrixparam{ppw01} ($\ppppbar\to W^-$), \matrixparam{ppwx01} ($\ppppbar\to W^+$)}
On-shell $W^\pm$-boson production has no process-specific cuts or dynamic scales. The process includes a non-trivial CKM matrix, 
which the user may modify, see \sct{sec:CKM}.

\paragraph{\matrixparam{ppeex02} ($\ppppbar\to e^-e^+$)}\label{sec:ppeex02}
Off-shell $Z$-boson production\footnote{Note that this process includes also off-shell photon contributions.} with decay to leptons includes the following predefined cuts:

\listvariable{M_leplep}{min_M_leplep, max_M_leplep}\\
\listvariable{R_leplep}{min_R_leplep}\\
\listvariable{lepton_cuts}{min_pT_lep_1st, min_pT_lep_2nd}

No process-specific dynamic scales are implemented.

If cuts are applied, this process may feature
a peculiarly strong dependence on the value of $\rcut{}$ in the \qt{}-subtraction procedure, see \sct{sec:qtsubtraction}. 
The $\ppppbar\to e^-e^+$ process therefore features a switch
\matrixparam{switch_qT_accuracy} in the file \matrixparam{parameter.dat}, which allows the user to decrease the uncertainty induced by the \qt{}-subtraction procedure at NNLO, at the cost of a slower numerical convergence:

\matrixparam{switch_qT_accuracy = 0}\quad Uses the default value $\rcut{}=0.15$\% with fast numerical convergence.\\
\matrixparam{switch_qT_accuracy = 1}\quad Uses $\rcut{}=0.05$\% with reduced uncertainty, but longer runtime.

We recommend to use \matrixparam{switch_qT_accuracy = 0} if the targeted precision of the extrapolated cross-section prediction ($\rcut\rightarrow0$) is of the order of $0.5\%-1\%$. To achieve results with numerical precision of $0.1\%-0.5\%$, \matrixparam{switch_qT_accuracy = 1} should be used.

\paragraph{\matrixparam{ppnenex02} ($\ppppbar\to \nu_e\bar\nu_e$)}
Off-shell $Z$-boson production with decay to neutrinos has no process-specific cuts or dynamic scales.

\paragraph{\matrixparam{ppenex02} ($\ppppbar\to e^-\bar{\nu}_e$), \matrixparam{ppexne02} ($\ppppbar\to e^+\nu_e$)}

Off-shell $W^\pm$-boson production has no process-specific cuts or dynamic scales. The process includes a non-trivial CKM matrix, 
which the user may modify, see \sct{sec:CKM}.

\subsubsection{Diphoton and vector-boson plus photon production}\label{sec:photonprocesses}

This group contains both the diphoton process and the $V\gamma$ processes with off-shell leptonic decays of the heavy vector bosons $V$. 

All processes with isolated photons in the final state have a peculiarly strong dependence on the value of $\rcut{}$ in the \qt{}-subtraction procedure, see \sct{sec:qtsubtraction}. 
For this reason the estimated uncertainty induced by finite $\rcut{}$ values is particularly large in these processes. The photon processes therefore feature a switch
\matrixparam{switch_qT_accuracy} in the file \matrixparam{parameter.dat}, which allows the user to decrease the uncertainty induced by the \qt{}-subtraction procedure at NNLO, at the cost of a slower numerical convergence:

\matrixparam{switch_qT_accuracy = 0}\quad Uses the default value $\rcut{}=0.15$\% with fast numerical convergence.\\
\matrixparam{switch_qT_accuracy = 1}\quad Uses $\rcut{}=0.05$\% with reduced uncertainty, but longer runtime.

We recommend to use \matrixparam{switch_qT_accuracy = 0} if the targeted precision of the extrapolated cross-section prediction ($\rcut\rightarrow0$) is of the order of $0.5\%-1\%$. To achieve results with numerical precision of $0.1\%-0.5\%$, \matrixparam{switch_qT_accuracy = 1} should be used.

\paragraph{\matrixparam{ppaa02} ($\ppppbar\to\gamma\gamma$)}
Diphoton production includes the following predefined cuts:

\listvariable{M_gamgam}{min_M_gamgam, max_M_gamgam}\\
\listvariable{pT_gam_1st}{min_pT_gam_1st}\\
\listvariable{gap_eta_gam}{gap_min_eta_gam, gap_max_eta_gam}\\
\listvariable{R_gamgam}{min_R_gamgam}

No process-specific dynamic scales are implemented.

\paragraph{\matrixparam{ppeexa03} ($\ppppbar\to e^-e^+\gamma$)}
$Z\gamma$ production \cite{Grazzini:2013bna,Grazzini:2015nwa} with $Z$-boson decay to 
charged leptons\footnote{Note that this process includes also $\gamma^\ast\gamma$ contributions, 
where one photon is off-shell and decays to leptons,
and $Z/\gamma^\ast$ production with a subsequent decay $Z/\gamma^\ast\to e^-e^+\gamma$.} 
includes the following predefined cuts:

\listvariable{M_leplep}{min_M_leplep}\\
\listvariable{M_lepgam}{min_M_lepgam}\\
\listvariable{R_leplep}{min_R_leplep}\\
\listvariable{R_lepgam}{min_R_lepgam}\\
\listvariable{R_lepjet}{min_R_lepjet}\\
\listvariable{R_gamjet}{min_R_gamjet}\\
\listvariable{pT_lep_1st}{min_pT_lep_1st}\\

\renewcommand\arraystretch{1.3}
\begin{table}[h]
The process facilitates the following additional predefined dynamic scales (symbols in \tab{tab:symbols}):\\[0.3cm]
\small
\begin{tabular}{lll}
\toprule
\matrixparam{dynamic_scale = 3:} && $\mu=m_{T,\gamma}\equiv p_{T,\gamma}$ \\
\matrixparam{dynamic_scale = 4:} && $\mu=m_{T,e^-e^+}$ \\
\matrixparam{dynamic_scale = 5:} && $\mu=\sqrt{m_{T,\gamma}\cdot m_{T,e^-e^+}}$ \\
\matrixparam{dynamic_scale = 6:} && $\mu=\sqrt{m_Z^2+m_{T,\gamma}^2}$ \\
\bottomrule
\end{tabular}
\renewcommand{\baselinestretch}{1.0}
\end{table}

\paragraph{\matrixparam{ppnenexa03} ($\ppppbar\to \nu_e\bar{\nu}_e\gamma$)}
$Z\gamma$ production \cite{Grazzini:2015nwa} with $Z$-boson decay to neutrinos includes the following predefined cuts:

\listvariable{R_gamjet}{min_R_gamjet}\\

\renewcommand\arraystretch{1.3}
\begin{table}[h]
The process facilitates the following additional predefined dynamic scales (symbols in \tab{tab:symbols}):\\[0.3cm]
\small
\begin{tabular}{lll}
\toprule
\matrixparam{dynamic_scale = 3:} && $\mu=m_{T,\gamma}\equiv p_{T,\gamma}$ \\
\matrixparam{dynamic_scale = 4:} && $\mu=m_{T,\nu_e\bar\nu_e}$ \\
\matrixparam{dynamic_scale = 5:} && $\mu=\sqrt{m_{T,\gamma}\cdot m_{T,\nu_e\bar\nu_e}}$ \\
\matrixparam{dynamic_scale = 6:} && $\mu=\sqrt{m_Z^2+m_{T,\gamma}^2}$ \\
\bottomrule
\end{tabular}
\renewcommand{\baselinestretch}{1.0}
\end{table}
\renewcommand\arraystretch{1.1}

\paragraph{\matrixparam{ppenexa03} ($\ppppbar\to e^-\bar{\nu}_e\gamma$), \matrixparam{ppexnea03} ($\ppppbar\to e^+\nu_e\gamma$)}
$W^\pm\gamma$ production \cite{Grazzini:2015nwa} with leptonic $W$-boson 
decay\footnote{Note that this process includes also contributions from $W^{\pm}$ production with a subsequent decay $W^+\to e^+\nu_e\gamma$ or $W^-\to e^-\bar\nu_e\gamma$, respectively.} 
includes the following predefined cuts:

\listvariable{R_lepgam}{min_R_lepgam}\\
\listvariable{R_lepjet}{min_R_lepjet}\\
\listvariable{R_gamjet}{min_R_gamjet}\\
\listvariable{mT_CMS}{min_mT_CMS}\\
\listvariable{gap_eta_gam}{gap_min_eta_gam, gap_max_eta_gam}\\

\renewcommand\arraystretch{1.3}
\begin{table}[h]
The process facilitates the following additional predefined dynamic scales (symbols in \tab{tab:symbols}):\\[0.3cm]
\small
\begin{tabular}{lll}
\toprule
\matrixparam{dynamic_scale = 3:} && $\mu=m_{T,\gamma}\equiv p_{T,\gamma}$ \\
\matrixparam{dynamic_scale = 4:} && $\mu=m_{T,e^\pm\nu_e}$ \\
\matrixparam{dynamic_scale = 5:} && $\mu=\sqrt{m_{T,\gamma}\cdot m_{T,e^\pm\nu_e}}$ \\
\matrixparam{dynamic_scale = 6:} && $\mu=\sqrt{m_W^2+m_{T,\gamma}^2}$  \\
\bottomrule
\end{tabular}
\renewcommand{\baselinestretch}{1.0}
\end{table}
\renewcommand\arraystretch{1.1}

\subsubsection{Vector-boson pair production}\label{sec:vectorbosonpairprocesses}

This group contains both the on-shell and the off-shell production of a vector-boson pair. 
The on-shell production of a \ww{} or a $ZZ$ pair allows selection cuts to be applied only on the vector bosons,
and distributions in the vector-boson kinematics can be studied.
The off-shell processes, on the other hand, give access to the full leptonic final states, 
i.e.\ they allow in principle arbitrary IR safe selection cuts on the leptons to be applied, and distributions in the kinematics of these leptons can be computed. 
Off-shell vector-boson pair production includes
processes with different-flavour (DF) and same-flavour (SF) leptons in the final state.
For the processes with two neutrinos and two leptons in the final state, the separation 
of DF ($e^-e^+\nu_\mu\bar\nu_\mu$) and SF ($e^-e^+\nu_e\bar\nu_e$) 
channels is done according to the underlying calculation, not to the experimental signature: 
For any analysis of 
two leptons plus missing transverse energy the predictions must be obtained by (incoherently) combining the DF and SF processes, i.e.\
\begin{align}
\begin{split}
\sigma(e^-e^++p_{\mathrm{T,miss}})&=\sigma(e^-e^+\nu_e\bar\nu_e)+\sigma(e^-e^+\nu_{\mu}\bar\nu_{\mu})+\sigma(e^-e^+\nu_{\tau}\bar\nu_{\tau})\\
&=\sigma(e^-e^+\nu_e\bar\nu_e)+2\times\sigma(e^-e^+\nu_{\mu}\bar\nu_{\mu})\,.
\end{split}
\end{align}

\subparagraph{Flavour-scheme choice and top-quark contamination in $\boldsymbol{WW}$ production}\label{sec:topcontamination}\quad

All processes including a pair of on- or off-shell $W$ bosons are subject to a contamination by off-shell 
top-quark contributions with $t\rightarrow Wb$ decays. Such contributions enter radiative corrections in both 
the four-flavour scheme (4FS), where bottom quarks are treated as massive, and the five-flavour scheme (5FS), 
where the bottom-quark mass is set to zero as all other light-quark masses. In case of \ww{} production, the 4FS has 
the advantage that the bottom quark appears only in the final state, and that the bottom-quark mass renders all
partonic subprocesses with bottom quarks in the final state separately finite.
Thus, the top-quark contamination is easily avoided by omitting bottom-quark emission subprocesses in this scheme, 
which are considered part of the (off-shell) top-pair background.
Consequently, we use this 4FS approach
as the default for any process that features 
an on- or off-shell \ww{} pair, namely by setting \ttt{flavour_scheme = 0} in the file \matrixparam{parameter.dat} 
in combination with $\mb\neq 0$ in the file \matrixparam{model.dat}. We note that 
this approach requires the use of consistent \pdf{} sets with $n_f=4$ light parton flavours.

Alternatively, one can use the 5FS by setting \ttt{flavour_scheme = 1}, $\mb=0$ and choosing $n_f=5$ \pdf{} sets. In this case, however, 
the top-quark contamination is not removed from the results. A numerical procedure to achieve a definition of the \ww{} 
cross section without top-quark contamination in the 5FS, which has been used in \citeres{Gehrmann:2014fva,Grazzini:2016ctr}, requires the 
repeated computation of the cross section for varying top-quark widths in order
to approach the limit $\Gamma_{t}\to0$ and thereby to isolate the contributions from single-top and top-pair production. 
As it has been shown in these 
references, the resulting top-subtracted \ww{} cross sections calculated in the 4FS and the 5FS prescription, respectively, agree within $1\%-2\%$, both at the inclusive level 
and with different sets of fiducial cuts applied. This justifies the use of the simpler 4FS computation for such processes.

\subparagraph{Off-shell $Z$ bosons in $ZZ$ production}\label{sec:offshell}\quad

For off-shell $ZZ$-production processes the cuts may be arranged in a way that
at least one of the $Z$ bosons is forced to be far in the off-shell region. For such 
cases these processes include an additional switch \ttt{switch_off_shell} in the file \matrixparam{parameter.dat} to improve 
the convergence of the computation in this phase-space region. This
is relevant, e.g., when studying the $ZZ$ background in Higgs boson measurements. 
The default choice \ttt{switch_off_shell = 0}
uses the standard setup for the grid generation (\gridrun{}, see \sct{sec:runinteractive}), which 
is suitable if both $Z$ bosons can simultaneously become resonant. Using \ttt{switch_off_shell = 1}
adapts the settings of the \gridrun{} for cases where at least one $Z$ boson is off-shell.

\paragraph{\matrixparam{ppzz02} ($\ppppbar\to ZZ$)}
On-shell \zz{} production \cite{Cascioli:2014yka} has no process-specific cuts or dynamic scales.

\paragraph{\matrixparam{ppwxw02} ($\ppppbar\to W^+W^-$)}
On-shell \ww{} production \cite{Gehrmann:2014fva,Grazzini:2016ctr} has no process-specific cuts or dynamic scales. 

By default (\ttt{flavour_scheme = 0}) this process is computed in the 4FS, dropping 
all diagrams with final-state bottom quarks to remove the top-quark contamination. 
The 5FS can be chosen by setting \ttt{flavour_scheme = 1}, where, however, 
the top-quark contamination is not removed from the results, since the recommended procedure is much more 
involved and requires several runs for the 5FS, see introduction of \sct{sec:vectorbosonpairprocesses} for further details.

\paragraph{\matrixparam{ppemexmx04} ($\ppppbar\to e^- \mu^- e^+ \mu^+$)}

Off-shell \zz{} production \cite{Grazzini:2015hta} with $Z$-boson decays to different-flavour (DF) 
leptons\footnote{Note that this process includes also $Z\gamma^\ast$ and $\gamma^\ast\gamma^\ast$ contributions with off-shell photons decaying to leptons,
as well as $Z/\gamma^\ast$ production with a subsequent decay $Z/\gamma^\ast\to e^- \mu^- e^+ \mu^+$.} 
includes the following predefined cuts:

\listvariable{M_leplep_OSSF}{min_M_leplep_OSSF, max_M_leplep_OSSF}\\
\hyperref[M_leplep_OSSF]{\hspace*{3.04cm}\ttt{min_M_Z1_OSSF, max_M_Z1_OSSF}}\\
\listvariable{R_leplep}{min_R_leplep}\\
\listvariable{pT_lep_1st}{min_pT_lep_1st}\\
\listvariable{pT_lep_2nd}{min_pT_lep_2nd}\\
\listvariable{M_4lep}{min_delta_M_4lep, max_delta_M_4lep, min_M_4lep, max_M_4lep}\\
\listvariable{lep_iso}{lep_iso_delta_0, lep_iso_epsilon}\\

\renewcommand\arraystretch{1.3}
\begin{table}[h]
The process facilitates the following additional predefined dynamic scales (symbols in \tab{tab:symbols}):\\[0.3cm]
\small
\begin{tabular}{lll}
\toprule
\matrixparam{dynamic_scale = 3:} && $\mu=\sqrt{m_{T,e^-e^+}\cdot m_{T,\mu^-\mu^+}}$ \\ 
\matrixparam{dynamic_scale = 4:} && $\mu=\sqrt{m_Z^2+p_{T,e^-e^+}^2}+\sqrt{m_Z^2+p_{T,\mu^-\mu^+}^2}$ \\
\bottomrule
\end{tabular}
\renewcommand{\baselinestretch}{1.0}
\end{table}
\renewcommand\arraystretch{1.1}

This process provides an additional switch \ttt{switch_off_shell} to speed up the grid-generation phase when considering at least one of the $Z$ bosons to be far off-shell; it should not be used otherwise, see introduction of \sct{sec:vectorbosonpairprocesses} for further details.

\paragraph{\matrixparam{ppeeexex04} ($\ppppbar\to e^-e^-e^+e^+$)}\label{sec:ZZSF}

Off-shell \zz{} production \cite{Grazzini:2015hta} with $Z$-boson decays to same-flavour (SF) leptons\footnote{Note that this process includes also $Z\gamma^\ast$ and $\gamma^\ast\gamma^\ast$ contributions with off-shell photons decaying to leptons,
as well as $Z/\gamma^\ast$ production with a subsequent decay $Z/\gamma^\ast\to e^- e^- e^+ e^+$.} 
includes the following predefined cuts:

\listswitch{lepton_identification}{}\\
\listvariable{M_Zrec}{min_M_Zrec, max_M_Zrec}\\
\listvariable{M_leplep_OSSF}{min_M_leplep_OSSF}\\
\listvariable{R_leplep}{min_R_leplep}\\
\listvariable{pT_lep_1st}{min_pT_lep_1st}\\
\listvariable{pT_lep_2nd}{min_pT_lep_2nd}\\
\listvariable{M_4lep}{min_delta_M_4lep, max_delta_M_4lep, min_M_4lep, max_M_4lep}\\
\listvariable{lep_iso}{lep_iso_delta_0, lep_iso_epsilon}\\

Since this process features four SF leptons, two of which are positively and two negatively charged, the leptons
cannot be unambiguously associated with the two parent $Z$ bosons as in the DF case. However, the experimental analyses 
often rely on cuts specific to (reconstructed) $Z$ bosons. Hence, in the SF channel an identification procedure is required  
to assign one opposite-charge same-flavour~(OSSF) lepton pair to each of the $Z$ bosons. The parameter \listswitch{lepton_identification}{} 
switches between such identification procedures of the $Z$ bosons as used by ATLAS and CMS. 
In both cases seven new particle groups are defined (see \sct{sec:fiducial_cuts} and the related \tab{tab:predefinedparticles} 
for the standard particle groups), which makes them available in the definition of cuts, scales and distributions: 
Particle group \ttt{Z1rec} contains the $Z$ boson reconstructed from the OSSF lepton pair
with its invariant mass closer to the $Z$-boson mass, labelled as $Z_{1,\rm rec}$,
and particle group \ttt{Z2rec} contains the remaining OSSF lepton pair, labelled as $Z_{2,\rm rec}$.
Particle group \ttt{Zrec} is filled with both reconstructed $Z$ bosons in the standard $\pT$-ordering. The particle groups \ttt{lmZ1}, \ttt{lmZ2}, \ttt{lpZ1} and \ttt{lpZ2} contain the negatively and positively charged leptons that belong to the corresponding reconstructed $Z$ bosons, respectively, i.e.\ each of these groups has by definition only a single entry. Examples of the usage of these particle groups can be found in 
the file \matrixparam{distribution.dat} of this process. Furthermore, the predefined cut \listvariable{M_Zrec}{min_M_Zrec, max_M_Zrec} uses 
the respective particle groups identified corresponding to the setting of the switch \ttt{lepton_identification}.\footnote{Note that the respective 
particle groups are also available within the C++ code, see \app{app:advancedstuff}.}

If \ttt{lepton_identification = 0} is set, the respective particle groups are not filled and thus cannot be used to define distributions.
Also cuts and dynamic scales depending on the identification must not be used in this case, such as \listvariable{M_Zrec}{min_M_Zrec, max_M_Zrec}. 
In the following we outline the predefined pairing prescriptions.

The ATLAS pairing (\ttt{lepton_identification = 1}) considers all possible (two, in the theoretical computation) combinations to associate two OSSF lepton pairs with $Z_{1}=e^-e^+$ and $Z_{2}=e^{-\prime}e^{+\prime}$. 
The criterion to decide on the pairing is the sum of the absolute differences of their invariant masses to 
the $Z$-boson mass, i.e.\ $\left| m_{e^-e^+}-m_Z\right|+\left|m_{e^{-\prime}e^{+\prime}}-m_Z\right|$,
and the assignment that
minimizes this sum is associated with the reconstructed $Z$ bosons $Z_{1,\rm rec}=Z_{1}$ and  $Z_{2,\rm rec}=Z_{2}$. The 
respective particle groups are filled accordingly.

The CMS pairing (\ttt{lepton_identification = 2}) selects the OSSF lepton pair among all possible pairings (four, in the theoretical computation) that minimizes the invariant-mass difference to the 
$Z$-boson mass, $\left| m_{e^-e^+}-m_Z\right|$. This pair is always identified as $Z_{1,\rm rec}$, while the remaining pair is defined as $Z_{2,\rm rec}$.

This process provides an additional switch \ttt{switch_off_shell} to speed up the grid-generation phase when considering at least one of the $Z$ bosons to be far off-shell; it should not be used otherwise, see introduction of \sct{sec:vectorbosonpairprocesses} for further details.

\renewcommand\arraystretch{1.3}
\begin{table}[h]
The process facilitates the following additional predefined dynamic scales (symbols in \tab{tab:symbols}):\\[0.3cm]
\small
\begin{tabular}{lll}
\toprule
\matrixparam{dynamic_scale = 3:} && $\mu=\sqrt{m_{T,Z_{1,\rm rec}}\cdot m_{T,Z_{2,\rm rec}}}$ \\ 
\matrixparam{dynamic_scale = 4:} && $\mu=\sqrt{m_Z^2+p_{T,Z_{1,\rm rec}}^2}+\sqrt{m_Z^2+p_{T,Z_{2,\rm rec}}^2}$ \\
\bottomrule
\end{tabular}
\renewcommand{\baselinestretch}{1.0}
\end{table}
\renewcommand\arraystretch{1.1}

\paragraph{\matrixparam{ppeexnmnmx04} ($\ppppbar\to e^- e^+ \nu_\mu \bar\nu_\mu$)}

Off-shell \zz{} production with $Z$-boson decays to leptons and neutrinos of different flavour\footnote{Note that this process includes also $Z\gamma^\ast$ contributions with the off-shell photon decaying to leptons, 
and $Z/\gamma^\ast$ production with a subsequent decay $Z/\gamma^\ast\to e^- e^+ \nu_\mu\bar\nu_\mu$
.} includes the following predefined cuts:

\listvariable{M_leplep}{min_M_leplep, max_M_leplep}\\
\listvariable{M_leplepnunu}{min_M_leplepnunu, max_M_leplepnunu,}\\
\hyperref[M_leplepnunu]{\hspace*{2.84cm}\ttt{min_delta_M_leplepnunu, max_delta_M_leplepnunu}}\\

\renewcommand\arraystretch{1.3}
\begin{table}[h]
The process facilitates the following additional predefined dynamic scale (symbols in \tab{tab:symbols}):\\[0.3cm]
\small
\begin{tabular}{lll}
\toprule
\matrixparam{dynamic_scale = 3:} && $\mu=\sqrt{m_Z^2+p_{T,e^-e^+}^2}+\sqrt{m_Z^2+p_{T,\nu_\mu\bar\nu_\mu}^2}$ \\
\bottomrule
\end{tabular}
\renewcommand{\baselinestretch}{1.0}
\end{table}
\renewcommand\arraystretch{1.1}

This process provides an additional switch \ttt{switch_off_shell} to speed up the grid-generation phase when considering at least one of the $Z$ bosons to be far off-shell; it should not be used otherwise, see introduction of \sct{sec:vectorbosonpairprocesses} for further details.

\paragraph{\matrixparam{ppemxnmnex04} ($\ppppbar\to e^- \mu^+ \nu_\mu \bar\nu_e$)}
Off-shell \ww{} production \cite{Gehrmann:2014fva,Grazzini:2016ctr} with $W$-boson decays to DF leptons and the corresponding neutrinos\footnote{Note that this process includes also $Z/\gamma^\ast$ production with a subsequent decay $Z/\gamma^\ast\to e^- \mu^+ \nu_\mu \bar\nu_e$.}  includes the following predefined cuts:

\listvariable{M_leplep}{min_M_leplep, max_M_leplep}\\
\listvariable{R_leplep}{min_R_leplep}\\
\listvariable{R_ejet}{min_R_ejet}\\
\listvariable{pT_leplep}{min_pT_leplep}\\
\listvariable{pT_lep_1st}{min_pT_lep_1st}\\
\listvariable{M_leplepnunu}{min_M_leplepnunu, max_M_leplepnunu,}\\
\hyperref[M_leplepnunu]{\hspace*{2.84cm}\ttt{min_delta_M_leplepnunu, max_delta_M_leplepnunu}}\\
\listvariable{gap_eta_e}{gap_min_eta_e, gap_max_eta_e}\\
\listvariable{rel_pT_miss}{min_rel_pT_miss}\\
\listvariable{phi_leplep}{max_phi_leplep}\\
\listvariable{phi_leplep_nunu}{min_phi_leplep_nunu}\\
\listvariable{pT_W}{min_pT_W, max_pT_W}\\

\renewcommand\arraystretch{1.3}
\begin{table}[h]
The process facilitates the following additional predefined dynamic scales (symbols in \tab{tab:symbols}):\\[0.3cm]
\small
\begin{tabular}{lll}
\toprule
\matrixparam{dynamic_scale = 3:} && $\mu=\sqrt{m_W^2+p_{T,e^-\bar\nu_e}^2}+\sqrt{m_W^2+p_{T,\mu^+\nu_\mu}^2}$ \\
\matrixparam{dynamic_scale = 4:} && $\mu=m_{T,e^-\bar\nu_e}+m_{T,\mu^+\nu_\mu}$ \\
\bottomrule
\end{tabular}
\renewcommand{\baselinestretch}{1.0}
\end{table}
\renewcommand\arraystretch{1.1}

By default (\ttt{flavour_scheme = 0}) this process is computed in the 4FS, dropping 
all diagrams with final-state bottom quarks to remove the top-quark contamination. 
The 5FS can be chosen by setting \ttt{flavour_scheme = 1}, where, however, 
the top-quark contamination is not removed from the results, since the recommended procedure is much more 
involved and requires several runs for the 5FS, see introduction of \sct{sec:vectorbosonpairprocesses} for further details.

\paragraph{\matrixparam{ppeexnenex04} ($\ppppbar\to e^- e^+ \nu_e \bar\nu_e$)}
Off-shell \ww{} and \zz{} production 
with decays to SF leptons and the corresponding neutrinos\footnote{Note that this process includes also $Z\gamma^\ast$ contributions with the off-shell photon decaying to leptons, 
and $Z/\gamma^\ast$ production with a subsequent decay $Z/\gamma^\ast\to e^- e^+ \nu_e \bar\nu_e$.}
includes the following predefined cuts:

\listvariable{M_leplep}{min_M_leplep, max_M_leplep}\\
\listvariable{R_leplep}{min_R_leplep}\\
\listvariable{R_ejet}{min_R_ejet}\\
\listvariable{pT_leplep}{min_pT_leplep}\\
\listvariable{pT_lep_1st}{min_pT_lep_1st}\\
\listvariable{M_leplepnunu}{min_M_leplepnunu, max_M_leplepnunu,}\\
\hyperref[M_leplepnunu]{\hspace*{2.84cm}\ttt{min_delta_M_leplepnunu, max_delta_M_leplepnunu}}\\
\listvariable{gap_eta_e}{gap_min_eta_e, gap_max_eta_e}\\
\listvariable{rel_pT_miss}{min_rel_pT_miss}\\
\listvariable{phi_leplep}{max_phi_leplep}\\
\listvariable{phi_leplep_nunu}{min_phi_leplep_nunu}\\
\listvariable{pT_W}{min_pT_W, max_pT_W}\\

\renewcommand\arraystretch{1.3}
\begin{table}[h]
The process facilitates the following additional predefined dynamic scales (symbols in \tab{tab:symbols}):\\[0.3cm]
\small
\begin{tabular}{lll}
\toprule
\matrixparam{dynamic_scale = 3:} && $\mu=\sqrt{m_W^2+p_{T,e^-\bar\nu_e}^2}+\sqrt{m_W^2+p_{T,e^+\nu_e}^2}$ \\
\matrixparam{dynamic_scale = 4:} && $\mu=m_{T,e^-\bar\nu_e}+m_{T,e^+\nu_e}$ \\
\bottomrule
\end{tabular}
\renewcommand{\baselinestretch}{1.0}
\end{table}
\renewcommand\arraystretch{1.1}

By default (\ttt{flavour_scheme = 0}) this process is computed in the 4FS, dropping 
all diagrams with final-state bottom quarks to remove the top-quark contamination. 
The 5FS can be chosen by setting \ttt{flavour_scheme = 1}, where, however, 
the top-quark contamination is not removed from the results, since the recommended procedure is much more 
involved and requires several runs for the 5FS, see introduction of \sct{sec:vectorbosonpairprocesses} for further details.

This process includes an additional switch \ttt{switch_off_shell} to speed up the grid-generation phase when considering at least one of the $Z$ bosons to be far off-shell; it should not be used otherwise, see introduction of \sct{sec:vectorbosonpairprocesses} for further details.

\paragraph{\matrixparam{ppemexnmx04} ($\ppppbar\to e^- \mu^- e^+ \bar\nu_\mu$), \matrixparam{ppeexmxnm04} ($\ppppbar\to e^- e^+ \mu^+ \nu_\mu$)}
Off-shell \wz{} production \cite{Grazzini:2016swo,Grazzini:2017ckn} with decays to one OSSF lepton pair, one DF lepton and one corresponding 
neutrino\footnote{Note that this process includes also $W^-/W^+$ production with a subsequent decay $W\to e^- e^+ \mu\nu_\mu$.} (DF channel) includes the following predefined cuts:

\listvariable{M_Zrec}{min_M_Zrec, max_M_Zrec}\\
\listvariable{delta_M_Zrec_MZ}{max_delta_M_Zrec_MZ}\\
\listvariable{delta_M_lepleplep_MZ}{min_delta_M_lepleplep_MZ}\\
\listvariable{R_leplep}{min_R_leplep}\\
\listvariable{R_lepZlepZ}{min_R_lepZlepZ}\\
\listvariable{R_lepZlepW}{min_R_lepZlepW}\\
\listvariable{electron_cuts}{min_pT_e_1st, min_pT_e_2nd}\\
\listvariable{muon_cuts}{min_pT_mu_1st, min_pT_mu_2nd}\\
\listvariable{lepton_cuts}{min_pT_lep_1st, min_pT_lep_2nd}\\
\listvariable{leading_lepton_cuts}{min_pT_lep_1st_if_e, min_pT_lep_1st_if_mu}\\
\listvariable{MT_Wrec}{min_MT_Wrec}\\

In the DF channel $W$ and $Z$ bosons can be unambiguously identified. In analogy to the SF case we define the following particle groups 
which can be accessed, e.g., in distributions: Particle group \ttt{Zrec} contains the $Z$ boson, reconstructed by 
the two electrons, and \ttt{Wrec} the $W$ boson, defined by the muon and the neutrino. \ttt{lepZ} contains the corresponding leptons 
of the $Z$ boson, ordered in their transverse momentum, and \ttt{lepW} the lepton of the $W$ boson.

\renewcommand\arraystretch{1.3}
\begin{table}[h]
The process facilitates the following additional predefined dynamic scales (symbols in \tab{tab:symbols}):\\[0.3cm]
\small
\begin{tabular}{lll}
\toprule
\matrixparam{dynamic_scale = 3:} && $\mu=\frac12\left(\sqrt{m_Z^2+p_{T,Z_{\rm rec}}}+\sqrt{m_W^2+p_{T,W_{\rm rec}^\pm}^2}\right)$ \\
\matrixparam{dynamic_scale = 4:} && $\mu=\frac12\left(m_{T,Z_{\rm rec}}+m_{T,W^\pm_{\rm rec}}\right)$ \\
\bottomrule
\end{tabular}
\renewcommand{\baselinestretch}{1.0}
\end{table}
\renewcommand\arraystretch{1.1}

\paragraph{\matrixparam{ppeeexnex04} ($\ppppbar\to e^- e^- e^+ \bar\nu_e$), \matrixparam{ppeexexne04} ($\ppppbar\to e^- e^+ e^+ \nu_e$)}\label{sec:WZSF}
Off-shell \wz{} production \cite{Grazzini:2016swo,Grazzini:2017ckn} with decays to three SF leptons and one corresponding neutrino\footnote{Note that this process includes also $W^-/W^+$ production with a subsequent decay $W\to e^- e^+ e\nu_e$.}  (SF channel) includes the following predefined cuts:

\listswitch{lepton_identification}{}\\
\listvariable{M_Zrec}{min_M_Zrec, max_M_Zrec}\\
\listvariable{M_leplep_OSSF}{min_M_leplep_OSSF}\\
\listvariable{delta_M_Zrec_MZ}{max_delta_M_Zrec_MZ}\\
\listvariable{delta_M_lepleplep_MZ}{min_delta_M_lepleplep_MZ}\\
\listvariable{R_leplep}{min_R_leplep}\\
\listvariable{R_lepZlepZ}{min_R_lepZlepZ}\\
\listvariable{R_lepZlepW}{min_R_lepZlepW}\\
\listvariable{lepW_cuts}{min_pT_lepW, max_eta_lepW}\\
\listvariable{lepZ_cuts}{min_pT_lepZ_1st, min_pT_lepZ_2nd}\\
\listvariable{lepton_cuts}{min_pT_lep_1st, min_pT_lep_2nd}\\
\listvariable{MT_Wrec}{min_MT_Wrec}\\

Since this process features three SF leptons, two of which have the same charge, and one neutrino, a-priori the leptons
are not unambiguously associated with the decays of the $Z$ and $W$ bosons. However, the experimental analyses 
often rely on cuts specific to (reconstructed) $Z$ and $W$ bosons. Hence, in the SF channel an identification procedure is required to
unambiguously assign one OSSF lepton pair to the $Z$ boson as well as the remaining lepton and the neutrino to the 
$W$ boson. The parameter \listswitch{lepton_identification}{} 
switches between two such predefined identification procedures, as used by ATLAS and CMS. 
In both cases four new particle groups are defined (see \sct{sec:fiducial_cuts} and the related \tab{tab:predefinedparticles} 
for the standard particle groups) to make them accessible in cuts, scales and distributions: 
Particle group \ttt{Zrec} contains the OSSF lepton pair that is reconstructed as a $Z$ boson, labelled $Z_{\rm rec}$. 
\ttt{Wrec} contains the lepton and the neutrino that are reconstructed as a $W$ boson, labelled $W_{\rm rec}$. 
\ttt{lepZ} is filled with the leptons corresponding to the reconstructed $Z$ boson in the standard $\pT$-ordering, and \ttt{lepW} with the lepton assigned to the $W$ boson.
By definition each of the other particle groups effectively contains only one particle, whereas \ttt{lepZ} contains two particles. Examples for the usage of these particle groups can be found in 
the file \matrixparam{distribution.dat} of this process. Furthermore, many of the predefined cuts, e.g.\ \listvariable{delta_M_Zrec_MZ}{max_delta_M_Zrec_MZ}, 
\listvariable{MT_Wrec}{min_MT_Wrec} or \listvariable{R_lepZlepW}{min_R_lepZlepW}, use 
the respective particle groups, which are filled according to the chosen value of the switch \ttt{lepton_identification}.\footnote{Note that the respective 
particle groups are also available within the C++ code, see \app{app:advancedstuff}.}

If \ttt{lepton_identification = 0} is set, the respective particle groups are not filled and thus cannot be used to define distributions. Also dynamic scales and 
cuts depending on the identification must not be used in this case. Finally, we outline the predefined pairing prescriptions:

ATLAS applies the so-called resonant-shape procedure \cite{Aad:2016ett} (\ttt{lepton_identification = 1}), where the assignment that maximizes the estimator
\begin{align}
\label{eq:pestimator}
P = \Bigg|\frac1{m^2_{e^-e^+}-m^2_Z+i\,\Gamma_Z\,m_Z}\Bigg|^2 \,\cdot\Bigg|\frac1{m^2_{e^{\pm\prime}\nu_e}-m^2_W+i\,\Gamma_W\,m_W}\Bigg|^2
\end{align}
is chosen to identify $Z_{\rm rec}=Z$ and  $W_{\rm rec}=W$, and the respective particles groups are filled accordingly.\footnote{We note that this definition requires the knowledge of the complete momentum of the neutrino. This variable can, of course, be used in the theoretical calculation, but cannot be directly extracted in the experimental analysis, where it must be reconstructed with the Monte Carlo.}

The CMS pairing (\ttt{lepton_identification = 2}) simply chooses the OSSF lepton pair that minimizes the invariant-mass difference to the 
$Z$-boson mass, i.e.\ $\left| m_{e^-e^+}-m_Z\right|$. This pair is identified as $Z_{\rm rec}$, and $W_{\rm rec}$ and the other particle groups are assigned accordingly.

\renewcommand\arraystretch{1.3}
\begin{table}[h]
The process facilitates the following additional predefined dynamic scales (symbols in \tab{tab:symbols}):\\[0.3cm]
\small
\begin{tabular}{lll}
\toprule
\matrixparam{dynamic_scale = 3:} && $\mu=\frac12\left(\sqrt{m_Z^2+p_{T,Z_{\rm rec}}}+\sqrt{m_W^2+p_{T,W_{\rm rec}^\pm}^2}\right)$ \\
\matrixparam{dynamic_scale = 4:} && $\mu=\frac12\left(m_{T,Z_{\rm rec}}+m_{T,W^\pm_{\rm rec}}\right)$ \\
\bottomrule
\end{tabular}
\renewcommand{\baselinestretch}{1.0}
\end{table}
\renewcommand\arraystretch{1.1}

\section{Phenomenological results}
\label{sec:physics}

In this Section we present results on integrated cross sections for all processes available in the first \Matrix{} release.
They are reported at LO, NLO and NNLO to study the impact of QCD radiative corrections.
We also discuss the impact of the loop-induced $gg$ contribution on the NNLO cross section, if applicable. 
The results in this Section are obtained with the \Matrix{} default setup for each of these processes. 
Their purpose is both to provide benchmark numbers for all processes that can be evaluated with \Matrix{},
and to give a reference for the user:
These benchmark results can be reproduced (on a statistical level) if no changes are applied to the default input cards (except for turning on the 
corresponding perturbative orders and the targeted precision the user is interested in).

\subsection{Settings}\label{sec:settings}

We consider proton--proton collisions at the 13\,TeV LHC. In terms of the input of the weak parameters, the $G_\mu$ scheme
is employed: When considering leptonic final state, which are always produced via off-shell EW vector bosons, we use the complex-mass scheme~\cite{Denner:2005fg} 
throughout, i.e.\ we use  complex $W$- and $Z$-boson masses and define the EW mixing angle as $\cos\theta_W^2=(m_W^2-i\Gamma_W\,m_W)/(m_Z^2-i\Gamma_Z\,m_Z)$
and $\alpha=\sqrt{2}\,G_\mu m_W^2\sin^2\theta_W/\pi$,
using the PDG~\cite{Olive:2016xmw} values $G_F = 1.16639\times 10^{-5}$\,GeV$^{-2}$, $m_W=80.385$\,GeV, $\Gamma_W=2.0854$\,GeV, $m_Z = 91.1876$\,GeV 
and $\Gamma_Z=2.4952$\,GeV. Furthermore, we set $m_H = 125$\,GeV and $\Gamma_H = 0.00407$\,GeV.
When considering on-shell single-boson production or on-shell production of heavy-boson pairs, the masses of the weak vector bosons and the weak mixing angle are consistently kept real by setting 
$\Gamma_W=\Gamma_Z=0$, and we also use a real Higgs boson mass, i.e.\ $\Gamma_H = 0$.
The number of heavy-quark flavours depends on the applied flavour scheme. 
As outlined in \sct{sec:topcontamination}, all processes involving \ww{} contributions use the 4FS as default to consistently remove top-quark contamination by dropping 
the (separately IR finite) partonic processes with real bottom-quark emissions. In the 4FS we use the on-shell bottom mass $m_b = 4.92$\,GeV.
All other processes apply the 5FS with a vanishing bottom mass $m_b = 0$.
The top quark is treated as massive and unstable throughout, 
and we set $m_t = 173.2$\,GeV as well as $\Gamma_t = 1.44262$\,GeV.\footnote{Massive 
top-quark contributions are neglected in the virtual two-loop corrections, but are kept anywhere else in the computations.}
We use the consistent NNPDF3.0~\cite{Ball:2014uwa} set of
parton distributions (PDFs) with $n_f=4$ or $n_f=5$ active quark flavours. 
In particular, N$^n$LO ($n=0,1,2$) predictions are obtained by using PDFs at the same perturbative order and 
the evolution of $\as$ at $(n+1)$-loop order, as provided by the corresponding PDF set.
The CKM matrix is set to unity except for the production of a single (on- or off-shell) $W^\pm$ boson. In that case we use the PDG SM values as reported in \citere{Olive:2016xmw}:
\begin{align}
V_{\rm CKM} =\left(\begin{array}{ccc} V_{ud} & V_{us} & V_{ub} \\  V_{cd} & V_{cs} & V_{cb} \\  V_{td} & V_{ts} & V_{tb} \end{array}\right) = \left(\begin{array}{ccc} 0.97417 & 0.2248 & 0.00409 \\  0.22 & 0.995 & 0.0405 \\  0.0082 & 0.04 & 1.009 \end{array}\right).
\end{align}

\newcommand{\ptjet}{\ensuremath{p_{T,j}}}

Our reference choice $\mu_0$ for renormalization ($\mu_R$) and factorization ($\mu_F$) scales as well as the set of cuts applied 
in our default setups depend on the individual process. Both are reported when discussing the results in the upcoming Section. 
Uncertainties from missing higher-order contributions are estimated in the usual way
by independently varying $\mu_R$ and $\mu_F$ in the range $0.5\mu_0\le \mu_R,\mu_F\le 2 \mu_0$, with the constraint $0.5\le \mu_R/\mu_F\le 2$. 
Unless specified otherwise, jets are defined by the anti-$k_T$ clustering algorithm, $R=0.4$, $\ptjet>25$\,GeV and $|\eta_j|<4.5$.

\subsection{Cross-section predictions}\label{sec:cs}

Reference predictions of the integrated cross sections for all processes that are available in \Matrix{} are reported in \tab{tab:results}. 
Note that the processes under consideration feature 
cross sections that may differ by several orders of magnitude, starting from a few fb up to several nb.

Besides results at LO, NLO and NNLO accuracy, a separate column refers to the 
absolute (and relative) size of the loop-induced $gg$ component $\sigma_{\textrm{loop}}$ ($\sigma_{\textrm{loop}}/\Delta\sigma_{\textrm{NNLO}}^{\textrm{ext}}$) of the NNLO corrections, where applicable. Two results are reported at NNLO: $\sigma_{\textrm{NNLO}}^{\rcut}$ denotes the NNLO cross 
section at a fixed $\rcut{}$ value; the default $\rcut=0.15$\% is used throughout for our 
reference results. Our best prediction is denoted as $\sigma_{\textrm{NNLO}}^{\textrm{extrapolated}}$, 
and it is determined by the $\rcut\rightarrow 0$ extrapolation of the \rcut{} dependence 
between $\rcut=0.15$\% and $\rcut=1\%$ (see \sct{sec:qtsubtraction} for details). Both NNLO predictions are provided at the end of 
every \Matrix{} run, and for each process the results in \tab{tab:results} are taken 
from the same \Matrix{} run. The relative uncertainties, automatically computed by the code,
refer to scale variations, as defined in \sct{sec:settings}.\footnote{The automatic evaluation of PDF uncertainties is not supported 
in the first release of \Matrix{}.}
The numerical uncertainty is reported in round brackets for all our predictions.
For $\sigma_{\textrm{NNLO}}^{\textrm{extrapolated}}$ this uncertainty 
is obtained by combining the statistical uncertainty from Monte Carlo integration with the systematic uncertainty induced by the \rcut{} dependence.
The absolute size of the NNLO contributions for the extrapolated result is defined as
$\Delta\sigma_{\textrm{NNLO}}^{\textrm{ext}}=\sigma_{\textrm{NNLO}}^{\textrm{extrapolated}}-\sigma_{\textrm{NLO}}$. 
Two additional columns refer to the relative size of the radiative corrections 
in terms of $K$ factors at NLO and NNLO, defined as
\begin{align}
K_{\textrm{NLO}} = \frac{\sigma_{\textrm{NLO}}}{\sigma_{\textrm{LO}}}\quad\textrm{and}\quad K_{\textrm{NNLO}} = \frac{\sigma_{\textrm{NNLO}}}{\sigma_{\textrm{NLO}}}\,.
\end{align}
The latter are computed from our best NNLO predictions, i.e.\ the extrapolated NNLO results.

\begin{table}
\begin{center}
\resizebox{\columnwidth}{!}{%
\begin{tabular}{c c c c c c c c}
\toprule
{\bf process}
& \multirow{ 2}{*}{$\sigma_{\textrm{LO}}$}
& \multirow{ 2}{*}{$\sigma_{\textrm{NLO}}$}
& $\sigma_{\textrm{loop}}$
& \multirow{ 2}{*}{$\sigma_{\textrm{NNLO}}^{\rcut}$}
& \multirow{ 2}{*}{$\sigma_{\textrm{NNLO}}^{\textrm{extrapolated}}$}
& \multirow{ 2}{*}{$K_{\textrm{NLO}}$}
& \multirow{ 2}{*}{$K_{\textrm{NNLO}}$}\\
{\bf (\matrixparam{\$\{process\_id\}})}
&
&
& ($\sigma_{\textrm{loop}}/\Delta\sigma_{\textrm{NNLO}}^{\textrm{ext}}$)
& 
& 
& 
& \\
\midrule
$pp \to H$ 
& \multirow{ 2}{*}{$15.42(0)_{-17\%}^{+22\%}$\,pb} 
& \multirow{ 2}{*}{$30.26(1)_{-15\%}^{+20\%}$\,pb} 
& \multirow{ 2}{*}{---} 
& \multirow{ 2}{*}{$39.93(3)_{-10\%}^{+11\%}$\,pb} 
& \multirow{ 2}{*}{$39.93(3)_{-10\%}^{+11\%}$\,pb} 
& \multirow{ 2}{*}{$+96.2\%$} 
& \multirow{ 2}{*}{$+32.0\%$} \\
(\ttt{pph21}) 
& 
& 
&  
& 
& 
& 
& \\
$pp \to Z$ 
& \multirow{ 2}{*}{$43.32(0)_{-13\%}^{+12\%}$\,nb} 
& \multirow{ 2}{*}{$54.20(1)_{-4.9\%}^{+3.1\%}$\,nb} 
& \multirow{ 2}{*}{---} 
& \multirow{ 2}{*}{$56.01(3)_{-1.1\%}^{+0.84\%}$\,nb} 
& \multirow{ 2}{*}{$55.99(3)_{-1.1\%}^{+0.84\%}$\,nb} 
& \multirow{ 2}{*}{$+25.1\%$} 
& \multirow{ 2}{*}{$+3.31\%$} \\
(\ttt{ppz01}) 
& 
& 
&  
& 
& 
& 
& \\
$pp \to W^-$ 
& \multirow{ 2}{*}{$60.15(0)_{-14\%}^{+13\%}$\,nb} 
& \multirow{ 2}{*}{$75.95(2)_{-5.3\%}^{+3.3\%}$\,nb} 
& \multirow{ 2}{*}{---} 
& \multirow{ 2}{*}{$78.36(3)_{-1.2\%}^{+0.98\%}$\,nb} 
& \multirow{ 2}{*}{$78.33(8)_{-1.2\%}^{+0.98\%}$\,nb} 
& \multirow{ 2}{*}{$+26.3\%$} 
& \multirow{ 2}{*}{$+3.14\%$} \\
(\ttt{ppw01}) 
& 
& 
&  
& 
& 
& 
& \\
$pp \to W^+$ 
& \multirow{ 2}{*}{$81.28(1)_{-14\%}^{+13\%}$\,nb} 
& \multirow{ 2}{*}{$102.2(0)_{-5.3\%}^{+3.4\%}$\,nb} 
& \multirow{ 2}{*}{---} 
& \multirow{ 2}{*}{$105.8(1)_{-1.3\%}^{+0.93\%}$\,nb} 
& \multirow{ 2}{*}{$105.8(1)_{-1.3\%}^{+0.93\%}$\,nb} 
& \multirow{ 2}{*}{$+25.7\%$} 
& \multirow{ 2}{*}{$+3.52\%$} \\
(\ttt{ppwx01}) 
& 
& 
&  
& 
& 
& 
& \\
$pp \to e^- e^+$ 
& \multirow{ 2}{*}{$592.8(1)_{-14\%}^{+14\%}$\,pb} 
& \multirow{ 2}{*}{$699.7(2)_{-4.5\%}^{+2.9\%}$\,pb} 
& \multirow{ 2}{*}{---} 
& \multirow{ 2}{*}{$728.4(3)_{-0.72\%}^{+0.48\%}$\,pb} 
& \multirow{ 2}{*}{$732.7(3.4)_{-0.79\%}^{+0.43\%}$\,pb} 
& \multirow{ 2}{*}{$+18.0\%$} 
& \multirow{ 2}{*}{$+4.72\%$} \\
(\ttt{ppeex02}) 
& 
& 
&  
& 
& 
& 
& \\
$pp \to \nu_e \bar\nu_e$ 
& \multirow{ 2}{*}{$2876(0)_{-13\%}^{+12\%}$\,pb} 
& \multirow{ 2}{*}{$3585(1)_{-4.9\%}^{+3.0\%}$\,pb} 
& \multirow{ 2}{*}{---} 
& \multirow{ 2}{*}{$3705(2)_{-1.1\%}^{+0.86\%}$\,pb} 
& \multirow{ 2}{*}{$3710(2)_{-1.1\%}^{+0.85\%}$\,pb} 
& \multirow{ 2}{*}{$+24.6\%$} 
& \multirow{ 2}{*}{$+3.48\%$} \\
(\ttt{ppnenex02}) 
& 
& 
&  
& 
& 
& 
& \\
$pp \to e^- \bar\nu_e$ 
& \multirow{ 2}{*}{$2972(0)_{-15\%}^{+14\%}$\,pb} 
& \multirow{ 2}{*}{$3674(1)_{-5.2\%}^{+3.1\%}$\,pb} 
& \multirow{ 2}{*}{---} 
& \multirow{ 2}{*}{$3772(2)_{-0.94\%}^{+0.89\%}$\,pb} 
& \multirow{ 2}{*}{$3768(3)_{-0.93\%}^{+0.90\%}$\,pb} 
& \multirow{ 2}{*}{$+23.6\%$} 
& \multirow{ 2}{*}{$+2.57\%$} \\
(\ttt{ppenex02}) 
& 
& 
&  
& 
& 
& 
& \\
$pp \to e^+ \nu_e$ 
& \multirow{ 2}{*}{$3964(0)_{-14\%}^{+14\%}$\,pb} 
& \multirow{ 2}{*}{$4855(1)_{-5.1\%}^{+3.0\%}$\,pb} 
& \multirow{ 2}{*}{---} 
& \multirow{ 2}{*}{$4986(2)_{-0.95\%}^{+0.88\%}$\,pb} 
& \multirow{ 2}{*}{$4986(3)_{-0.95\%}^{+0.88\%}$\,pb} 
& \multirow{ 2}{*}{$+22.5\%$} 
& \multirow{ 2}{*}{$+2.70\%$} \\
(\ttt{ppexne02}) 
& 
& 
&  
& 
& 
& 
& \\
$pp \to \gamma\gamma$ 
& \multirow{ 2}{*}{$5.592(1)_{-11\%}^{+10\%}$\,pb} 
& \multirow{ 2}{*}{$25.75(1)_{-7.5\%}^{+8.8\%}$\,pb} 
& $2.534(1)_{-17\%}^{+24\%}$\,pb 
& \multirow{ 2}{*}{$40.86(2)_{-7.2\%}^{+8.7\%}$\,pb} 
& \multirow{ 2}{*}{$40.28(30)_{-7.0\%}^{+8.7\%}$\,pb} 
& \multirow{ 2}{*}{$+361\%$} 
& \multirow{ 2}{*}{$+56.4\%$} \\
(\ttt{ppaa02}) 
& 
& 
& ($17.4\%$) 
& 
& 
& 
& \\
$pp \to e^- e^+ \gamma$ 
& \multirow{ 2}{*}{$1469(0)_{-12\%}^{+12\%}$\,fb} 
& \multirow{ 2}{*}{$2119(1)_{-4.6\%}^{+2.9\%}$\,fb} 
& $16.02(1)_{-18\%}^{+24\%}$\,fb 
& \multirow{ 2}{*}{$2326(1)_{-1.3\%}^{+1.2\%}$\,fb} 
& \multirow{ 2}{*}{$2316(5)_{-1.2\%}^{+1.1\%}$\,fb} 
& \multirow{ 2}{*}{$+44.3\%$} 
& \multirow{ 2}{*}{$+9.29\%$} \\
(\ttt{ppeexa03}) 
& 
& 
& ($8.14\%$) 
& 
& 
& 
& \\
$pp \to \nu_e \bar\nu_e \gamma$ 
& \multirow{ 2}{*}{$63.61(1)_{-3.5\%}^{+2.7\%}$\,fb} 
& \multirow{ 2}{*}{$98.75(2)_{-2.7\%}^{+3.3\%}$\,fb} 
& $2.559(2)_{-19\%}^{+26\%}$\,fb 
& \multirow{ 2}{*}{$114.7(1)_{-2.6\%}^{+3.2\%}$\,fb} 
& \multirow{ 2}{*}{$113.5(6)_{-2.4\%}^{+2.9\%}$\,fb} 
& \multirow{ 2}{*}{$+55.2\%$} 
& \multirow{ 2}{*}{$+15.0\%$} \\
(\ttt{ppnenexa03}) 
& 
& 
& ($17.3\%$) 
& 
& 
& 
& \\
$pp \to e^- \bar\nu_e \gamma$ 
& \multirow{ 2}{*}{$726.1(1)_{-12\%}^{+11\%}$\,fb} 
& \multirow{ 2}{*}{$1850(1)_{-5.3\%}^{+6.6\%}$\,fb} 
& \multirow{ 2}{*}{---} 
& \multirow{ 2}{*}{$2286(1)_{-3.7\%}^{+4.0\%}$\,fb} 
& \multirow{ 2}{*}{$2256(15)_{-3.5\%}^{+3.7\%}$\,fb} 
& \multirow{ 2}{*}{$+155\%$} 
& \multirow{ 2}{*}{$+22.0\%$} \\
(\ttt{ppenexa03}) 
& 
& 
&  
& 
& 
& 
& \\
$pp \to e^+ \nu_e \gamma$ 
& \multirow{ 2}{*}{$861.7(1)_{-11\%}^{+10\%}$\,fb} 
& \multirow{ 2}{*}{$2187(1)_{-5.3\%}^{+6.6\%}$\,fb} 
& \multirow{ 2}{*}{---} 
& \multirow{ 2}{*}{$2707(3)_{-3.8\%}^{+4.1\%}$\,fb} 
& \multirow{ 2}{*}{$2671(35)_{-3.6\%}^{+3.8\%}$\,fb} 
& \multirow{ 2}{*}{$+154\%$} 
& \multirow{ 2}{*}{$+22.1\%$} \\
(\ttt{ppexnea03}) 
& 
& 
&  
& 
& 
& 
& \\
$pp \to Z Z$ 
& \multirow{ 2}{*}{$9.845(1)_{-6.3\%}^{+5.2\%}$\,pb} 
& \multirow{ 2}{*}{$14.10(0)_{-2.4\%}^{+2.9\%}$\,pb} 
& $1.361(1)_{-19\%}^{+25\%}$\,pb 
& \multirow{ 2}{*}{$16.68(1)_{-2.6\%}^{+3.2\%}$\,pb} 
& \multirow{ 2}{*}{$16.67(1)_{-2.6\%}^{+3.2\%}$\,pb} 
& \multirow{ 2}{*}{$+43.3\%$} 
& \multirow{ 2}{*}{$+18.2\%$} \\
(\ttt{ppzz02}) 
& 
& 
& ($52.9\%$) 
& 
& 
& 
& \\
$pp \to W^+ W^-$ 
& \multirow{ 2}{*}{$66.64(1)_{-6.7\%}^{+5.7\%}$\,pb} 
& \multirow{ 2}{*}{$103.2(0)_{-3.1\%}^{+3.9\%}$\,pb} 
& $4.091(3)_{-19\%}^{+27\%}$\,pb 
& \multirow{ 2}{*}{$117.1(1)_{-2.2\%}^{+2.5\%}$\,pb} 
& \multirow{ 2}{*}{$117.1(1)_{-2.2\%}^{+2.5\%}$\,pb} 
& \multirow{ 2}{*}{$+54.9\%$} 
& \multirow{ 2}{*}{$+13.4\%$} \\
(\ttt{ppwxw02}) 
& 
& 
& ($29.5\%$) 
& 
& 
& 
& \\
$pp \to e^- \mu^- e^+ \mu^+$ 
& \multirow{ 2}{*}{$11.34(0)_{-7.3\%}^{+6.3\%}$\,fb} 
& \multirow{ 2}{*}{$16.87(0)_{-2.5\%}^{+3.0\%}$\,fb} 
& $1.971(1)_{-18\%}^{+25\%}$\,fb 
& \multirow{ 2}{*}{$20.30(1)_{-2.9\%}^{+3.5\%}$\,fb} 
& \multirow{ 2}{*}{$20.30(1)_{-2.9\%}^{+3.5\%}$\,fb} 
& \multirow{ 2}{*}{$+48.8\%$} 
& \multirow{ 2}{*}{$+20.3\%$} \\
(\ttt{ppemexmx04}) 
& 
& 
& ($57.6\%$) 
& 
& 
& 
& \\
$pp \to e^- e^- e^+ e^+$ 
& \multirow{ 2}{*}{$5.781(1)_{-7.4\%}^{+6.3\%}$\,fb} 
& \multirow{ 2}{*}{$8.623(3)_{-2.5\%}^{+3.1\%}$\,fb} 
& $0.9941(4)_{-18\%}^{+25\%}$\,fb 
& \multirow{ 2}{*}{$10.37(1)_{-3.0\%}^{+3.5\%}$\,fb} 
& \multirow{ 2}{*}{$10.37(1)_{-3.0\%}^{+3.5\%}$\,fb} 
& \multirow{ 2}{*}{$+49.2\%$} 
& \multirow{ 2}{*}{$+20.2\%$} \\
(\ttt{ppeeexex04}) 
& 
& 
& ($56.9\%$) 
& 
& 
& 
& \\
$pp \to e^- e^+ \nu_\mu \bar\nu_\mu$ 
& \multirow{ 2}{*}{$22.34(0)_{-6.4\%}^{+5.3\%}$\,fb} 
& \multirow{ 2}{*}{$33.90(1)_{-2.7\%}^{+3.3\%}$\,fb} 
& $3.212(1)_{-19\%}^{+25\%}$\,fb 
& \multirow{ 2}{*}{$40.39(2)_{-2.8\%}^{+3.5\%}$\,fb} 
& \multirow{ 2}{*}{$40.38(2)_{-2.8\%}^{+3.5\%}$\,fb} 
& \multirow{ 2}{*}{$+51.7\%$} 
& \multirow{ 2}{*}{$+19.1\%$} \\
(\ttt{ppeexnmnmx04}) 
& 
& 
& ($49.6\%$) 
& 
& 
& 
& \\
$pp \to e^- \mu^+ \nu_\mu \bar\nu_e $ 
& \multirow{ 2}{*}{$232.9(0)_{-7.6\%}^{+6.6\%}$\,fb} 
& \multirow{ 2}{*}{$236.1(1)_{-2.4\%}^{+2.8\%}$\,fb} 
& $26.93(1)_{-19\%}^{+27\%}$\,fb 
& \multirow{ 2}{*}{$264.7(1)_{-1.4\%}^{+2.2\%}$\,fb} 
& \multirow{ 2}{*}{$264.6(2)_{-1.4\%}^{+2.2\%}$\,fb} 
& \multirow{ 2}{*}{$+1.34\%$} 
& \multirow{ 2}{*}{$+12.1\%$} \\
(\ttt{ppemxnmnex04}) 
& 
& 
& ($94.3\%$) 
& 
& 
& 
& \\
$pp \to e^- e^+ \nu_e \bar\nu_e$ 
& \multirow{ 2}{*}{$115.0(0)_{-7.3\%}^{+6.3\%}$\,fb} 
& \multirow{ 2}{*}{$203.4(1)_{-3.8\%}^{+4.7\%}$\,fb} 
& $12.62(1)_{-19\%}^{+26\%}$\,fb 
& \multirow{ 2}{*}{$240.8(1)_{-3.0\%}^{+3.4\%}$\,fb} 
& \multirow{ 2}{*}{$240.7(1)_{-3.0\%}^{+3.4\%}$\,fb} 
& \multirow{ 2}{*}{$+76.9\%$} 
& \multirow{ 2}{*}{$+18.4\%$} \\
(\ttt{ppeexnenex04}) 
& 
& 
& ($33.8\%$) 
& 
& 
& 
& \\
$pp \to e^- \mu^- e^+ \bar\nu_\mu $ 
& \multirow{ 2}{*}{$11.50(0)_{-6.8\%}^{+5.7\%}$\,fb} 
& \multirow{ 2}{*}{$23.55(1)_{-4.5\%}^{+5.5\%}$\,fb} 
& \multirow{ 2}{*}{---} 
& \multirow{ 2}{*}{$26.17(1)_{-2.1\%}^{+2.2\%}$\,fb} 
& \multirow{ 2}{*}{$26.17(2)_{-2.1\%}^{+2.2\%}$\,fb} 
& \multirow{ 2}{*}{$+105\%$} 
& \multirow{ 2}{*}{$+11.1\%$} \\
(\ttt{ppemexnmx04}) 
& 
& 
&  
& 
& 
& 
& \\
$pp \to e^- e^- e^+ \bar\nu_e $ 
& \multirow{ 2}{*}{$11.53(0)_{-6.8\%}^{+5.7\%}$\,fb} 
& \multirow{ 2}{*}{$23.63(1)_{-4.5\%}^{+5.5\%}$\,fb} 
& \multirow{ 2}{*}{---} 
& \multirow{ 2}{*}{$26.27(1)_{-2.1\%}^{+2.3\%}$\,fb} 
& \multirow{ 2}{*}{$26.25(2)_{-2.1\%}^{+2.3\%}$\,fb} 
& \multirow{ 2}{*}{$+105\%$} 
& \multirow{ 2}{*}{$+11.1\%$} \\
(\ttt{ppeeexnex04}) 
& 
& 
&  
& 
& 
& 
& \\
$pp \to e^- e^+ \mu^+ \nu_\mu$ 
& \multirow{ 2}{*}{$17.33(0)_{-6.3\%}^{+5.3\%}$\,fb} 
& \multirow{ 2}{*}{$34.14(1)_{-4.3\%}^{+5.3\%}$\,fb} 
& \multirow{ 2}{*}{---} 
& \multirow{ 2}{*}{$37.74(2)_{-2.0\%}^{+2.2\%}$\,fb} 
& \multirow{ 2}{*}{$37.74(4)_{-2.0\%}^{+2.2\%}$\,fb} 
& \multirow{ 2}{*}{$+97.0\%$} 
& \multirow{ 2}{*}{$+10.6\%$} \\
(\ttt{ppeexmxnm04}) 
& 
& 
&  
& 
& 
& 
& \\
$pp \to e^- e^+ e^+ \nu_e$ 
& \multirow{ 2}{*}{$17.37(0)_{-6.3\%}^{+5.3\%}$\,fb} 
& \multirow{ 2}{*}{$34.21(2)_{-4.3\%}^{+5.3\%}$\,fb} 
& \multirow{ 2}{*}{---} 
& \multirow{ 2}{*}{$37.85(2)_{-2.0\%}^{+2.3\%}$\,fb} 
& \multirow{ 2}{*}{$37.84(3)_{-2.0\%}^{+2.3\%}$\,fb} 
& \multirow{ 2}{*}{$+96.9\%$} 
& \multirow{ 2}{*}{$+10.6\%$} \\
(\ttt{ppeexexne04}) 
& 
& 
&  
& 
& 
& 
& \\
\bottomrule

\end{tabular}}
\end{center}
\renewcommand{\baselinestretch}{1.0}
\caption{\label{tab:results} Integrated cross sections for all available processes in \Matrix{} using the default setups.}
\end{table}

For all production processes involving massive on-shell bosons ($H$, $Z$, $W^\pm$, $W^+W^-$ and $ZZ$ production), \tab{tab:results} reports fully inclusive cross sections, i.\,e. no phase-space cuts are applied. For all remaining processes, phase-space cuts are applied on 
the final-state leptons, neutrinos and photons in order to simulate a realistic selection in a fiducial volume.
The respective sets of cuts for each of these processes are discussed below.
For detailed studies of phenomenological results we refer to dedicated publications on the respective processes. We restrict ourselves to summarizing
basic features of the calculations and the overall effect of the higher-order QCD corrections.

\subparagraph{Higgs boson production}\quad

Radiative corrections to Higgs boson production are known to be particularly large. The corresponding cross sections in \tab{tab:results} have been computed 
with fixed renormalization and factorization scales set to $\mu_0=\mH$. The results have been checked to be in perfect agreement within the quoted numerical 
uncertainty with the analytic code {\sc SusHi} \cite{Harlander:2012pb}. We find $K_{\textrm{NLO}}=1.96$ and $K_{\textrm{NNLO}}=1.32$ for the NLO and NNLO $K$ factors, 
respectively. As it is well known \cite{Harlander:2002wh,Anastasiou:2002yz,Ravindran:2003um}, scale variations significantly decrease upon inclusion of radiative corrections, but at LO and NLO they do not reflect the actual size of missing higher-order 
contributions.

\newcommand{\ptgamma}{\ensuremath{p_{T,\gamma}}}
\newcommand{\ptgammaone}{\ensuremath{p_{T,\gamma_1}}}
\newcommand{\ptgammatwo}{\ensuremath{p_{T,\gamma_2}}}
\newcommand{\ptlep}{\ensuremath{p_{T,{\ell}}}}
\newcommand{\etalep}{\ensuremath{|\eta_{\ell}|}}
\newcommand{\etagamma}{\ensuremath{|\eta_{\gamma}|}}
\newcommand{\mll}{\ensuremath{m_{\ell^-\ell^+}}}
\newcommand{\ptmiss}{\ensuremath{p_{T}^{\text{miss}}}}
\newcommand{\ptmissrel}{\ensuremath{p_{T}^{\text{miss,rel}}}}
\newcommand{\dRlepjet}{\ensuremath{\Delta R_{\ell j}}}
\newcommand{\dRgammajet}{\ensuremath{\Delta R_{\gamma j}}}
\newcommand{\dRlepgamma}{\ensuremath{\Delta R_{\ell\gamma}}}
\newcommand{\dRleplep}{\ensuremath{\Delta R_{\ell\ell}}}
\newcommand{\mww}{\ensuremath{m_{WW}}}
\newcommand{\mtatlas}{\ensuremath{m_T^{\rm ATLAS}}}
\newcommand{\ptww}{\ensuremath{p_{T,WW}}}
\newcommand{\ptw}{\ensuremath{p_{T,W}}}
\newcommand{\ptwone}{\ensuremath{p_{T,W_1}}}
\newcommand{\ptwtwo}{\ensuremath{p_{T,W_2}}}
\newcommand{\ptwp}{\ensuremath{p_{T,W^-}}}
\newcommand{\ptwm}{\ensuremath{p_{T,W^+}}}
\newcommand{\ptlone}{\ensuremath{p_{T,\ell_1}}}
\newcommand{\ptltwo}{\ensuremath{p_{T,\ell_2}}}
\newcommand{\Etlone}{\ensuremath{E_{T,\ell_1}}}
\newcommand{\Etltwo}{\ensuremath{E_{T,\ell_2}}}
\newcommand{\ptli}{\ensuremath{p_{T,\ell_i}}}
\newcommand{\ptll}{\ensuremath{p_{T,\ell\ell}}}
\newcommand{\ymu}{\ensuremath{y_\mu}}
\newcommand{\ye}{\ensuremath{y_e}}
\newcommand{\dphillnunu}{\ensuremath{\Delta\phi_{\ell\ell,\nu\nu}}}
\newcommand{\dphill}{\ensuremath{\Delta\phi_{\ell\ell}}}
\newcommand{\dRll}{\ensuremath{\Delta R_{\ell\ell}}}
\newcommand{\dRej}{\ensuremath{\Delta R_{e,j}}}
\newcommand{\dRllp}{\ensuremath{\Delta R_{\ell\ell'}}}
\newcommand{\etal}{\ensuremath{\eta_{\ell}}}
\newcommand{\etalp}{\ensuremath{\eta_{\ell'}}}

\newcommand{\mlll}{\ensuremath{m_{\ell\ell\ell}}}
\newcommand{\mwz}{\ensuremath{m_{WZ}}}
\newcommand{\mtw}{\ensuremath{m_{T,W}}}
\newcommand{\ptwz}{\ensuremath{p_{T,WZ}}}
\newcommand{\ptz}{\ensuremath{p_{T,Z}}}
\newcommand{\ptl}{\ensuremath{p_{T,\ell}}}
\newcommand{\ptlp}{\ensuremath{p_{T,\ell'}}}
\newcommand{\ptlsub}{\ensuremath{p_{T,\ell_{\ge2}}}}
\newcommand{\lw}{\ensuremath{\ell_{\textrm{w}}}}
\newcommand{\lpw}{\ensuremath{\ell^+_{\textrm{w}}}}
\newcommand{\lmw}{\ensuremath{\ell^-_{\textrm{w}}}}
\newcommand{\lpmw}{\ensuremath{\ell^{\pm}_{\textrm{w}}}}

\newcommand{\lz}{\ensuremath{\ell_{\textrm{z}}}}
\newcommand{\lpz}{\ensuremath{\ell^+_{\textrm{z}}}}
\newcommand{\lmz}{\ensuremath{\ell^-_{\textrm{z}}}}
\newcommand{\lpmz}{\ensuremath{\ell^{\pm}_{\textrm{z}}}}
\newcommand{\lzlead}{\ensuremath{\ell_{\textrm{z},1}}}
\newcommand{\lzsubl}{\ensuremath{\ell_{\textrm{z},2}}}

\newcommand{\ptlz}{\ensuremath{p_{T,\lz}}}
\newcommand{\ptlw}{\ensuremath{p_{T,\lw}}}
\newcommand{\ptlzlead}{\ensuremath{p_{T,\lzlead}}}
\newcommand{\ptlzsubl}{\ensuremath{p_{T,\lzsubl}}}

\subparagraph{Drell--Yan production}\quad

On-shell Drell--Yan production is another well-studied process, and it was
the first hadron-collider process for which NNLO corrections were computed~\cite{Hamberg:1990np,Harlander:2002wh}.
The results reported in \tab{tab:results} are obtained with renormalization and factorization scales set to $\mu_0=m_Z$ and $\mu_0=m_W$ for $pp\to Z$ and $pp\to W^\pm$,
respectively. The same fixed scales are applied to the corresponding off-shell processes. The Drell--Yan cross section is more than three orders of magnitude larger than the one 
of Higgs boson production, and the impact of radiative corrections is known to be smaller: 
NLO corrections increase the LO result by about 25\%, and NNLO corrections amount to a further +3\% effect.
$pp\to Z$ has been checked numerically against the analytic result of \citere{Hamberg:1990np,Harlander:2002wh}, and we have validated the CKM implementation by 
finding full agreement at the level of the numerical errors for $pp\to e^\pm\nu$ with \FEWZ{}~\cite{Gavin:2012sy} and {\sc DYNNLO}~\cite{Catani:2009sm}.

The cross sections of the charged-current and neutral-current Drell--Yan processes correspond
in a first approximation to the on-shell $W$ or $Z$ production cross sections times the respective
leptonic branching ratios.
Consequently, they decrease by at least one order of magnitude with respect to the on-shell case.

\renewcommand\arraystretch{1.5}
\begin{table}
\begin{center}
\resizebox{\columnwidth}{!}{%
\begin{tabular}{l | c | c | c}
\toprule
& $pp\to e^-e^+$
& $pp\to e^-\bar\nu_e/pp\to e^+\nu_e$
& $pp\to \gamma\gamma$\\
\midrule
\multirow{ 2}{*}{\bf lepton cuts}
& $\ptlep > 25$\,GeV, $\etalep<2.47$
\multirow{ 2}{*}& $\ptlep > 25$\,GeV, $\etalep<2.47$ 
\multirow{ 2}{*}& ------ \\[-0.1cm]
&$66{\rm\,GeV}<\mll<116$\,GeV
&& \\
\multirow{ 2}{*}{\bf photon cuts} 
& \multirow{ 2}{*}{------} 
& \multirow{ 2}{*}{------} 
& $\ptgammaone > 40$\,GeV, $\ptgammatwo > 25$\,GeV \\[-0.1cm]
&&& $\etagamma<2.5$ \\
{\bf neutrino cuts}
& ------ 
& $\ptmiss >20$\,GeV 
& ------ \\[-0.5cm]
&&& \\
\multirow{ 3}{*}{\bf photon isolation} 
& \multirow{ 3}{*}{------} 
& \multirow{ 3}{*}{------} 
& Frixione isolation with\\[-0.1cm]
&&& $n=1$, $\varepsilon = 0.5$ and $\delta_0=0.4$ \\[-0.5cm] 
\\
{\bf jet definition}
& \multicolumn{3}{c}{anti-$k_T$ algorithm with $R=0.4$;\; $\ptjet>25$\,GeV, $|\eta_j|<4.5$}\\
\bottomrule
\end{tabular}}
\end{center}
\renewcommand{\baselinestretch}{1.0}
\caption{\label{tab:WZgammagamma} Default setup of fiducial cuts for $Z$, $W^\pm$ and $\gamma\gamma$ production processes.}
\vspace{0.75cm}
\end{table}

The following sets of cuts, which are also summarized in \refta{tab:WZgammagamma}, are applied to these processes: 
Every final-state lepton is required to have a minimum transverse momentum of $\ptlep{} > 25$\,GeV and a maximal pseudo-rapidity $\etalep{} < 2.47$. Neutrinos originating 
from a $W$-boson decay are restricted by a minimal requirement on the total missing transverse momentum, $\ptmiss > 20$\,GeV.
Additionally, we require $66$\,GeV$<\mll<116$\,GeV for the invariant mass of the two leptons in $pp\to e^-e^+$.
The lower cut separates the leptons to avoid singularities arising from the photon-mediated contributions to this process.
The process $pp\to \nu_e\bar\nu_e$, which is only relevant as a technical check, is calculated
without any phase-space cuts.
We find that, except for the $pp\to e^-e^+$ process, which is affected by perturbative 
instabilities as discussed below, 
the off-shell contributions and additional phase-space cuts hardly have 
any effect on radiative corrections, which remain $K_{\rm NLO}\sim 1.25$
and  $K_{\rm NNLO}\sim 1.03$ as in the on-shell case. 
The ratio of $W^-$ and $W^+$ cross sections does not significantly differ
between on-shell and 
off-shell $W^\pm$ production: We find roughly $\sigma_{W^-}/\sigma_{W^+}\sim 0.75$, essentially independent of the perturbative order.

We note that the $pp\to e^-e^+$ process has a peculiarly large \rcut{} dependence at NNLO, 
similar to the processes involving final-state photons, thereby leading to a rather large systematic uncertainty.
The large \rcut{} dependence is due to the choice of symmetric \pt{} thresholds on the 
leptons,
which causes perturbative instabilities in the integrated cross section,
as first observed in \citere{Frixione:1997ks} (for a recent discussion of this problem, see \citere{Alioli:2016fum}).
However, choosing asymmetric \pt{} cuts on the harder and softer lepton does not 
reduce this dependence significantly. On the contrary,
we find that if the asymmetric cuts are separately applied on the negatively and positively charged leptons
(instead of applying them on the harder and softer lepton)
the ensuing \rcut{} behaviour is extremely flat and therefore a small 
$\rcut\rightarrow 0$ extrapolation uncertainty is obtained.
The \rcut{} behaviour of the $pp\to e^-e^+$ process is 
discussed in \sct{sec:qtsubtraction}, where we also present a comparison of our results with \FEWZ{}. 
We stress that more accurate results for this process can be obtained through the setting of \matrixparam{switch_qT_accuracy = 1} in the file \matrixparam{parameter.dat} by using 
a minimal value of $\rcut=0.05$\% (default is $\rcut=0.15$\%) for the extrapolation range,
see \sct{sec:ppeex02}.

\subparagraph{Diphoton and vector-boson plus photon production}\quad

For diphoton production we choose the invariant mass of the photon pair 
as the central scale, i.e.\ $\mu_0=m_{\gamma\gamma}$, 
Frixione isolation with $n=1$, $\varepsilon = 0.5$ and $\delta_0=0.4$ (see \eqn{eq:frixione}), and the following fiducial cuts, which are also summarized in \refta{tab:WZgammagamma}: 
The isolated photons are required to have a pseudo-rapidity $\etagamma<2.5$, 
and the transverse momentum of the (sub)leading photon must fulfil $\ptgamma>40(25)$\,GeV.
Our predictions show the importance of QCD corrections for this process: With $K_{\rm NLO} = 4.61$ and $K_{\rm NNLO} = 1.56$, higher-order 
effects are enormous~\cite{Catani:2011qz,Campbell:2016yrh} and not at all reflected by the estimated scale uncertainties at lower orders. 
This process entails a loop-induced $gg$ component in the NNLO cross section. 
With only a $17$\% contribution to the NNLO correction it has a rather moderate 
impact though. Our results have been compared with the results of \citere{Catani:2011qz}, which have been calculated with the {\sc 2$\gamma$NNLO} code, and we find numerical agreement within the respective uncertainties.

Next, we consider the associated production of an off-shell vector boson with a photon, 
i.e.\ the leptonic final states
$e^-e^+\gamma$/$\nu_e\bar\nu_e\gamma$ (summarized as $Z\gamma$ production) and 
$e^+\nu_e\gamma$/$e^-\bar\nu_e\gamma$ (summarized as $W\gamma$ production).\footnote{We note again that $Z\gamma$ and $W\gamma$ are only used as shorthand notations here. 
The full amplitudes for the leptonic final states are used throughout without any approximation, including off-shell effects and spin correlations.}
Our setup is adopted from \citere{Grazzini:2015nwa}:
The dynamical scale $\mu_0=\sqrt{m_V^2+p_{T,\gamma}^2}$ is chosen as central value for both 
renormalization and factorization scales, where $m_V=m_Z$ for $Z\gamma$ and $m_V=m_W$ for $W\gamma$ production. 
As for diphoton production, Frixione isolation with $n=1$, $\varepsilon = 0.5$ and $\delta_0=0.4$ (see \eqn{eq:frixione}) is used to identify photons.
The fiducial cuts include standard cuts on leptons, photons and the missing transverse momentum, as well as lepton--photon, lepton--jet and photon--jet separations in $R$.
The numerical values of these cuts are summarized in \tab{tab:Vgamma}.
\renewcommand\arraystretch{1.5}
\begin{table}
\begin{center}
\resizebox{\columnwidth}{!}{%
\begin{tabular}{l | c | c | c}
\toprule
& $pp\to e^-e^+\gamma$
& $pp\to \nu_e\bar\nu_e\gamma$
& $pp\to e^-\bar\nu_e\gamma/pp\to e^+\nu_e\gamma$\\
\midrule
\multirow{ 2}{*}{\bf lepton cuts} 
& $\ptlep > 25$\,GeV, $\etalep<2.47$
& \multirow{ 2}{*}{------} 
& \multirow{ 2}{*}{$\ptlep > 25$\,GeV, $\etalep<2.47$} \\[-0.1cm]
& $\mll > 40$\,GeV
&& \\
{\bf photon cuts}
& $\ptgamma > 15$\,GeV, $\etagamma<2.37$
& $\ptgamma > 100$\,GeV, $\etagamma<2.37$
& $\ptgamma > 15$\,GeV, $\etagamma<2.37$ \\
{\bf neutrino cuts}
&  ------ 
& $\ptmiss >90$\,GeV & $\ptmiss >35$\,GeV\\
\multirow{ 2}{*}{\bf separation cuts} 
& $\dRlepjet>0.3$, $\dRgammajet >0.3$,
& \multirow{ 2}{*}{$\dRgammajet>0.3$}
& $\dRlepjet>0.3$, $\dRgammajet >0.3$, \\[-0.1cm]
& $\dRlepgamma>0.7$ 
&& $\dRlepgamma>0.7$ \\[-0.5cm]
\\
{\bf photon isolation}
& \multicolumn{3}{c}{Frixione isolation with $n=1$, $\varepsilon = 0.5$ and $\delta_0=0.4$}\\[0.2cm] 
{\bf jet definition}
& \multicolumn{3}{c}{anti-$k_T$ algorithm with $R=0.4$;\; $\ptjet>30$\,GeV, $|\eta_j|<4.4$}\\
\bottomrule
\end{tabular}}
\end{center}
\renewcommand{\baselinestretch}{1.0}
\caption{\label{tab:Vgamma} Default setup of fiducial cuts for $Z\gamma$ and $W^\pm\gamma$ production processes.}
\vspace{0.75cm}
\end{table}

The $Z\gamma$ processes feature large corrections, $K_{\rm NLO}(K_{\rm NNLO})=1.44(1.09)$ for $pp\to e^-e^+\gamma$ and 
$K_{\rm NLO}(K_{\rm NNLO})=1.55(1.15)$ for $pp\to \nu_e\bar\nu_e\gamma$. 
For $W\gamma$ production, radiative corrections are known to be huge due to a radiation zero at LO \cite{Mikaelian:1979nr}:
At NLO the $W^\pm\gamma$ cross section is increased by more than +150\%, and the NNLO corrections have a further effect of $+22$\%. 
The ratio between the $W^-\gamma$ and $W^+\gamma$ cross sections
is roughly $\sigma_{W^-\gamma}/\sigma_{W^+\gamma}\sim 0.75$, widely independent of the perturbative order and very similar to the ratio of the charged Drell--Yan processes. 

One should bear in mind that all processes with isolated photons in the final state have 
a relatively large uncertainty at NNLO ($\sim 0.5\%-1$\%) even after the $\rcut\rightarrow 0$ extrapolation (which in the default setup is based on the \rcut{} dependence between $\rcut=0.15$\% and $\rcut=1$\%).
More accurate results can be obtained by setting \matrixparam{switch_qT_accuracy = 1} in the file \matrixparam{parameter.dat}, which uses a minimal value of $\rcut=0.05$\% for the 
extrapolation range,
see \sct{sec:photonprocesses}.
We refer to the discussion in \sct{sec:qtsubtraction} for details.

\subparagraph{Vector-boson pair production}\quad

The on-shell \zz{} and \ww{} results in \tab{tab:results} correspond to the inclusive cross sections of \citere{Cascioli:2014yka} and \citere{Gehrmann:2014fva}, respectively, with an updated set of input parameters. We have explicitly checked 
 that \Matrix{} reproduces the results of \citeres{Cascioli:2014yka,Gehrmann:2014fva} when adjusting the setup accordingly. Consistent with these studies, we have used fixed renormalization and factorization
scales of $\mu_0=m_Z$ and $\mu_0=m_W$ for \zz{} and \ww{} production, respectively. Radiative corrections are large for both processes: The NLO corrections amount to $+43$\% at NLO and still 
$+18$\% at NNLO in the case of \zz{} production, where a bit more than half of the NNLO corrections originates from the loop-induced $gg$ channel, though. The predicted \ww{} 
cross section receives NLO corrections of $+55$\%, and NNLO corrections lead to a further increase by $+13$\%, a third of which results from the loop-induced $gg$ contribution. 
For both processes the corrections exceed by far the perturbative uncertainties estimated by scale variations at lower orders. This is caused, in part, by the 
additional contribution from the $gg$ component, which is not covered by NLO scale variations.
The purely gluon-induced NLO corrections to the $gg$ channel, which are part of a complete N$^3$LO calculation, have been
computed in \citeres{Caola:2015psa,Caola:2015rqy}.

Several leptonic channels originate from off-shell \zz{} production. 
They involve the SF and DF four-lepton channels, $4\ell$ and $2\ell2\ell'$, respectively,
which have been studied at NNLO QCD in \citere{Grazzini:2015hta}.
On the other hand, one of the $Z$ bosons may decay to two neutrinos instead. 
In that case the SF channel is defined as the one where 
the neutrino flavour matches the lepton flavour ($2\ell2\nu$), while the 
DF flavour channel is defined as the one where the lepton and neutrino
flavours are different ($2\ell2\nu'$).\footnote{We note that both final states contain
an OSSF lepton pair and (possibly) missing transverse momentum from the 
two neutrinos that cannot be detected. Our distinction into SF and DF final states
is motivated more by the underlying technical calculations than by their phenomenology
in this case.}
The SF $2\ell2\nu$ final state is special since 
it receives contributions from both resonant \zz{} and \ww{} sub-topologies, 
which mix the two processes. 
From an experimental viewpoint, in the \zz{} or \ww{} analyses the two production mechanisms  
are isolated by using suitable cuts that enhance the 
respective process in its signal region. Since we include all resonant and 
non-resonant topologies leading to such final states, our computation 
of $2\ell2\nu$ is applicable to both \zz{} and \ww{} studies
by simply applying the corresponding cuts.
NNLO cross sections for the $2\ell2\nu$ and $2\ell2\nu'$ channels 
are reported here for the first time. A detailed study of these processes 
will be presented elsewhere.

For the off-shell $ZZ$ processes we use the setup of \citere{Grazzini:2015hta}: 
The renormalization and factorization scales are fixed to $\mu_0=m_Z$. 
The fiducial cuts are summarized in \tab{tab:ZZ}. 
\renewcommand\arraystretch{1.5}
\begin{table}
\begin{center}
\resizebox{\columnwidth}{!}{%
\begin{tabular}{l | c | c }
\toprule
& $pp \to e^- \mu^- e^+ \mu^+$/$pp \to e^- e^- e^+ e^+$
& $pp \to e^- e^+ \nu_e \bar\nu_e$/$pp \to e^- e^+ \nu_\mu \bar\nu_\mu$ \\
\midrule
\multirow{ 2}{*}{\bf lepton cuts}
& $\ptlep > 7$\,GeV, $\etalep<2.7$
& $\ptlep > 7$\,GeV, $\etalep<2.7$ \\[-0.1cm]
& $66{\rm\,GeV}<\mll<116$\,GeV
& $66{\rm\,GeV}<\mll<116$\,GeV \\
{\bf neutrino cuts}
& ------
& $\ptmiss >30$\,GeV \\
{\bf separation cuts}
& $\dRleplep>0.2$
& ------ \\[-0.5cm]
\\
{\bf jet definition}
& \multicolumn{2}{c}{anti-$k_T$ algorithm with $R=0.4$;\; $\ptjet>25$\,GeV, $|\eta_j|<4.5$}\\
\bottomrule
\end{tabular}}
\end{center}
\renewcommand{\baselinestretch}{1.0}
\caption{\label{tab:ZZ} Default setup of fiducial cuts for \zz{} and \zz{}/\ww{} production processes.}
\vspace{0.75cm}
\end{table}

They involve standard 
transverse-momentum and rapidity thresholds for the leptons,
and a lepton--lepton separation in $R$.
In the $2\ell2\ell'$ channel, the invariant mass of OSSF lepton pairs is required to be
in a mass window around the $Z$ peak. In the 
$4\ell$ channel, there are two possible combinations of OSSF lepton pairs that 
can be associated with the parent $Z$ bosons. We choose the combination 
which minimizes 
$\left| m_{\ell^-\ell^+}-m_Z\right|+\left|m_{\ell^{-\prime}\ell^{+\prime}}-m_Z\right|$, see \sct{sec:ZZSF} for details,
and apply the invariant mass cuts only on the corresponding 
lepton pairs. Since no dedicated phenomenological studies of
the $2\ell2\nu$/$2\ell2\nu'$
signatures at NNLO exist, we simply adopt
the $2\ell2\ell'$ setup, while removing the $R$ separation of the leptons and adding 
a loose cut on the missing transverse energy of the neutrinos. This choice provides 
a generic benchmark scenario for these processes.

Comparing the SF $4\ell$ process $pp \to e^- e^- e^+ e^+$ and the DF $2\ell2\ell'$ process 
$pp \to e^- \mu^- e^+ \mu^+$ 
in \tab{tab:results}, it is obvious that they give very similar results,
taking into account the relative combinatorial factor of one-half in the SF channel
(if the two SF channels $pp \to e^- e^- e^+ e^+$ and $pp \to \mu^- \mu^- \mu^+ \mu^+$ 
are added, DF and SF channels would be of the same size again).
It is not surprising that the 
$K$ factors for the $4\ell$, $2\ell 2\ell'$ and $2\ell 2\nu'$ channels are 
very close, given the fact that very similar cuts are applied and that the dominant 
contribution results from resonant \zz{} production in all these processes.
The NLO corrections amount to 
roughly +50\%, and the cross sections are increased by further $\sim+20\%$ at NNLO, i.e.\ 
radiative corrections in the fiducial regions are even a bit larger than for the 
inclusive ZZ cross section. 
For the $4\ell$ and $2\ell 2\ell'$ channels, 
the loop-induced $gg$ component has a slightly bigger impact ($\sim 57\%$ of the NNLO corrections) in the fiducial phase space 
than in the fully inclusive case ($\sim 53\%$), whereas it contributes a bit less  
for the $2\ell 2\nu'$ channel ($\sim 50$\%). 

The SF $2\ell 2\nu$ channel, 
on the other hand, shows a rather different behaviour due to the large impact of its 
\ww{} topologies, which are expected to dominate by about an order of magnitude
due to the involved EW couplings and branching ratios. Under the quite loose cuts,
compared to a dedicated $ZZ\to\ell\ell\nu\nu$ selection, the cross section at LO is still
about a factor of five larger than for the $2\ell 2\nu'$ process, due to the dominance
of \ww{} contributions. This cross section receives somewhat larger corrections than 
the ``pure'' \zz{} processes, namely $K_{\rm NLO}=1.77$ and $K_{\rm NNLO}=1.18$, where 
the $gg$ component contributes $33$\% of the NNLO corrections, comparable to on-shell 
\ww{} production. 

The off-shell \ww{} process with DF leptons ($\ell\nu\ell'\nu'$), 
namely $pp \to e^- \mu^+ \nu_\mu \bar\nu_e$, has been studied 
at NNLO in \citere{Grazzini:2016ctr}. We adopt the fixed scale choice of $\mu_0=m_W$ 
and the fiducial cuts used in that study. The latter are summarized in \tab{tab:WWandWZ}:
\renewcommand\arraystretch{1.5}
\begin{table}
\begin{center}
\resizebox{\columnwidth}{!}{%
\begin{tabular}{l | c | c}
\toprule
& $pp \to e^- \mu^+ \nu_\mu \bar\nu_e $ 
& $pp\to \ell'^\pm{\nu}_{\ell^\prime} \ell^+\ell^-,\quad \ell,\ell'\in\{e,\mu\}$ \\
\midrule
\multirow{ 3}{*}{\bf lepton cuts} 
& $\ptlone >25$\,GeV, $\ptltwo >20$\,GeV 
& $\ptlz>15$\,GeV, $\ptlw>20$\,GeV \\[-0.1cm]
& $|\eta_e|<2.47$, $|\eta_e|\notin[1.37;1.52]$
& $|\etal|<2.5$ \\[-0.1cm]
& $|\eta_\mu|<2.4$, $\mll>10$\,GeV 
& $|m_{\lz\lz}-m_Z|<10$\,GeV \\
{\bf neutrino cuts}
& $\ptmiss >30$\,GeV, $\ptmissrel >15$\,GeV 
& $\mtw> 30$\,GeV \\
{\bf separation cuts}
& $\dRleplep>0.1$ 
& $\Delta R_{\lz\lz} >0.2$, $\Delta R_{\lz\lw}>0.3$ \\
{\bf jet cuts}
& $N_{\mathrm{jets}}=0$ & ------\\[-0.5cm]
\\
{\bf jet definition} & \multicolumn{2}{c}{anti-$k_T$ algorithm with $R=0.4$;\; $\ptjet>25$\,GeV, $|\eta_j|<4.5$} \\
\bottomrule
\end{tabular}}
\end{center}
\renewcommand{\baselinestretch}{1.0}
\caption{\label{tab:WWandWZ} Default setup of fiducial cuts for \ww{} and \wz{} production processes.}
\vspace{0.75cm}
\end{table}

Besides standard cuts like transverse momentum thresholds, rapidity ranges and 
different isolation criteria, the selection cut with the largest impact on the size of 
higher-order corrections is a jet veto, which is required in \ww{} analyses to suppress 
top-quark backgrounds. 
As a consequence of the jet veto, and in contrast 
to the inclusive \ww{} cross section, the fiducial cross section receives 
very small radiative corrections, only $+1.3$\% at NLO. The NNLO 
corrections amount to $+12$\%, but they are almost entirely due to the loop-induced $gg$ 
component.
This component (at its leading order, which in terms of power counting
belongs to the NNLO corrections of the complete process) has Born-level kinematics and is
thus not affected by the jet veto, whereas real-radiation corrections are significantly 
suppressed.
However, higher-order corrections to the $gg$ contribution
are affected by the jet veto, i.e.\ similar to the radiative corrections to the $q\bar{q}$ channel, they are significantly reduced with respect to an inclusive calculation.
Hence, due to the suppression of radiative corrections by the jet veto, 
and the fact that no further new channels 
open up beyond NNLO, scale variations should provide a reasonable estimate of the 
uncertainties due to yet un-calculated higher-order QCD contributions.
The purely gluon-induced NLO corrections to the $gg$ channel have been
computed in \citere{Caola:2015rqy}.

With \wz{} production \cite{Grazzini:2016swo,Grazzini:2017ckn}, the last diboson process
has recently been computed at NNLO accuracy. 
Four different processes with three leptons and one neutrino are associated with 
\wz{} production: $W^-Z$ and $W^+Z$ production can each be split into a SF and a DF channel.
Since these processes have charged final 
states, no loop-induced $gg$ component contributes at NNLO. Following 
\citere{Grazzini:2017ckn} we set $\mu_0=(m_Z+m_W)/2$ for the central value
of renormalization and factorization scales and use 
the fiducial cuts summarized in \tab{tab:WWandWZ}: The lepton transverse-momentum 
thresholds distinguish between leptons associated with the $Z$- and the $W$-boson decays.
The lepton pair associated with the $Z$-boson decay is required to have an invariant 
mass close to the $Z$-boson mass, and the transverse mass of the $W$ boson, defined through the lepton 
associated with the $W$-boson decay and the transverse missing-energy vector (see \app{app:cuts}), is 
restricted from below. Furthermore, leptons are required to be separated in $R$, where 
the separation depends on whether the respective leptons are both associated with the $Z$-boson 
decay or with the decays of two different heavy bosons. In the SF channel there is an ambiguity 
how to assign the leptons to the $Z$- and $W$-boson decays, and we follow the 
resonant-shape identification procedure of \citere{Aad:2016ett} (see also \sct{sec:WZSF} for details).
Since \citere{Grazzini:2017ckn} uses the most recent input parameters 
corresponding to the default 
\Matrix{} settings, the 13\,TeV results of the fiducial cross sections are exactly 
(within the numerical uncertainties) reproduced. 
Radiative corrections in that process are known to be large because of an approximate 
radiation zero \cite{Baur:1994ia} in the Born scattering amplitudes, which is broken beyond LO.
We find $K_{\rm NLO}=2.05(1.97)$ and $K_{\rm NNLO}=1.11(1.11)$ for $W^-Z$ ($W^+Z$), both for SF and DF channels.
The $\sigma_{W^-Z}/\sigma_{W^+Z}$ ratio is about 0.69, both at NLO and NNLO, i.e.\ slightly smaller than what is found for the
$\sigma_{W^-}/\sigma_{W^+}$ ratio in the charged-current Drell--Yan process.

\section{Systematic uncertainties of $\boldsymbol{\qT}$ subtraction}\label{sec:qtsubtraction}

As pointed out before, NLO and NNLO 
cross sections computed with the $\qT$-subtraction formalism
exhibit a residual dependence on the cut-off \rcut{} in the slicing parameter $r=\qt/M$, where 
$\qt$ is the transverse momentum and $M$ the mass of the colourless system.
This residual dependence is due to power-suppressed terms, which are left
after the subtraction of the IR singular contribution at finite 
values of \rcut{} and vanish only in the limit $\rcut{}\rightarrow 0$.
The \rcut{} dependence of the cross sections computed with the \qt{}-subtraction method has been discussed in some detail for 
the \wgamma{}, off-shell \ww{} and off-shell \wz{} production processes in \citeres{Grazzini:2015nwa,Grazzini:2016ctr,Grazzini:2017ckn}, to which we refer the reader interested in these specific processes.
In the following, we study the systematic uncertainties of our results for a representative set of processes 
available in \Matrix{}, using the corresponding default setup of each process.

\Matrix{} performs an extrapolation $\rcut{}\rightarrow 0$ 
for total rates computed by means of the \qt{}-subtraction procedure, i.e.\ at NNLO, and 
possibly at NLO if the $\qT$-subtraction method is applied. 
A conservative estimate of the extrapolation uncertainty is included in the numerical error of 
this extrapolated result, which is considered our best prediction at the corresponding 
perturbative order and printed on screen at the end of each run.
To perform the extrapolation, \Matrix{} automatically computes the cross section at fixed values of $\rcut$ in 
the interval $[\rcutmin;1\%]$ using steps of $0.01\%$. 
Unless stated otherwise (see the process-specific information in \sct{sec:processes}),  
the default value \Matrix{} uses is $\rcutmin{}=0.15\%$.
The extrapolation procedure, which is discussed below, has been tested to work extremely well at NLO, where \rcut{}-independent results 
are available. Note that already the cross section at the lowest calculated value 
$\rcutmin{}=0.15\%$ (actually also for higher $\rcut$ values up to at least $\rcut\sim1\%$) 
provides a very reasonable prediction in cases where the \rcut{} dependence is small, 
and thus the result at $\rcutmin{}$ is also printed on screen at the end of each run.
A comparison of the extrapolated cross section and the result at the fixed value 
$\rcutmin$ indicates at which level of accuracy the fixed-$\rcut$ result can be trusted:
In case of a significant $\rcut$ dependence of the total rate, we recommend to correct the 
kinematic distributions by the ratio $\sigma^\textrm{extrapolated}_\textrm{NNLO}/\sigma^{r_{\rm cut}}_\textrm{NNLO}$. In the first release of \Matrix{},
distributions are indeed always calculated at $\rcut=\rcutmin$. We note that such reweighting should not 
be applied to distributions that are trivial at LO: For example, the transverse-momentum 
of the colourless system vanishes at LO, and its high-$\pT$ region is not affected 
by a finite \rcut{} value.
Given that we have not observed any significant $\rcut$ dependence of our NNLO results for kinematic distributions 
in various dedicated studies (see, e.g., \citere{Grazzini:2017ckn}) we consider this 
procedure sufficiently accurate, and leave a proper extrapolation procedure of 
distributions for a future update of \Matrix{}.

The $\rcut{}\rightarrow 0$ extrapolation of the cross section is
obtained using a 
simple quadratic least $\chi^2$ fit.
Such fit is repeated varying the upper bound of the fit interval, starting from a minimum 
upper bound of $0.5\%$ ($0.25\%$ for dilepton production or processes involving photons with $\rcutmin{}=0.15\%$; $0.15\%$ for the same processes with $\rcutmin{}=0.05\%$), and the result with the lowest $\chi^2/\text{degrees-of-freedom}$ value is kept as the best fit. 
The extrapolation uncertainty is determined by comparing the result of the best fit with the results obtained
by variations of the upper bound of the fit interval. To be conservative, a lower bound on this uncertainty is introduced, corresponding 
to half of the difference between the $\rcut{}\rightarrow 0$ result and the 
cross section at $\rcutmin{}$. This extrapolation error is combined quadratically with the numerical 
error, which is determined by extrapolating also statistical errors at finite \rcut{} values to  $\rcut{}=0$.

\begin{figure}
\begin{center}
\includegraphics[width=0.50\textwidth]{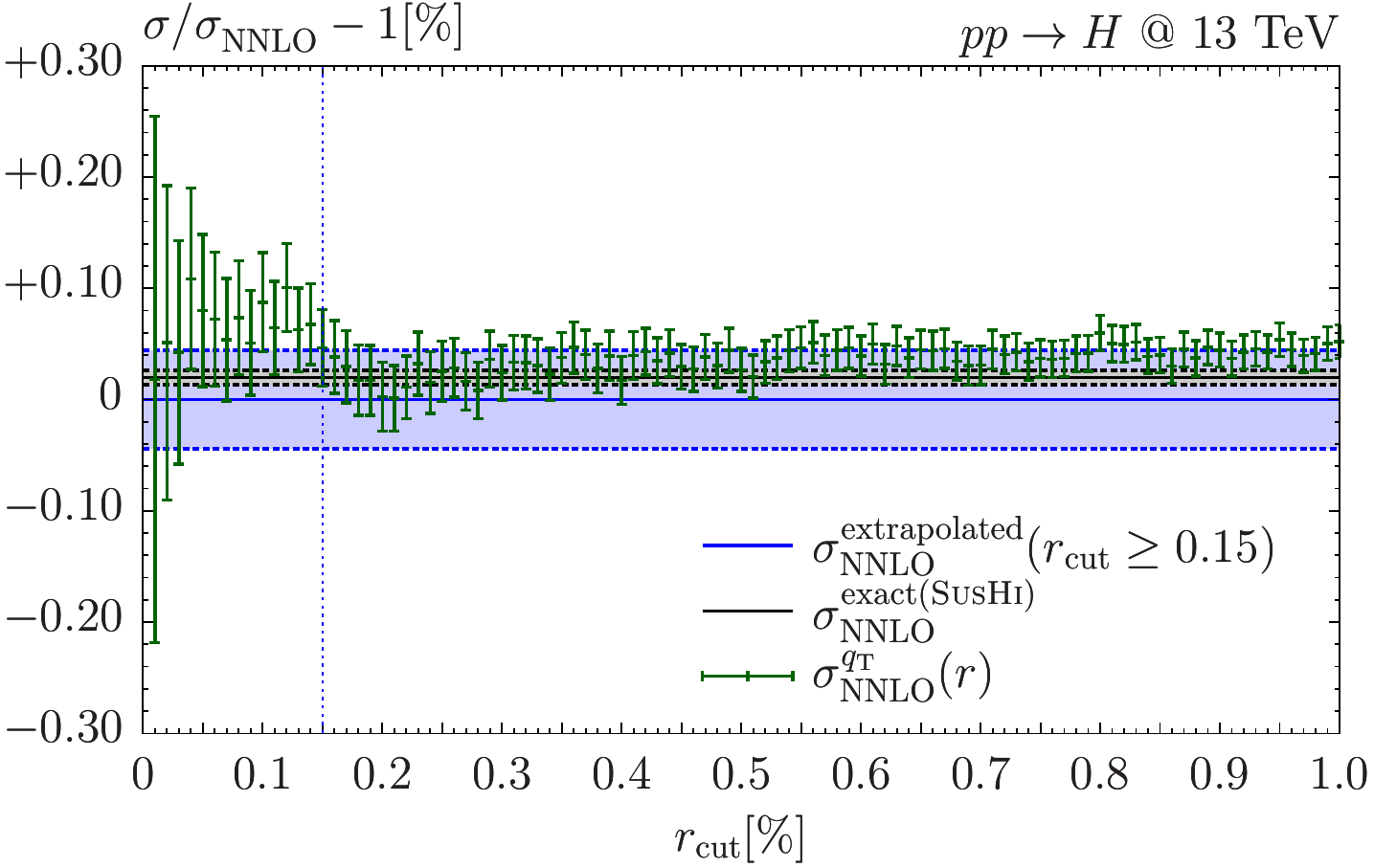}\hfill
\includegraphics[width=0.50\textwidth]{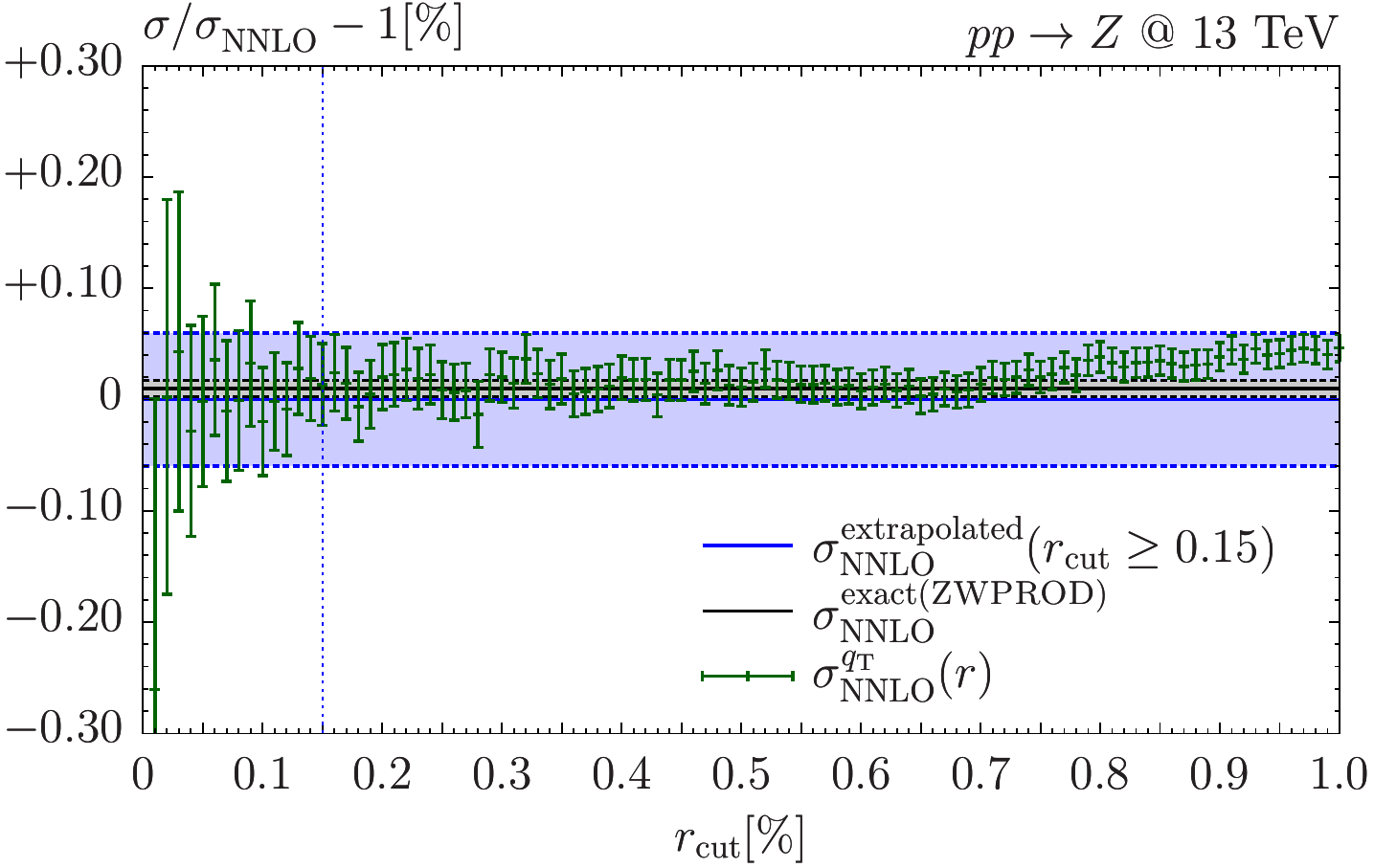}\\
\includegraphics[width=0.50\textwidth]{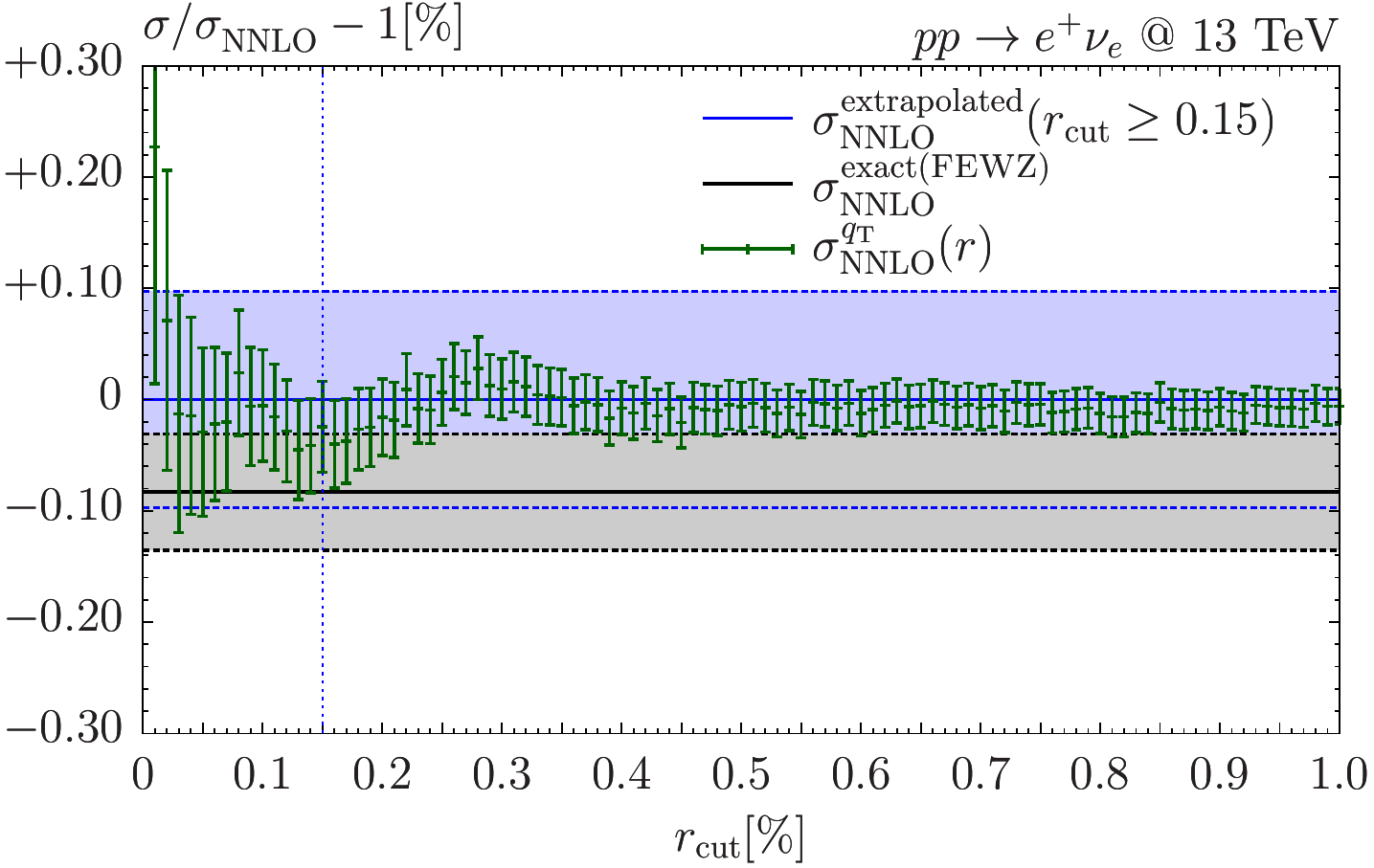}\hfill
\includegraphics[width=0.50\textwidth]{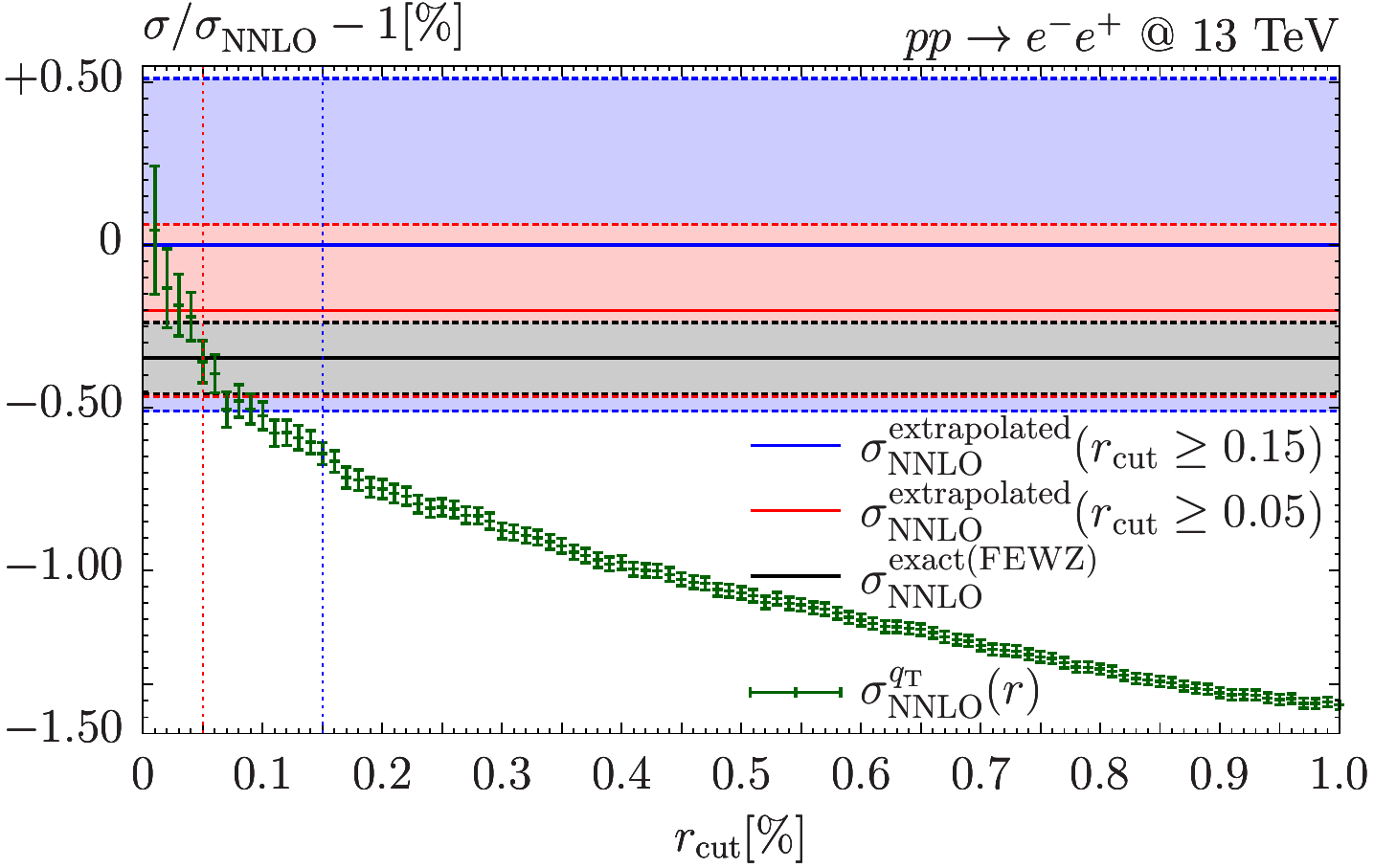}\\
\includegraphics[width=0.50\textwidth]{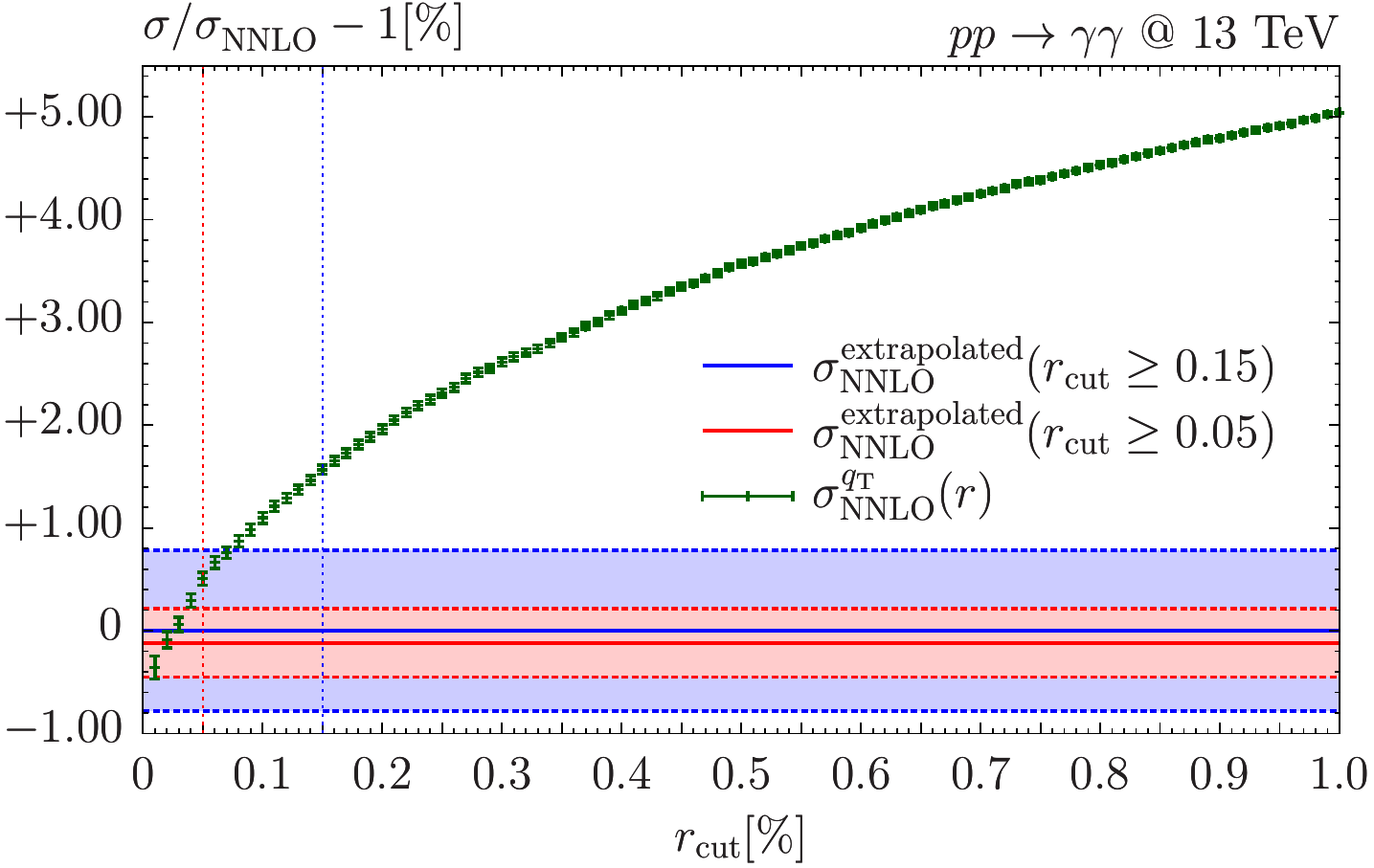}\hfill
\includegraphics[width=0.50\textwidth]{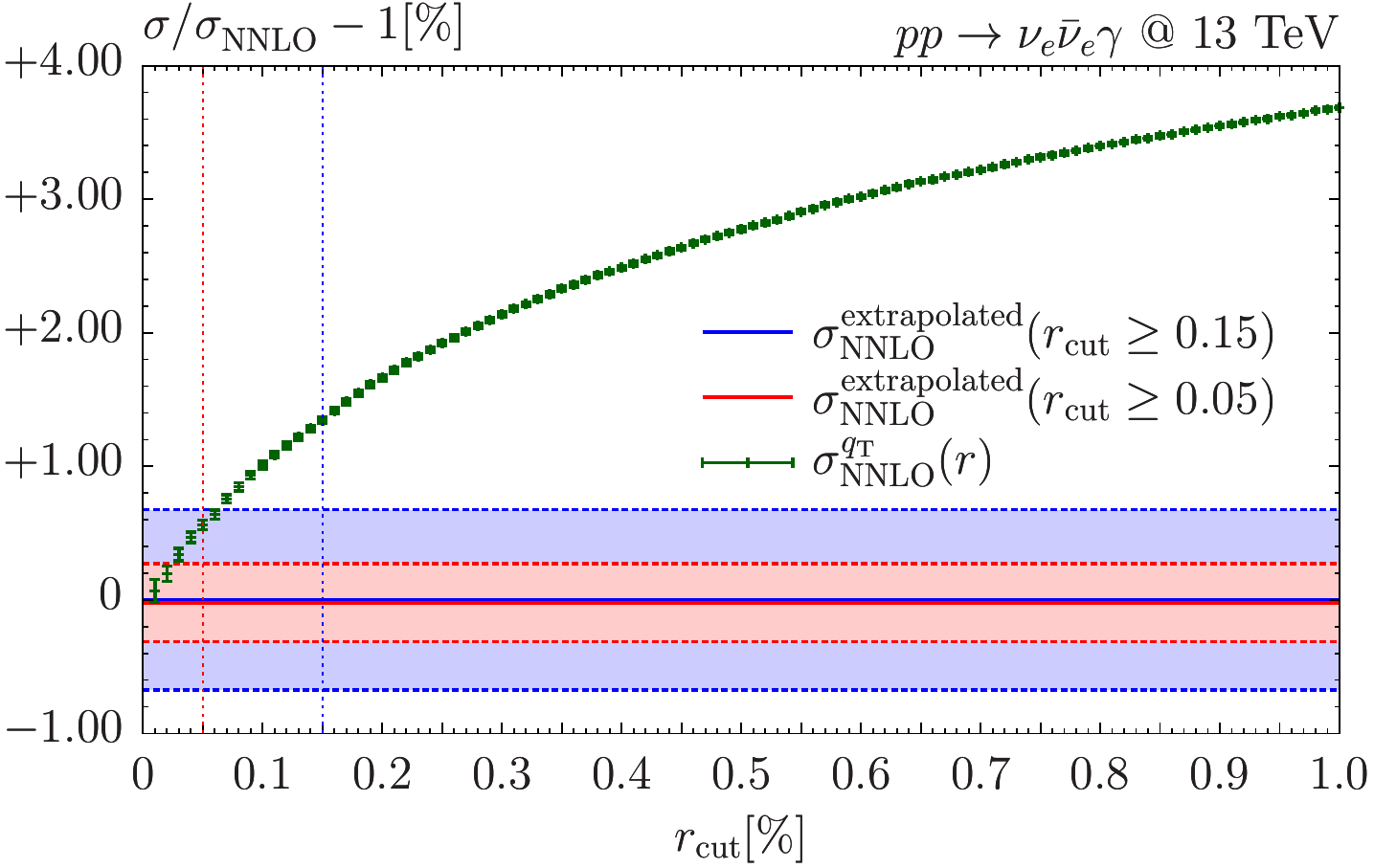}\\
\includegraphics[width=0.50\textwidth]{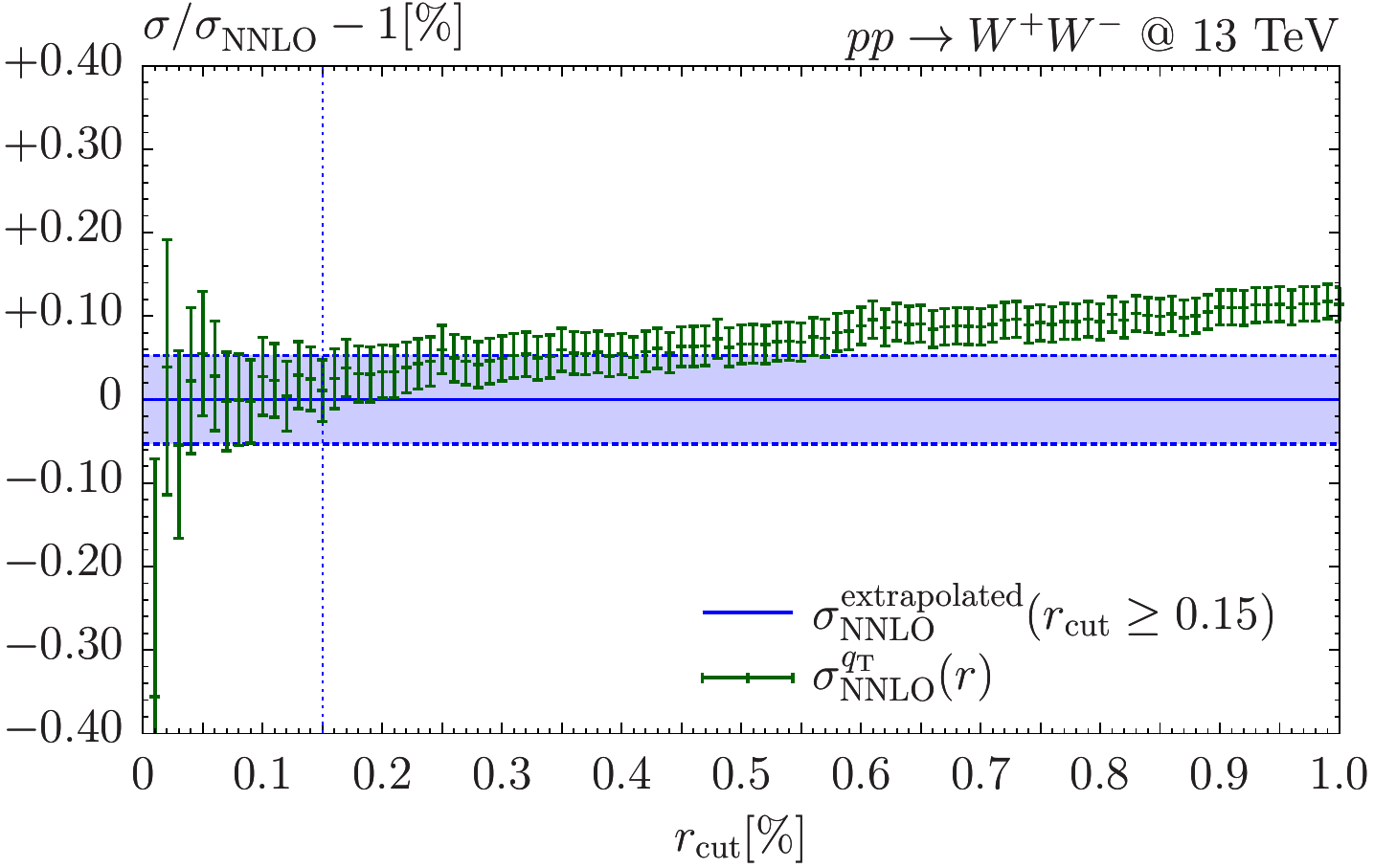}\hfill
\includegraphics[width=0.50\textwidth]{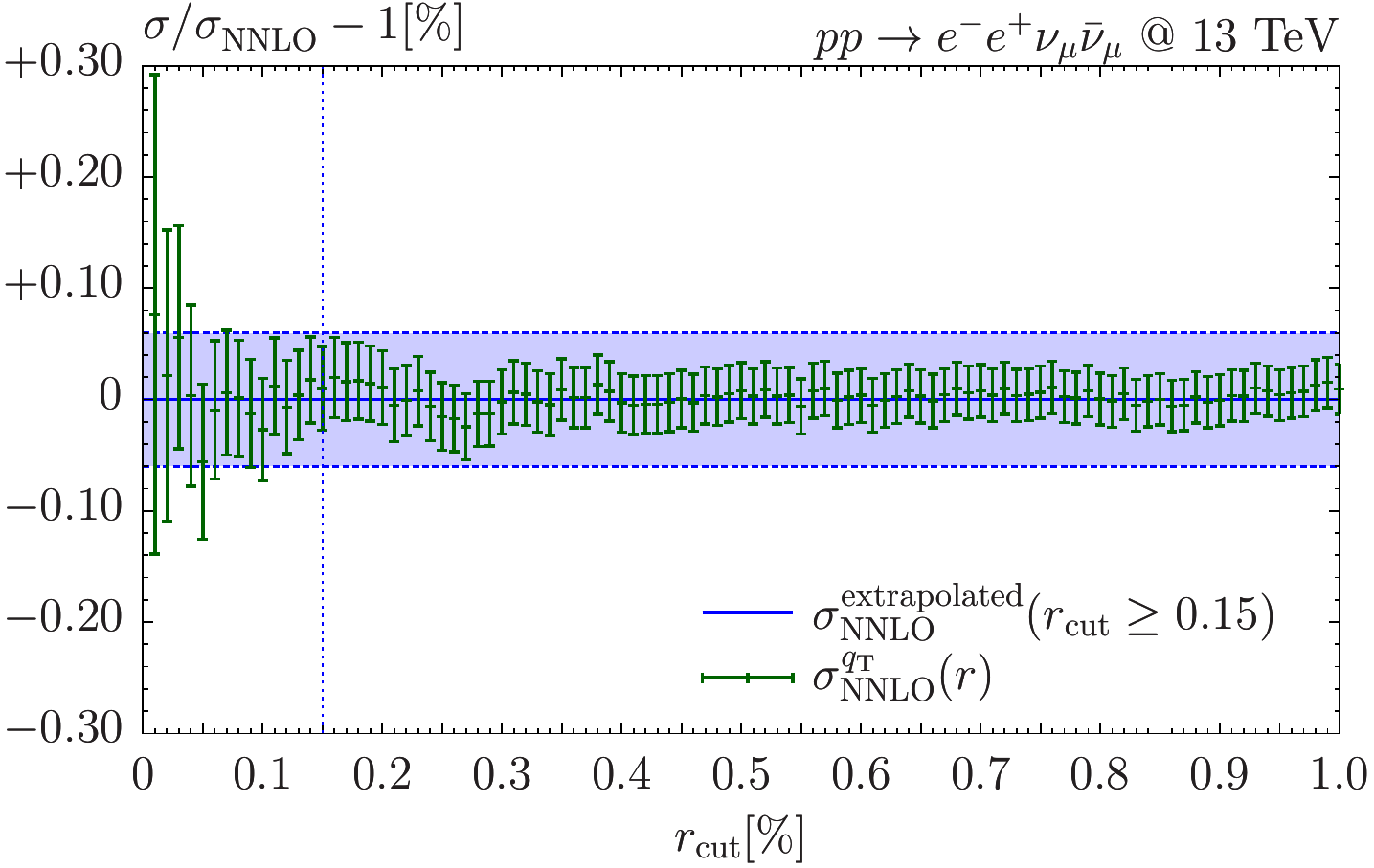}
\vspace{-.6cm}
\caption[]{\label{fig:stability}{
Dependence of the NNLO cross sections on $\rcut{}$ for various processes.
The NNLO results at fixed values of \rcut{} are normalized to the
$\rcut\to 0$ extrapolation obtained
by using $\rcut\ge 0.15$\%. The 
blue band represents the combined numerical and extrapolation uncertainty.
For processes with a large \rcut{} dependence, the extrapolated result and uncertainty 
obtained by using $\rcut\ge 0.05$\% is shown in red. 
Where available, \rcut{}-independent reference results are black. 
}
}
\end{center}
\end{figure}

Our results for the \rcut{} dependence of a representative set of processes are shown in \fig{fig:stability}.
Before commenting the various plots we provide some general explanation.
The central values of the green bars represent the NNLO cross section calculated at the respective fixed \rcut{} 
values, $\rcut{}\in[0.01\%;1\%]$ in steps of $0.01\%$, and their sizes denote the numerical uncertainties. 
Our reference prediction, computed with the default \Matrix{} setup, is the $\rcut{}\rightarrow0$ extrapolation
obtained from the values $\rcut{}\ge0.15\%$, shown as a blue solid line.
A vertical blue dotted line at $\rcut{}=0.15\%$ indicates the lowest value used for this extrapolation. 
The blue uncertainty band is obtained by combining the numerical and extrapolation uncertainties and corresponds to the
on-screen output of \Matrix{}.
When the $\rcut$ dependence is strong, we also show the \Matrix{} result extrapolated from 
$\rcut{}\ge0.05\%$ values as a red solid line with a red band, with its uncertainty computed analogously to the blue band.
Where available, NNLO results obtained either from analytical calculations or from
alternative NNLO numerical programs are reported as black lines, while the grey band shows their numerical
integration error.
All the results are reported as relative deviations from the reference prediction in percent. 

We start our discussion from the first two plots in \fig{fig:stability},
which refer to the inclusive on-shell production of a Higgs (left) and a $Z$ boson (right), respectively.
In both cases the NNLO cross sections turn out to be extremely stable with respect to \rcut{}:
Almost all fixed-$\rcut$ results deviate by less than one permille from the default $\rcut{}\rightarrow 0$ 
prediction, and all fixed-$\rcut$ results are compatible with the extrapolated result 
within their numerical uncertainties.
The high stability with respect to $\rcut$ in these cases would justify choosing essentially 
any $\rcut{}$ value in the given range to provide a reasonable prediction for the NNLO cross section.
Our default choice to use $\rcut{}\ge0.15$\% to obtain the $\rcut\rightarrow 0$ extrapolation appears 
to be a sound compromise between a sufficiently low \rcut{} value and a good numerical convergence. 
The blue band constitutes a reasonable estimate of the remaining extrapolation 
uncertainty of our reference result. Since no cuts have been applied,
our results can be compared with available
analytic computations of the inclusive Higgs and $Z$ production cross sections.
Such results are obtained with
{\sc SusHi}~\cite{Harlander:2012pb} for Higgs boson production, and with {\sc ZWPROD}~\cite{Hamberg:1990np,Harlander:2002wh}
for $Z$-boson production, and are reported in \fig{fig:stability} by the black 
solid lines with grey numerical error bands. The agreement is excellent and confirms that with \Matrix{}
we control these computations at the sub-permille level.

The next process we consider is $pp\rightarrow e^+\nu_e$ via an off-shell $W^+$ boson (third plot in \fig{fig:stability}).  
The \rcut{} dependence is similar to the case of on-shell Higgs and $Z$ production with all fixed-$\rcut$ results deviating not more 
than about one permille from the extrapolated result. 
We therefore conclude that the stability fully justifies our reference prediction and that its 
blue uncertainty band, which is slightly larger than for the on-shell processes discussed before, gives a 
reasonable estimate of the uncertainty due to the \rcut{} dependence of the 
cross section.
In the same plot we also report the result obtained with \FEWZ{} \cite{Gavin:2012sy},
depicted by a solid black line with grey error bands. The \Matrix{} and \FEWZ{} results are fully consistent within the respective numerical uncertainties.
A similar level of agreement is obtained with the program \NNLOjet{} \cite{Ridder:2016nkl}.\footnote{Note that we have set the CKM matrix to 
unity here, see \sct{sec:CKM}, in order to be able to compare against the results of \NNLOjet{}. 
The CKM input does not have any impact on the \rcut{} dependence beyond statistical uncertainties, 
which is why the discussion above is valid irrespective of the chosen CKM settings.}
Since both \FEWZ{} and \NNLOjet{} are based on fully independent subtraction schemes, the above agreement can be
considered as an important mutual consistency check of the three NNLO calculations.

Next, we discuss the dilepton production process $pp\rightarrow e^-e^+$, which, in our default setup,
exhibits a rather large \rcut{} dependence.
This is illustrated in the fourth plot of \fig{fig:stability}.
As discussed in \refse{sec:physics}, the strong \rcut{} dependence can be traced back to the presence
of perturbative instabilities \cite{Frixione:1997ks}
affecting the fixed-order computation in the case of symmetric $p_T$ cuts.
We have explicitly tested 
that if we change our default setup from $\ptl>25$\,GeV to $p_{T,e^-}>25$\,GeV and $p_{T,e^+}>24$\,GeV, we get a stable 
\rcut{} dependence. If we change this setup by letting the cut on $p_{T,e^+}$ approach $25$\,GeV, the \rcut{} dependence of the cross section becomes 
increasingly stronger. We also point out that, if we remove the lepton \pt{} thresholds completely or choose them as small as $\ptl>5$\,GeV, 
we obtain a flat \rcut{} dependence of the cross section.
The sizeable \rcut{} dependence implies a larger uncertainty in the $\rcut\to 0$ extrapolation.
We indeed see that by using the default setup the estimated uncertainty, represented by the blue band, is about $\pm 0.5\%$. By reducing the minimum \rcut{} value to $0.05\%$,
we reduce the extrapolation uncertainty by about a factor of 2 (red band) and we obtain a fully consistent result with the one obtained from $\rcut\geq 0.15\%$.
This is a strong indication that the extrapolation procedure is robust and provides a reasonable estimate of the ensuing uncertainty.
In the same plot we report the result obtained with \FEWZ{}: the agreement with the extrapolated results is excellent.
A similar level of agreement is obtained with \NNLOjet{}. As in the previous cases, the agreement of NNLO predictions obtained with fully independent methods
confirms the robustness of the results that can be obtained with \Matrix{}. We remark that this conclusion holds
also in a case, like the one of dilepton production with symmetric cuts,
in which a fixed-order computation is challenged.

\enlargethispage{\baselineskip}

In terms of the \rcut{} dependence of their cross sections, processes 
with an isolated photon in the final state suffer from large power-suppressed corrections.
Since the case of $W\gamma$ production was discussed in \citere{Grazzini:2015nwa},
here we consider diphoton and $Z\gamma$ production with the $Z$ boson decaying to a neutrino pair.
The corresponding results are shown in the fifth and sixth plot in \fig{fig:stability},
respectively.
Looking at our extrapolated reference results for $\rcut{}\ge 0.15$\% and $\rcut{}\ge 0.05$\%,
we see that
they are nicely consistent with the behaviour of the numerical results below the respective minimal \rcut{} value and that they are in neat 
mutual agreement.
In particular, the result at the lowest \rcut{} value $\rcut{}=0.01$\% is consistent with both the red and blue bands in all cases. 

It is clear that, for dilepton, diphoton and $Z\gamma$ production,
the extrapolation  $\rcut\rightarrow 0$ of runs with $\rcut{}\ge 0.15$\% allows us to control the uncertainty of our 
NNLO predictions at the $0.5-1$\% level. If the minimal \rcut{} value is decreased to $\rcutmin{}= 0.05$\%,
the ensuing uncertainty is reduced to few permille.
We have explicitly checked that this picture is common to all processes involving photons. We conclude 
that, taking into account the estimated \rcut{} uncertainties, we obtain consistent predictions for all these 
processes.

The last two plots in \fig{fig:stability} are representative \rcut{}-dependence plots for on- and off-shell 
diboson production: The first one shows on-shell \ww{} production, and the second one 
off-shell \zz{} production in the $2\ell 2\nu'$ decay channel. These plots
feature to a large extent the same behaviour as 
observed for on-shell Higgs and vector-boson production processes: In general, the NNLO cross section is very 
stable with respect to \rcut{} over two orders of magnitude. The \rcut{} dependence of the 
\ww{} cross section seems to have a very slight slope, which, however, is perfectly modelled by the extrapolation.
Our default choice of the minimal \rcut{} value leads to a reasonable reference result with the estimated uncertainties being meaningful 
and in some cases even a bit conservative. 

In summary, for most of the processes implemented in the first release of \Matrix{}, NNLO predictions
can generally be controlled at the one permille level (or better). For processes with a large \rcut{} dependence like those involving photons or Drell--Yan dilepton pairs, fiducial cross sections
can be computed with uncertainties of few permille by using the setting \matrixparam{switch_qT_accuracy = 1}.
This accuracy should be sufficient for all practical purposes.

\section{Summary}
\label{sec:summary}
In this paper we have introduced the new computational framework \Matrix{}\cite{MATRIX}, 
which allows a user to produce NNLO QCD predictions for a wide class of 
hadron-collider processes. Using the \qt{}-subtraction formalism, our computations are 
fully differential in the phase space of the final-state particles and of the associated QCD radiation,
thereby enabling the evaluation 
of arbitrary IR safe observables. Since our 
implementation is completely general, it is applicable to the computation of NNLO 
corrections to any process with colourless final states. The list of available 
processes is therefore limited only by the availability of two-loop amplitudes 
to the Born-level processes. The first \Matrix{} release involves  $2\to 1$ and $2\to 2$ hadronic 
reactions with Higgs and vector bosons in the final state. 
In particular, we consider final states with two, three and four leptons (plus missing energy)
from the decays of the vector bosons, and we account for all resonant and non-resonant diagrams with off-shell 
effects and spin correlations. This enables the evaluation of cross 
sections with realistic cuts to define any fiducial volume.

We have presented a detailed description of the first release of 
the \Matrix{} code. 
Besides the theoretical framework and the tools \Matrix{} is based on, we have 
focused on the technical aspects relevant for a user to produce fully differential 
NNLO results. \Matrix{} features automatic compilation and running through dedicated scripts. The most 
popular clusters are preconfigured and can be simply selected by the user. Having specified 
a target precision and a desired runtime per job, the code automatically determines
the required parallelization in each run. In combination with the fast numerical multi-channel 
integration offered by \Munich{}, this allows us
to obtain accurate NNLO results even for the most complicated 
of the available processes on a middle-sized cluster in less than a couple of days, 
simpler processes being significantly faster. With every run \Matrix{} provides not 
only the central prediction, but automatically evaluates the effect of independent factorization- and renormalization-scale 
variations in order to obtain an estimate of the perturbative uncertainties at each order.
Furthermore, by simultaneously computing NNLO cross sections at several values of the \qt{}-subtraction 
parameter \rcut{}, \Matrix{} performs an extrapolation $\rcut\rightarrow 0$ of the integrated cross section 
in order to provide its final prediction that includes an extrapolation uncertainty.
Such procedure allows us to offer a robust estimate of the systematic uncertainty due to the \qt{}-subtraction procedure.
Both scale variations and the \rcut{} extrapolation procedure are fully automated within \Matrix{} 
without the need of extra computing power.

In particular for processes with a large \rcut{} dependence, such as dilepton production
or processes with isolated photons, 
\Matrix{} is able to significantly improve over predictions computed at fixed \rcut{} values by performing the $\rcut\to 0$ extrapolation. 
Besides an improved accuracy in the central prediction, our procedure includes a conservative estimate 
of systematic uncertainties, which allows the user to control the precision of these processes at the level 
of few permille, when using corresponding settings.

We have discussed in detail all 
relevant \Matrix{} input cards accessible to the user. Besides standard settings 
applicable to all processes, information specific to each individual process 
has been provided, such as suitable dynamic scales which are predefined for 
certain processes and process-specific cuts. The latter facilitate the restriction 
of the phase space to fiducial volumes as defined by the LHC experiments. 
In combination with a general way to define distributions, this enables 
the possibility to compute fiducial cross sections and distributions that
can be directly compared to unfolded experimental data.

Reference predictions for the integrated cross sections
of all processes available in \Matrix{} have been provided at LO, NLO and NNLO in the default setups. 
For the NNLO cross section we have quoted predictions for a fixed value $\rcut=0.15$\% and the final NNLO result
after performing the extrapolation $\rcut\rightarrow 0$. We have studied the impact of radiative 
corrections for each of these processes as well as the impact of the 
loop-induced $gg$ component, where applicable.
The impact of NLO and NNLO QCD corrections is generally large. While NLO corrections typically range between 
$30\%-100\%$, NNLO corrections are still as large as $3\%-30\%$ for the processes 
and scenarios we have considered. 
The size of radiative corrections is typically widely un-affected if only fiducial cuts
for particle identification, like transverse-momentum thresholds, geometric (pseudo-)rapidity ranges
or isolation cuts are applied (see, e.g., inclusive on-shell $ZZ$ production
and off-shell $ZZ\to 4\ell$ production in a $ZZ$ signal region).
This is in general no longer true if the considered cuts modify the dominant resonance structures
(e.g.\ $ZZ\to 4\ell$ production in the $H\to ZZ$ background region).
If real radiation is restricted, in particular by a veto against jets,
the size of higher-order corrections is strongly suppressed, and 
NLO and NNLO $K$ factors can be very different 
as compared to the inclusive phase-space 
(see, e.g., inclusive on-shell \ww{} production and off-shell $WW\to 2\ell2\nu$ production, which requires a jet veto to
suppress the top-quark backgrounds).

Scale variations tend to underestimate the uncertainty from missing higher-order contributions at LO, and also widely at NLO. 
This is due to the fact that vector-boson and vector-boson
pair production processes are driven by $q\bar q$ initial states at LO 
($gg$ initial states in case of Higgs boson production). The $gq$ channel opens up only at NLO, 
and NNLO is the first order where all partonic channels are contributing.
As a consequence, in most of the cases NNLO uncertainties obtained 
through scale variations should provide the correct order of magnitude of yet un-calculated perturbative QCD contributions.
When NNLO corrections are particularly large, as in the case of diphoton production, a more conservative
estimate of missing higher-order contributions can be obtained by considering the difference with the previous order.

All the vector-boson pair production processes with an electrically neutral final state 
include a loop-induced $gg$ contribution at NNLO. Its size strongly 
depends on the considered process and can range between roughly $10\%$ and $60\%$ of the NNLO corrections. In cases where radiative corrections are suppressed by 
a jet veto, since the $gg$ component is not affected due to its Born-level kinematics,
it may provide the dominant NNLO contribution.
As the $gg$ component is effectively only LO accurate, scale variations might 
underestimate the actual size of its missing perturbative corrections in some cases.

To conclude, we believe that the numerical tool presented in this paper will be highly valuable for
the high-energy community. 
Several experimental studies have already used the results obtained with \Matrix{} for data--theory 
comparison in \sm{} measurements (see, e.g., \citeres{Aad:2016sau, %
CMS:2016xzm, Sirunyan:2017lvq, %
Aad:2015rka, Aaboud:2016urj, Aaboud:2017rwm, %
CMS:2016kxu, CMS:2015fnj, Khachatryan:2016txa, Sirunyan:2017zjc, %
Aad:2016wpd, Aaboud:2016mrt, Aaboud:2017qkn, %
Khachatryan:2015sga, CMS:2016vww, %
Aad:2016ett, Aaboud:2016yus, ATLAS:2016qzn, %
Khachatryan:2016poo, Khachatryan:2016tgp}) 
and for background estimates in various new-physics searches.
\Matrix{}\cite{MATRIX} can be used to produce benchmark
predictions for a wide range of processes relevant both for SM measurements 
and as backgrounds to Higgs and new-physics searches.
Extensions
of the code to include additional processes, the inclusion of further perturbative contributions (e.g.\ NLO corrections to the $gg$ channel or EW corrections) as well as the
resummation of certain classes of logarithmic contributions are left for future work.

\subparagraph{Acknowledgements.}

We are grateful to Dirk Rathlev for his contributions at the beginning of the \Matrix{} project.
We would like to thank Stefano Catani for comments on the manuscript and Alexander Huss for providing cross checks for the Drell--Yan processes with {\sc NNLOjet}.
We are indebted to Jonas Lindert, Philipp Maierh\"ofer and Stefano Pozzorini for their continuous support on \OpenLoops{}.
We thank Thomas Gehrmann, Andreas von Manteuffel and
Lorenzo Tancredi for providing us with private implementations of the two-loop amplitudes
to on-shell \ww{} and $ZZ$ production.
This research was supported in part by the Munich Institute for Astro- and Particle Physics (MIAPP) of the DFG cluster of excellence "Origin and Structure of the Universe``,
by the Swiss National Science Foundation (SNF) under 
contracts CRSII2-141847, 200020-169041 and by 
the Research Executive Agency (REA) of the European Union under the Grant Agreement 
number PITN--GA--2012--316704 ({\it HiggsTools}). 
The work of MW is supported by the ERC Consolidator Grant 614577 HICCUP.

\newpage
\appendix

\gdef\thesection{Appendix \Alph{section}}
\section{Predefined cuts}\label{app:cuts}
\gdef\thesection{\Alph{section}}

In this Appendix the process-dependent cuts introduced in \refse{sec:processes}
are explained in more detail. It can be used as a dictionary since 
the respective cuts in \refse{sec:processes} contain interactive links  
to their explanations in \refta{tab:predefinedcuts}.

\newcommand{\firstuserswitch}[2]{\ttt{#1} & \multicolumn{2}{p{11.25cm}}{#2}}
\newcommand{\userswitch}[2]{\phantomsection\label{#1}\\\midrule\ttt{#1} & \multicolumn{2}{p{11.25cm}}{#2}}
\newcommand{\usercut}[2]{\quad\ttt{#1} & \multicolumn{2}{p{11.25cm}}{#2}}

\setlength\LTleft{-0.2cm}
{
\footnotesize 


\begin{center}
\begin{longtable}{llll}
\toprule
\ttt{user_switch}  & \multicolumn{2}{p{11.25cm}}{ {\multirow{ 2}{*}{\bf description}}}     \\
\quad\ttt{user_cut} & \phantomsection\label{lepton_identification}\\
\midrule
\endfirsthead
\toprule
\ttt{user_switch}  & \multicolumn{2}{p{11.25cm}}{ {\multirow{ 2}{*}{\bf description}}}     \\
\quad\ttt{user_cut} & \\
\midrule
\endhead 
\multicolumn{3}{r@{}}{continued on next page \ldots}\\
\bottomrule
\endfoot
\endlastfoot
\firstuserswitch{lepton_identification}{Switch to change between different identification procedures of $Z$ and $W$ bosons in the same-flavour channels of $ZZ$ and $WZ$ production. See \sct{sec:ZZSF} and \sct{sec:WZSF} for details.}
\userswitch{M_leplep}{Switch for cuts on invariant mass $m_{\ell\ell}$ of all possible lepton pairs.}\\*
\usercut{min_M_leplep}{Minimal requirement $m_{\ell\ell}> $ \ttt{min_M_leplep} for all lepton pairs.}\\*
\usercut{max_M_leplep}{Maximal requirement $m_{\ell\ell}< $ \ttt{max_M_leplep} for all lepton pairs.}
\userswitch{M_leplep_OSSF}{Switch for cuts on invariant mass $m_{\ell^-\ell^+}$ of opposite-sign, same-flavour (OSSF) lepton pairs.}\\*
\usercut{min_M_leplep_OSSF}{Minimal requirement $m_{\ell^-\ell^+}> $ \ttt{min_M_leplep_OSSF} for OSSF pairs.}\\*
\usercut{max_M_leplep_OSSF}{Maximal requirement $m_{\ell^-\ell^+}< $ \ttt{max_M_leplep_OSSF} for OSSF pairs.}\\*
\usercut{min_M_Z1_OSSF}{Minimal requirement $m_{\ell^-\ell^+}> $ \ttt{min_M_Z1_OSSF} for OSSF pair closer to $m_Z$.}\\*
\usercut{max_M_Z1_OSSF}{Maximal requirement $m_{\ell^-\ell^+}< $ \ttt{max_M_Z1_OSSF} for OSSF pair closer to $m_Z$.}
\userswitch{M_Zrec}{Switch for cuts on invariant mass $m_{\ell^-\ell^+}$ of lepton pairs associated with $Z$ bosons.}\\*
\usercut{min_M_Zrec}{Minimal requirement $m_{\ell^-\ell^+}> $ \ttt{min_M_Zrec} for (reconstructed) $Z$ bosons.}\\*
\usercut{max_M_Zrec}{Maximal requirement $m_{\ell^-\ell^+}< $ \ttt{max_M_Zrec} for (reconstructed) $Z$ bosons.}
\userswitch{delta_M_Zrec_MZ}{Switch for cuts on absolute difference between invariant mass $m_{\ell^-\ell^+}$ of lepton pairs associated with $Z$ bosons and the $Z$-boson mass.}\\*
\usercut{max_delta_M_Zrec_MZ}{Maximal requirement $|m_{\ell^-\ell^+}-m_Z|< $ \ttt{max_delta_M_Zrec_MZ} for (reconstructed) $Z$ bosons.}
\userswitch{delta_M_lepleplep_MZ}{Switch for cuts on absolute difference between invariant mass $m_{\ell\ell\ell}$ of 3-lepton system and the $Z$-boson mass.}\\*
\usercut{min_delta_M_lepleplep_MZ}{Minimal requirement $|m_{\ell\ell\ell}-m_Z|> $ \ttt{min_delta_M_lepleplep_MZ}.}
\userswitch{M_4lep}{Switch for cuts on invariant mass $m_{\ell\ell\ell\ell}$ of 4-lepton system.}\\*
\usercut{min_M_4lep}{Minimal requirement $m_{\ell\ell\ell\ell}> $ \ttt{min_M_4lep} for 4-lepton system.}\\*
\usercut{max_M_4lep}{Maximal requirement $m_{\ell\ell\ell\ell}< $ \ttt{max_M_4lep} for 4-lepton system.}\\*
\usercut{min_delta_M_4lep}{Minimal requirement $|m_{\ell\ell\ell\ell}-m_Z|> $ \ttt{min_delta_M_4lep} for 4-lepton system.}\\*
\usercut{max_delta_M_4lep}{Maximal requirement $|m_{\ell\ell\ell\ell}-m_Z|< $ \ttt{max_delta_M_4lep} for 4-lepton system.}
\userswitch{M_leplepnunu}{Switch for cuts on invariant mass $m_{\ell\ell\nu\nu}$ of 2-lepton--2-neutrino system.}\\*
\usercut{min_M_leplepnunu}{Minimal requirement $m_{\ell\ell\nu\nu}> $ \ttt{min_delta_M_leplepnunu} for 2$\ell$2$\nu$ system.}\\*
\usercut{max_M_leplepnunu}{Maximal requirement $m_{\ell\ell\nu\nu}< $ \ttt{max_delta_M_leplepnunu} for 2$\ell$2$\nu$ system.}\\*
\usercut{min_delta_M_leplepnunu}{Min. requirement $|m_{\ell\ell\nu\nu}-m_Z|> $ \ttt{min_delta_M_leplepnunu} for 2$\ell$2$\nu$ system.}\\*
\usercut{max_delta_M_leplepnunu}{Max. requirement $|m_{\ell\ell\nu\nu}-m_Z|< $ \ttt{max_delta_M_leplepnunu} for 2$\ell$2$\nu$ system.}
\userswitch{pT_leplep}{Switch for cuts on transverse momentum of lepton pair $p_{T,\ell\ell}$.}\\*
\usercut{min_pT_leplep}{Minimal requirement $p_{T,\ell\ell} >$ \ttt{min_pT_leplep} for lepton pair.}
\userswitch{pT_lep_1st}{Switch for cuts on transverse momentum of hardest lepton $p_{T,\ell_1}$.}\\*
\usercut{min_pT_lep_1st}{Minimal requirement $p_{T,\ell_1} >$ \ttt{min_pT_lep_1st} for hardest lepton.}
\userswitch{pT_lep_2nd}{Switch for cuts on transverse momentum of second-hardest lepton $p_{T,\ell_2}$.}\\*
\usercut{min_pT_lep_2nd}{Minimal requirement $p_{T,\ell_2} >$ \ttt{min_pT_lep_2nd} for second-hardest lepton.}
\userswitch{lepton_cuts}{Switch for cuts on transverse momentum of (second-)hardest lepton $p_{T,\ell_1}$($p_{T,\ell_2}$).}\\*
\usercut{min_pT_lep_1st}{Minimal requirement $p_{T,\ell_1} >$ \ttt{min_pT_lep_1st} for hardest lepton.}\\*
\usercut{min_pT_lep_2nd}{Minimal requirement $p_{T,\ell_2} >$ \ttt{min_pT_lep_2nd} for second-hardest lepton.}
\userswitch{leading_lepton_cuts}{Switch for flavour-dependent cuts on hardest-lepton transverse momentum $p_{T,\ell_1}$.}\\*
\usercut{min_pT_1st_if_e}{Minimal requirement $p_{T,\ell_1} >$ \ttt{min_pT_1st_if_e} if hardest lepton is electron.}\\*
\usercut{min_pT_1st_if_mu}{Minimal requirement $p_{T,\ell_1} >$ \ttt{min_pT_1st_if_mu} if hardest lepton is muon.}
\userswitch{lepZ_cuts}{Switch for cuts on transverse momentum of (second-)hardest lepton associated with $Z$-boson decay $p_{T,\ell_{Z,1}}\, (p_{T,\ell_{Z,2}})$.}\\*
\usercut{min_pT_lepZ_1st}{Minimal requirement $p_{T,\ell_{Z,1}} >$ \ttt{min_pT_lepZ_1st} for hardest lepton of $Z$.}\\*
\usercut{min_pT_lepZ_2nd}{Minimal requirement $p_{T,\ell_{Z,2}} >$ \ttt{min_pT_lepZ_2nd} for second-hardest lepton of $Z$.}
\userswitch{lepW_cuts}{Switch for cuts on lepton associated with $W$-boson decay.}\\*
\usercut{min_pT_lepW}{Minimal requirement $p_{T,\ell_W} >$ \ttt{min_pT_lepW} for lepton from $W$-boson decay.}\\*
\usercut{max_eta_lepW}{Minimal requirement $\eta_{\ell_W} <$ \ttt{max_eta_lepW} for lepton from $W$-boson decay.}
\userswitch{R_leplep}{Switch for cuts on lepton separation in $\dRleplep=\sqrt{\Delta y_{\ell\ell}^2 + \Delta\phi_{\ell\ell}^2}$.}\\*
\usercut{min_R_leplep}{Minimal requirement $\dRleplep> $ \ttt{min_R_leplep} for all lepton pairs.}
\userswitch{R_lepZlepZ}{Switch for cuts on separation between leptons associated with $Z$-boson decay in $\Delta R_{\ell_Z\ell_Z}=\sqrt{\Delta y_{\ell_Z\ell_Z}^2 + \Delta\phi_{\ell_Z\ell_Z}^2}$.}\\*
\usercut{min_R_lepZlepZ}{Minimal requirement $\Delta R_{\ell_Z\ell_Z}> $ \ttt{min_R_lepZlepZ} for leptons of $Z$ decay.}
\userswitch{R_lepZlepW}{Switch for cuts on separation between leptons associated with $Z$-boson decay and lepton associated with $W$-boson decay in $\Delta R_{\ell_Z\ell_W}=\sqrt{\Delta y_{\ell_Z\ell_W}^2 + \Delta\phi_{\ell_Z\ell_W}^2}$.}\\*
\usercut{min_R_lepZlepW}{Minimal requirement $\Delta R_{\ell_Z\ell_W}> $ \ttt{min_R_lepZlepW} for leptons of $Z$ and $W$ decay.}
\userswitch{phi_leplep}{Switch for cuts on azimuthal separation $\Delta \phi_{\ell\ell}$ between all lepton pairs.}\\*
\usercut{min_phi_leplep}{Minimal requirement $\Delta \phi_{\ell\ell}> $ \ttt{min_phi_leplep} for all lepton pairs.}
\userswitch{phi_leplep_nunu}{Switch for cuts on azimuthal separation $\Delta \phi_{\ell\ell,\nu\nu}$ between the transverse-momentum vectors of the 2-lepton system 
$\boldsymbol p_{\rm \boldsymbol T,\ell\ell}$ and the missing energy $\mathbf\ptmiss$.}\\*
\usercut{min_phi_leplep_nunu}{Minimal requirement $\Delta \phi_{\ell\ell,\nu\nu}> $ \ttt{min_phi_leplep_nunu}.}
\userswitch{electron_cuts}{Switch for cuts on transverse momentum of (second-)hardest electron $p_{T,e_1}$($p_{T,e_2}$).}\\*
\usercut{min_pT_e_1st}{Minimal requirement $p_{T,e_1} >$ \ttt{min_pT_e_1st} for hardest electron.}\\*
\usercut{min_pT_e_2nd}{Minimal requirement $p_{T,e_2} >$ \ttt{min_pT_e_2nd} for second-hardest electron.}
\userswitch{muon_cuts}{Switch for cuts on transverse momentum of (second-)hardest muon $p_{T,\mu_1}$($p_{T,\mu_2}$).}\\*
\usercut{min_pT_mu_1st}{Minimal requirement $p_{T,\mu_1} >$ \ttt{min_pT_mu_1st} for hardest muon.}\\*
\usercut{min_pT_mu_2nd}{Minimal requirement $p_{T,\mu_2} >$ \ttt{min_pT_mu_2nd} for second-hardest muon.}
\userswitch{gap_eta_e}{Switch for detector gap in absolute pseudo-rapidity $|\eta_e|$ of electrons.}\\*
\usercut{gap_min_eta_e}{Start of the gap, keeping only events with $|\eta_e|<$ \ttt{gap_min_eta_e}.}\\*
\usercut{gap_max_eta_e}{End of the gap, keeping only events with $|\eta_e|>$ \ttt{gap_max_eta_e}.}
\userswitch{M_gamgam}{Switch for cuts on invariant mass $m_{\gamma\gamma}$ of photon pairs.}\\*
\usercut{min_M_gamgam}{Minimal requirement $m_{\gamma\gamma}> $ \ttt{min_M_gamgam} for photon pairs.}\\*
\usercut{max_M_gamgam}{Maximal requirement $m_{\gamma\gamma}< $ \ttt{min_M_gamgam} for photon pairs.}
\userswitch{pT_gam_1st}{Switch for cuts on transverse momentum of hardest photon $p_{T,\gamma_1}$.}\\*
\usercut{min_pT_gam_1st}{Minimal requirement $p_{T,\gamma_1} >$ \ttt{min_pT_gam_1st} for hardest photon.}
\userswitch{R_gamgam}{Switch for cuts on photon separation in $\Delta R_{\gamma\gamma}=\sqrt{\Delta y_{\gamma\gamma}^2 + \Delta\phi_{\gamma\gamma}^2}$.}\\*
\usercut{min_R_gamgam}{Minimal requirement $\Delta R_{\gamma\gamma}> $ \ttt{min_R_gamgam} for photon pairs.}
\userswitch{gap_eta_gam}{Switch for detector gap in absolute pseudo-rapidity $|\eta_\gamma|$ of photons.}\\*
\usercut{gap_min_eta_gam}{Start of the gap, keeping only events with $|\eta_\gamma|<$ \ttt{gap_min_eta_gam}.}\\*
\usercut{gap_max_eta_gam}{End of the gap, keeping only events with $|\eta_\gamma|>$ \ttt{gap_max_eta_gam}.}
\userswitch{M_lepgam}{Switch for cuts on invariant mass $m_{\ell\gamma}$ of lepton--photon pairs.}\\*
\usercut{min_M_lepgam}{Minimal requirement $m_{\ell\gamma}> $ \ttt{min_M_lepgam} for lepton--photon pairs.}
\userswitch{R_lepgam}{Switch for cuts on lepton--photon separation in $\dRlepgamma=\sqrt{\Delta y_{\ell\gamma}^2 + \Delta\phi_{\ell\gamma}^2}$.}\\*
\usercut{min_R_lepgam}{Minimal requirement $\dRlepgamma> $ \ttt{min_R_lepgam} for lepton--photon pairs.}
\userswitch{R_lepjet}{Switch for cuts on lepton--jet separation in $\dRlepjet=\sqrt{\Delta y_{\ell j}^2 + \Delta\phi_{\ell j}^2}$.}\\*
\usercut{min_R_lepjet}{Minimal requirement $\Delta R_{\ell j}> $ \ttt{min_R_lepjet} for lepton--jet pairs.}
\userswitch{lep_iso}{Switch for special isolation of lepton--lepton and lepton--parton pairs, as used for example in the $ZZ$ Higgs background, see \citere{deFlorian:2016spz}. For each lepton $i$ we compute the sum of the transverse momenta over all leptons and partons in a certain $R$ cone around $i$ and take the ratio to its transverse momentum. 
All events are discarded where this ratio is below a certain threshold: \[\left.\sum_{\substack{j\in\{\ell,\text{partons}\}\\* \text{ with } \Delta R_{ij}< \delta_0}}p_{T,j} \middle/ p_{T,i}\right. < \epsilon\]}\vspace{-0.3cm}\\*
\usercut{lep_iso_delta_0}{$\delta_0$ in the formula above.}\\*
\usercut{lep_iso_epsilon}{$\epsilon$ in the formula above.}
\userswitch{R_ejet}{Switch for cuts on electron--jet separation in $\Delta R_{e j}=\sqrt{\Delta y_{e j}^2 + \Delta\phi_{e j}^2}$.}\\*
\usercut{min_R_ejet}{Minimal requirement $\Delta R_{ej}> $ \ttt{min_R_ejet} for electron--jet pairs.}
\userswitch{R_gamjet}{Switch for cuts on photon--jet separation in $\dRgammajet=\sqrt{\Delta y_{\ell j}^2 + \Delta\phi_{\ell j}^2}$.}\\*
\usercut{min_R_gamjet}{Minimal requirement $\dRgammajet> $ \ttt{min_R_gamjet} for photon--jet pairs.}
\userswitch{mT_CMS}{Switch for cuts on transverse mass $m_{T,W}$ of $W$ boson as defined by CMS \cite{Chatrchyan:2013fya}.}\\*
\usercut{min_mT_CMS}{Minimal requirement $m_{T,W}> $ \ttt{min_mT_CMS} for the $W$ boson.}
\userswitch{pT_W}{Switch for cuts on transverse momentum $p_{T,\ell\nu_\ell}$ of (identified) $W$ bosons.}\\*
\usercut{min_pT_W}{Minimal requirement $p_{T,\ell\nu_\ell}>$ \ttt{min_pT_W} for (identified) $W$ bosons.}\\*
\usercut{max_pT_W}{Maximal requirement $p_{T,\ell\nu_\ell}<$ \ttt{max_pT_W} for (identified) $W$ bosons.}
\userswitch{MT_Wrec}{Switch for cuts on transverse mass $m_{T,\ell\nu_\ell}$ of (identified) $W$ bosons, defined by
\[m_{T,\ell\nu_\ell} =\sqrt{ \left(E_{T,\ell}+E_{T,\nu_{\ell}}\right)^2 - \left(p_{T,\ell\nu_{\ell}}\right)^2}\quad \mathrm{with} \quad (E_{T,x})^2=m_x^2+(p_{T,x})^2 \]\vspace{-0.4cm}}\\*
\usercut{min_MT_Wrec}{Minimal requirement $m_{T}^{\ell\nu_\ell}>$ \ttt{min_mT_Wrec} for (identified) $W$ bosons.}
\userswitch{rel_pT_miss}{Switch for cuts on the relative missing transverse momentum $\ptmissrel$, which is defined as $\ptmiss\times \sin|\Delta\phi|$, 
where $\Delta\phi$ is the azimuthal separation between $\mathbf\ptmiss$ and the momentum of the closest lepton, see \citere{Aad:2016wpd}.}\\*
\usercut{min_rel_pT_miss}{Minimal requirement $\ptmissrel >$ \ttt{min_rel_pT_miss}.}\\
\bottomrule\\

\caption{Explanations of the pre-defined cuts available in the various \Matrix{} processes. See \sct{sec:processes} for which cuts belong to which process.}
\label{tab:predefinedcuts}
\end{longtable}
\end{center}
\renewcommand{\baselinestretch}{1.0}}
\normalsize

\gdef\thesection{Appendix \Alph{section}}
\section{Modifications of the C++ code (advanced user)}\label{app:advancedstuff}
\gdef\thesection{\Alph{section}}

The user is generally advised to contact the authors if he/she is interested in changing any parts of the C++ code in order 
to define new dynamic scales, fiducial cuts or distributions. Below we provide some general guidance on how 
such implementations can be realized in the C++ code. Any changes of the C++ code require recompilation of the relevant 
process. This can be done with the \matrixparam{matrix} script, but in certain cases it might be simpler to use directly 
the \matrixparam{Makefile} the \matrixparam{matrix} script has created during the initial compilation of a process with ID \matrixparam{\$\{process_id\}}, by typing
\lstset{basicstyle=\small, frame=none}
{\tt
\begin{lstlisting}[language=bash]
 $ make ${process_id}
\end{lstlisting}
}
This enables recompilation without cleaning the whole process and without checking again whether all the relevant libraries are 
correctly installed.

\subsection{General C++ commands}
\subsubsection{Access to particle groups}

In the C++ routines for the user-defined scales and cuts one has access to all particle groups listed in \fig{tab:predefinedparticles} 
and the ones defined by the user, see below. As discussed before, the particle groups are ordered in 
the transverse momentum of the particles and can be accessed via \matrixparam{PARTICLE("$\{particle_group\}")[index]}, 
where \matrixparam{$\{particle_group\}} 
is one of the particle groups defined in \refta{tab:predefinedparticles} or the user-defined 
ones, and \matrixparam{index} indicates the position in the $\pt{}$-ordering of the group starting from the hardest 
one at \matrixparam{index = 0}.

For example, the hardest lepton can be accessed via

\definecolor{listinggray}{gray}{0.9}
\definecolor{lbcolor}{rgb}{0.96,0.96,0.96}
\definecolor{Darkgreen}{rgb}{0,0.55,0}
\lstset{language=C++,
backgroundcolor=\color{lbcolor},
                basicstyle=\scriptsize,
                keywordstyle=\color{blue},
                morekeywords={particle,fourvector,xdistribution },
                stringstyle=\color{Darkgreen},
                commentstyle=\color{red},
                identifierstyle=\color{black},
                morecomment=[l][\color{magenta}]{\#}
}

\lstdefinestyle{InputStyle} {
language=,
backgroundcolor=,
basicstyle=,
keywordstyle=,
stringstyle=,
commentstyle=,
basicstyle=\scriptsize,
frame=single
}

{\tt
\begin{lstlisting}                                                                                                                                                                                                                                                             
particle hardest_lepton = PARTICLE("lep")[0];
\end{lstlisting}
}

The particle class itself has various predefined class variables that can be directly used. For example,
the \pt{} or squared invariant mass can be determined as follows:

{\tt
\begin{lstlisting}                                                                                                                                                                                                                                                             
double pT_hardest_lepton = PARTICLE("lep")[0].pT;
double m2_hardest_lepton = PARTICLE("lep")[0].m2;
\end{lstlisting}
}

These can also determined via the full four-vector that has certain predefined functions and can be used as follows:

{\tt
\begin{lstlisting}                                                                                                                                                                                                                                                             
fourvector fourvector_of_hardest_lepton = PARTICLE("lep")[0].momentum;
double pT_hardest_lepton = fourvector_of_hardest_lepton.pT();
double m2_hardest_lepton = fourvector_of_hardest_lepton.m2();
\end{lstlisting}
}

or similar functions for other observables, such as rapidity and pseudo-rapidity

{\tt
\begin{lstlisting}                                                   
double y_hardest_lepton = PARTICLE("lep")[0].momentum.rapidity();
double eta_hardest_lepton = PARTICLE("lep")[0].momentum.eta();
\end{lstlisting}
}

It is important to note that the four-vectors of the particles can be simply added to 
define a new four-vector, where the same functions can be used. For example, 
the transverse mass ($\sqrt{m^2+p_T^2}$) of the system of the lepton pair (hardest and second-hardest lepton) 
can be simply computed by

{\tt
\begin{lstlisting}                                                   
fourvector fourvector_of_leplep = PARTICLE("lep")[0].momentum+PARTICLE("lep")[1].momentum;
double mT_leplep = fourvector_of_leplep.ET();
\end{lstlisting}
}

Similarly, one may also directly sum two objects of type {\tt particle}
to define a new particle whose momentum corresponds to the sum. Thus, it would be equivalent
to compute the transverse mass of the system of the lepton pair by using
{\tt
\begin{lstlisting}                                                   
particle leplep = PARTICLE("lep")[0] + PARTICLE("lep")[1];
double mT_leplep = leplep.ET;
\end{lstlisting}
}

Finally, if a certain observable one may want to compute is neither predefined in the {\tt particle} 
nor the {\tt fourvector} class, one can always access the momenta directly by using

{\tt
\begin{lstlisting}                                                   
fourvector fourvector_of_hardest_lep = PARTICLE("lep")[0].momentum;
double E_hardest_lep = fourvector_of_hardest_lepton.x0();
double x_hardest_lep = fourvector_of_hardest_lepton.x1();
double y_hardest_lep = fourvector_of_hardest_lepton.x2();
double z_hardest_lep = fourvector_of_hardest_lepton.x3();
\end{lstlisting}
}

and compute the desired observable from the explicit momentum components.

\subsubsection{Access to user-defined parameters}

In the file \matrixparam{parameter.dat} three types of user-defined parameters can be added, which are available in the C++ code and can be directly accessed throughout 
the process-specific C++ files inside the folder \Matrixversion{}\matrixparam{/prc/\$\{process_id\}/user}.
\begin{itemize}
\item An integer-valued user parameter is added via
{\tt
\begin{lstlisting}[style=InputStyle]
user_switch my_integer_parameter = 1
\end{lstlisting}
}
to the file \matrixparam{parameter.dat} and is accessed in the C++ code by
{\tt
\begin{lstlisting}                 
static int my_integer = USERSWITCH("my_integer_parameter");
\end{lstlisting}
}
Such switches are useful in many respects, two already used examples are to turn on and off cuts, or to choose between different identification procedures.

\item A real-valued user parameter is added via
{\tt
\begin{lstlisting}[style=InputStyle]
user_cut my_real_parameter = 1.23
\end{lstlisting}
}
to the file \matrixparam{parameter.dat} and is accessed in the C++ code by
{\tt
\begin{lstlisting}                 
static double my_real = USERCUT("my_real_parameter");
\end{lstlisting}
}
Such real parameters are useful in many respects, the most important example is their use to define and implement cuts that can be changed later from the 
file \matrixparam{parameter.dat} without recompilation of the code.

\item Finally, a new particle (group) can be defined by adding
{\tt
\begin{lstlisting}[style=InputStyle]
user_particle my_own_particle = my_own_particle
\end{lstlisting}
}
to the file \matrixparam{parameter.dat}. Only if defined this way, it can be 
filled in the C++ code,
{\tt
\begin{lstlisting}
USERPARTICLE("my_own_particle").push_back(PARTICLE("lep")[0] + PARTICLE("lep")[1]);
\end{lstlisting}
}
which would add a particle whose momentum is the sum of the hardest and second-hardest lepton 
to the user-defined particle group \matrixparam{my_own_particle}. If a user-defined particle group is filled with more than one particle, 
the usual $\pT$-ordering is done automatically before the respective particle group is used in scales, cuts or distributions. 
Note that a \matrixparam{USERPARTICLE} may only be filled in a certain position of the code,
see below in Appendix~\app{app:particlegroup}, and that it can be accessed later 
like all other particle groups via the container \matrixparam{PARTICLE}.

\end{itemize}

\subsection{Definition of a new particle group}\label{app:particlegroup}

The predefined particle groups are sufficient for most practical cases. However, the user
is allowed to define his own particle group by filling the respective four-vectors. This can 
be very useful if an intermediate particle cannot be unambiguously reconstructed, like in the case
of the SF channel in \zz{}(or \wz{}) production, where an identification procedure is needed for 
the $Z$ (and the $W$) bosons. As described in \scts{sec:ZZSF} and \ref{sec:WZSF}, such definition of process-specific 
particle groups is done intrinsically for these two processes following different identification procedures 
used by ATLAS and CMS. In the following we describe the necessary steps for a user to add his own
definition of a particle group to any process.

The relevant C++ file is \Matrixversion{}\matrixparam{/prc/\$\{process_id\}/user/specify.particles.cxx} to fill user-defined particles.
As described above, one has to add the definition of a new particle group to the file \matrixparam{parameter.dat}. After that 
the respective particle group exists as an empty array of objects of type {\tt particle} inside the C++ code, which must 
be filled by appending at least one object of type {\tt particle} to the array.

Let us give a simple example: A new particle group should be filled depending on the value of an integer switch with either the 
hardest lepton or the second-hardest lepton. The relevant input in the file \matrixparam{parameter.dat} would look like

{\tt
\begin{lstlisting}[style=InputStyle]
user_particle relevant_lepton = relevant_lepton  #  lepton, depending on switch_lepton
user_switch switch_lepton = 0                    #  (0) hardest lepton, (1) second-hardest lepton
\end{lstlisting}
}

and the relevant C++ code in the \matrixparam{specify.particles.cxx} would be

{\tt
\begin{lstlisting}     
...
  static int switch_lepton = USERSWITCH("switch_lepton");
  if (switch_lepton == 0){
    USERPARTICLE("relevant_lepton").push_back(PARTICLE("lep")[0]);
  }
  else if (switch_lepton == 1){
    USERPARTICLE("relevant_lepton").push_back(PARTICLE("lep")[1]);
  }
  else {
    logger << LOG_ERROR << "ERROR: switch_lepton = " << switch_lepton << "; allowed values: 0, 1" << endl;
    assert(false);
  }
...
\end{lstlisting}
}

Bear in mind that after definition of a \matrixparam{USERPARTICLE}, the respective 
particle group is automatically filled afterwards. If the user-defined particle group contains more than one particle, the usual $\pT$-ordering is applied. The
new particle group can then be accessed via the standard 
\matrixparam{PARTICLE} container, e.g.\

{\tt
\begin{lstlisting}     
... 
  particle the_relevant_lepton = PARTICLE("relevant_lepton")[0]
...
\end{lstlisting}
}

At this point the definition of the new particle has no practical effect yet, but one could now use the new particle group 
in the definition of a cut or for a distribution, and then decide via the switch in the input file whether it uses the hardest lepton 
or the second-hardest lepton. Such simple example may not appear to be extremely useful, however, this 
changes drastically if such cut or distribution is done according to a more complicated identification of a particle. In that 
case the identification procedure can be switched on-the-fly without the need of recompilation and without having to 
reimplement the same cuts and distributions for every new identification procedure. We refer the interested reader 
to \Matrixversion{}\matrixparam{/prc/ppeeexex04/user/specify.particles.cxx} for a sample implementation.

\subsection{Implementation of a new dynamic scale}
The relevant C++ file is \Matrixversion{}\matrixparam{/prc/\$\{process_id\}/user/specify.scales.cxx} to add a new user-defined dynamic scale.
All processes have at least two dynamic scales already implemented, and one can follow these implementations.
In principle, one is free to code whatever one desires
in that file, without taking care of the existing structure. One only has to make sure that 
in the end the variable \matrixparam{temp_mu_central} is set to the correct value. However, we recommend to follow the existing structure of the 
{\tt if} and {\tt else if} blocks to keep with the functionality of choosing different dynamic scales in the file \matrixparam{parameter.dat}.

Let us give a simple example, for completeness: If we want to add a dynamic scale 123 that computes the sum of the $Z$-boson mass 
and the transverse momentum of the hardest lepton, we would set

{\tt
\begin{lstlisting}[style=InputStyle]
dynamic_scale   =  123         #  dynamic ren./fac. scale
\end{lstlisting}
}

in the file \matrixparam{parameter.dat} and add an {\tt else if} block to the \matrixparam{specify.scales.cxx} file:

{\tt
\begin{lstlisting}
...
  else if (sd == 123){                                                                                                                                                                                                                                                                       
    // sum of Z-boson mass and pT of hardest lepton
    double m_Z = osi_msi.M_Z;
    double pT_hardest_lepton = PARTICLE("lep")[0].pT;
    temp_mu_central = m_Z + pT_hardest_lepton;                                                                                                                                        
  }
...
\end{lstlisting}
}

\subsection{Implementation of a new user-defined cut}

The relevant C++ file is
\Matrixversion{}\matrixparam{/prc/\$\{process_id\}/user/specify.cuts.cxx} to add a new user-defined cut.
Most processes already have several process-specific cuts implemented, and one can in principle follow these implementations.
Also in this file the user is essentially free to implement whatever he deserves without taking care of the existing structure. 
The only relevant information is that under whatever conditions one requires an event to be discarded, one sets 

{\tt
\begin{lstlisting}
...
  osi_cut_ps[i_a] = -1;
  return;
...
\end{lstlisting}
}

in order to cut the current phase-space point. The relevant momenta at each event are accessed via the particle groups 
as explained above. Nevertheless, we recommend to keep the existing structure by defining cuts via the \matrixparam{user_switch} and 
\matrixparam{user_cut} parameters that can be interactively changed afterwards in the file \matrixparam{parameter.dat} without recompilation 
of the C++ code, instead of hard-coding such information in the file \matrixparam{specify.cuts.cxx}.

As a simple example we consider a lower cut on the absolute rapidity difference between the hardest and second-hardest lepton. 
Such cuts are added to the file \matrixparam{parameter.dat},

{\tt
\begin{lstlisting}[style=InputStyle]
user_switch dy_lep1lep2 = 1     #  switch to turn on (1) and off (0) cuts on absolute dy of leptons
user_cut min_dy_lep1lep2 = 0.5  #  requirement on absolute rapidity difference of leptons (lower cut)
\end{lstlisting}
}

and implemented into the C++ code as follows:

{\tt
\begin{lstlisting}
...
  // get settings for cut on absolute rapidity difference of leptons
  static int switch_dy_lep1lep2     = USERSWITCH("dy_lep1lep2 ");
  static double cut_min_dy_lep1lep2 = USERCUT("min_dy_lep1lep2");
    
  // perform cut on absolute rapidity difference of leptons according to settings
  if (switch_dy_lep1lep2 == 1){
    double y_lep1 = PARTICLE("lep")[0].rapidity;
    double y_lep2 = PARTICLE("lep")[1].rapidity;
    double dy_lep1lep2 = y_lep1 - y_lep2;
    if (abs(dy_lep1lep2) < cut_min_dy_lep1lep2) {
      osi_cut_ps[i_a] = -1;  // cut phase-space point
      return;
    }
  }
...
\end{lstlisting}
}

\subsection{Implementation of a new observable for distributions}

The relevant C++ file is \Matrixversion{}{\tt /src-MUNICH/classes/xdistribution.cpp} to add a new user-defined distribution. 
Note that this part of the code is not specific to a certain process, and any observable implemented here can in principle be used in all processes.
The relevant routine of the {\tt xdistribution} class is
{\tt
\begin{lstlisting}
void xdistribution::computeObservable(...) {
  ...
}
\end{lstlisting}
}

A rather comprehensive description of how to add a new distribution can be found commented inside this routine. We summarize the most 
important information. As pointed out in \sct{sec:distributions}, each observable has a certain type identifier \matrixparam{distributiontype} 
set in the file \matrixparam{distribution.dat}, which must be specified in every distribution block. Inside the \matrixparam{computeObservable} 
routine of the {\tt xdistribution.cpp} file, we can add a new distribution-type by extending the {\tt if} and {\tt else if} blocks for 
\matrixparam{xdistribution_type}, which corresponds to the string set for \matrixparam{distributiontype} in the file \matrixparam{distribution.dat}.
The sum of the momenta of {\tt particle i} defined for each distribution in the file \matrixparam{distribution.dat} is saved to an array with 
entries of type {\tt fourvector} called \matrixparam{reconstructedParticles[i]} inside the C++ code. The distributions can now be defined using these 
particles, by setting the variable \matrixparam{observable} to the value of the 
observable that should be binned for the desired \matrixparam{xdistribution_type}.

Let us consider a simple example where we want to plot the distribution of events in the sum of the rapidities of the two hardest leptons
by defining a new \matrixparam{xdistribution_type}. The definition of the distribution in the file \matrixparam{distribution.dat} would look like
{\tt
\begin{lstlisting}[style=InputStyle]
distributionname  =  y_lep1_plus_y_lep2
distributiontype  =  sum_of_y
particle 1        =  lep 1
particle 2        =  lep 2
startpoint        =  0.
endpoint          =  10.
binwidth          =  0.2
\end{lstlisting}
}
where the name and the specific definition of the binning has relevance for this example. The C++ code for the distribution type 
\matrixparam{sum_of_y} can be implemented in a general way for an arbitrary number of \matrixparam{patricle i} definitions by adding an {\tt else if} 
block to the \matrixparam{computeObservable} routine in the file {\tt xdistribution.cpp}:
{\tt
\begin{lstlisting}
...
  else if (xdistribution_type == "sum_of_y}") {
    double sum_y = 0;
    for (int group = 0; group < particles.size(); group++) {
      fourvector fourvector_of_current_reconstructed_particle = reconstructedParticles[group]
      sum_y = sum_y + fourvector_of_current_reconstructed_particle.y();
    }
    observable = sum_y;  
  }     
...
\end{lstlisting}
}

\gdef\thesection{Appendix \Alph{section}}
\section{Troubleshooting}\label{app:Troubleshooting}
\gdef\thesection{\Alph{section}}

\lstset{language=,
backgroundcolor=,
basicstyle=,
keywordstyle=,
stringstyle=,
commentstyle=,
frame=
}

\subsection{Compiling on lxplus}
There is a problem when compiling \OpenLoops{} on the lxplus cluster due to an outdated Fortran version. Furthermore, when 
using the window manager \matrixparam{screen}, the compilers/executables might not be working (including Python). In both cases you need to execute
{\tt
\begin{lstlisting}[language=bash]
 $ source /afs/cern.ch/sw/lcg/hepsoft/0.9/x86_64-slc6-gcc48-opt/setup.sh
\end{lstlisting}
}
before compiling \OpenLoops{}.\footnote{In general, it is a good idea to 
add it to your \matrixparam{.bashrc}/\matrixparam{.bash\_profile} (and/or your \matrixparam{.screenrc}) to avoid having to retype it for each new session/screen.}

\subsection{Using a window manager on lxplus}
Since lxplus grants read/write permissions via kerberos tickets, which 
are valid only for 24 hours, it is 
not trivial to employ a window manager. In particular, the 
standard option \matrixparam{screen} does not work properly. We 
recommend \matrixparam{tmux} on lxplus instead, which can be 
used as follows:

First, create a session with a kerberos ticket
{\tt
\begin{lstlisting}[language=bash]
krenew -b -t -- tmux new-session -d -s my_tmux_session
\end{lstlisting}
}
and enter the session with 

{\tt
\begin{lstlisting}[language=bash]
tmux attach
\end{lstlisting}
}

Change the directory to a \Matrix{} process folder and start a run inside the 
\matrixparam{tmux} session. The session can now be detached 
(\keystroke{Ctrl+b} \keystroke{d}) and the run will continue.
However, read/write permission will end after 24 hours. In order to 
maintain them, the kerberos ticket must be renewed inside the
\matrixparam{tmux} session. To do so, enter the \matrixparam{tmux} 
session again and open a second window inside the same 
session (\keystroke{Ctrl+b} \keystroke{c}). Now, enter

{\tt
\begin{lstlisting}[language=bash]
kinit
\end{lstlisting}
}

and type your CERN password to renew the kerberos ticket. 
Change between the two \matrixparam{tmux} windows (\keystroke{Ctrl+b} \keystroke{n})
and get back to the output of the \Matrix{} run. Before further 24 hours have 
passed, the kerberos ticket needs to be renewed again. In principle,
one could have a script take care of these renewals. 
However, it is not secure to safe the CERN password within a human-readable executable.

\subsection{Problems with libquadmath}
If you encounter
{\tt
\begin{lstlisting}[language=bash]
error while loading shared libraries:\
 libquadmath.so.0: cannot open shared object file: No such file or directory
\end{lstlisting}
}
at runtime, implying that dynamic linking to \matrixparam{libquadmath} failed, you can set \matrixparam{path\_to\_libgfortran} in the file \matrixparam{MATRIX\_configuration} to the path where \matrixparam{libquadmath} is installed on your system.

\linespread{0}\selectfont
\bibliographystyle{UTPstyle}
\bibliography{matrix_release}

\providecommand{\href}[2]{#2}\begingroup\raggedright\begin{thebibliography}{100}

\bibitem{MATRIX}
{\Matrix{} is available for download from: \url{http://matrix.hepforge.org/}}.

\bibitem{Hamberg:1990np}
R.~Hamberg, W.~L. van Neerven, and T.~Matsuura, {\it {A complete calculation of
  the order $\alpha_s^{2}$ correction to the Drell-Yan $K$ factor}},  {\em
  Nucl. Phys.} {\bf B359} (1991) 343--405. [Erratum: Nucl.
  Phys.B644,403(2002)].

\bibitem{Harlander:2002wh}
R.~V. Harlander and W.~B. Kilgore, {\it {Next-to-next-to-leading order Higgs
  production at hadron colliders}},  {\em Phys. Rev. Lett.} {\bf 88} (2002)
  201801, [\href{http://xxx.lanl.gov/abs/hep-ph/0201206}{{\tt
  hep-ph/0201206}}].

\bibitem{Anastasiou:2002yz}
C.~Anastasiou and K.~Melnikov, {\it {Higgs boson production at hadron colliders
  in NNLO QCD}},  {\em Nucl. Phys.} {\bf B646} (2002) 220--256,
  [\href{http://xxx.lanl.gov/abs/hep-ph/0207004}{{\tt hep-ph/0207004}}].

\bibitem{Ravindran:2003um}
V.~Ravindran, J.~Smith, and W.~L. van Neerven, {\it {NNLO corrections to the
  total cross-section for Higgs boson production in hadron hadron collisions}},
   {\em Nucl. Phys.} {\bf B665} (2003) 325--366,
  [\href{http://xxx.lanl.gov/abs/hep-ph/0302135}{{\tt hep-ph/0302135}}].

\bibitem{Anastasiou:2003ds}
C.~Anastasiou, L.~J. Dixon, K.~Melnikov, and F.~Petriello, {\it {High precision
  QCD at hadron colliders: Electroweak gauge boson rapidity distributions at
  NNLO}},  {\em Phys. Rev.} {\bf D69} (2004) 094008,
  [\href{http://xxx.lanl.gov/abs/hep-ph/0312266}{{\tt hep-ph/0312266}}].

\bibitem{Anastasiou:2005qj}
C.~Anastasiou, K.~Melnikov, and F.~Petriello, {\it {Fully differential Higgs
  boson production and the di-photon signal through next-to-next-to-leading
  order}},  {\em Nucl. Phys.} {\bf B724} (2005) 197--246,
  [\href{http://xxx.lanl.gov/abs/hep-ph/0501130}{{\tt hep-ph/0501130}}].

\bibitem{Melnikov:2006kv}
K.~Melnikov and F.~Petriello, {\it {Electroweak gauge boson production at
  hadron colliders through $\mathcal{O}(\alpha_S^2)$}},  {\em Phys. Rev.} {\bf
  D74} (2006) 114017, [\href{http://xxx.lanl.gov/abs/hep-ph/0609070}{{\tt
  hep-ph/0609070}}].

\bibitem{Catani:2007vq}
S.~Catani and M.~Grazzini, {\it {An NNLO subtraction formalism in hadron
  collisions and its application to Higgs boson production at the LHC}},  {\em
  Phys. Rev. Lett.} {\bf 98} (2007) 222002,
  [\href{http://xxx.lanl.gov/abs/hep-ph/0703012}{{\tt hep-ph/0703012}}].

\bibitem{Anastasiou:2007mz}
C.~Anastasiou, G.~Dissertori, and F.~St{\"o}ckli, {\it {NNLO QCD predictions
  for the $H \to WW \to \ell \nu \ell \nu$ signal at the LHC}},  {\em JHEP}
  {\bf 09} (2007) 018, [\href{http://xxx.lanl.gov/abs/0707.2373}{{\tt
  arXiv:0707.2373}}].

\bibitem{Grazzini:2008tf}
M.~Grazzini, {\it {NNLO predictions for the Higgs boson signal in the $H\to WW
  \to l\nu l\nu$ and $H\to ZZ \to 4l$ decay channels}},  {\em JHEP} {\bf 02}
  (2008) 043, [\href{http://xxx.lanl.gov/abs/0801.3232}{{\tt
  arXiv:0801.3232}}].

\bibitem{Catani:2009sm}
S.~Catani, L.~Cieri, G.~Ferrera, D.~de~Florian, and M.~Grazzini, {\it {Vector
  boson production at hadron colliders: a fully exclusive QCD calculation at
  NNLO}},  {\em Phys. Rev. Lett.} {\bf 103} (2009) 082001,
  [\href{http://xxx.lanl.gov/abs/0903.2120}{{\tt arXiv:0903.2120}}].

\bibitem{Ferrera:2011bk}
G.~Ferrera, M.~Grazzini, and F.~Tramontano, {\it {Associated $WH$ production at
  hadron colliders: a fully exclusive QCD calculation at NNLO}},  {\em Phys.
  Rev. Lett.} {\bf 107} (2011) 152003,
  [\href{http://xxx.lanl.gov/abs/1107.1164}{{\tt arXiv:1107.1164}}].

\bibitem{Ferrera:2014lca}
G.~Ferrera, M.~Grazzini, and F.~Tramontano, {\it {Associated $ZH$ production at
  hadron colliders: the fully differential NNLO QCD calculation}},  {\em Phys.
  Lett.} {\bf B740} (2015) 51--55,
  [\href{http://xxx.lanl.gov/abs/1407.4747}{{\tt arXiv:1407.4747}}].

\bibitem{Ferrera:2017zex}
G.~Ferrera, G.~Somogyi, and F.~Tramontano, {\it {Associated production of a
  Higgs boson decaying into bottom quarks at the LHC in full NNLO QCD}},
  \href{http://xxx.lanl.gov/abs/1705.10304}{{\tt arXiv:1705.10304}}.

\bibitem{Campbell:2016jau}
J.~M. Campbell, R.~K. Ellis, and C.~Williams, {\it {Associated production of a
  Higgs boson at NNLO}},  {\em JHEP} {\bf 06} (2016) 179,
  [\href{http://xxx.lanl.gov/abs/1601.00658}{{\tt arXiv:1601.00658}}].

\bibitem{Harlander:2003ai}
R.~V. Harlander and W.~B. Kilgore, {\it {Higgs boson production in bottom quark
  fusion at next-to-next-to leading order}},  {\em Phys. Rev.} {\bf D68} (2003)
  013001, [\href{http://xxx.lanl.gov/abs/hep-ph/0304035}{{\tt
  hep-ph/0304035}}].

\bibitem{Harlander:2010cz}
R.~V. Harlander, K.~J. Ozeren, and M.~Wiesemann, {\it {Higgs plus jet
  production in bottom quark annihilation at next-to-leading order}},  {\em
  Phys. Lett.} {\bf B693} (2010) 269--273,
  [\href{http://xxx.lanl.gov/abs/1007.5411}{{\tt arXiv:1007.5411}}].

\bibitem{Harlander:2011fx}
R.~Harlander and M.~Wiesemann, {\it {Jet-veto in bottom-quark induced Higgs
  production at next-to-next-to-leading order}},  {\em JHEP} {\bf 04} (2012)
  066, [\href{http://xxx.lanl.gov/abs/1111.2182}{{\tt arXiv:1111.2182}}].

\bibitem{Buehler:2012cu}
S.~B{\"u}hler, F.~Herzog, A.~Lazopoulos, and R.~M{\"u}ller, {\it {The fully
  differential hadronic production of a Higgs boson via bottom quark fusion at
  NNLO}},  {\em JHEP} {\bf 07} (2012) 115,
  [\href{http://xxx.lanl.gov/abs/1204.4415}{{\tt arXiv:1204.4415}}].

\bibitem{Marzani:2008az}
S.~Marzani, R.~D. Ball, V.~Del~Duca, S.~Forte, and A.~Vicini, {\it {Higgs
  production via gluon-gluon fusion with finite top mass beyond next-to-leading
  order}},  {\em Nucl. Phys.} {\bf B800} (2008) 127--145,
  [\href{http://xxx.lanl.gov/abs/0801.2544}{{\tt arXiv:0801.2544}}].

\bibitem{Harlander:2009mq}
R.~V. Harlander and K.~J. Ozeren, {\it {Finite top mass effects for hadronic
  Higgs production at next-to-next-to-leading order}},  {\em JHEP} {\bf 11}
  (2009) 088, [\href{http://xxx.lanl.gov/abs/0909.3420}{{\tt
  arXiv:0909.3420}}].

\bibitem{Harlander:2009my}
R.~V. Harlander, H.~Mantler, S.~Marzani, and K.~J. Ozeren, {\it {Higgs
  production in gluon fusion at next-to-next-to-leading order QCD for finite
  top mass}},  {\em Eur. Phys. J.} {\bf C66} (2010) 359--372,
  [\href{http://xxx.lanl.gov/abs/0912.2104}{{\tt arXiv:0912.2104}}].

\bibitem{Pak:2009dg}
A.~Pak, M.~Rogal, and M.~Steinhauser, {\it {Finite top quark mass effects in
  NNLO Higgs boson production at LHC}},  {\em JHEP} {\bf 02} (2010) 025,
  [\href{http://xxx.lanl.gov/abs/0911.4662}{{\tt arXiv:0911.4662}}].

\bibitem{Neumann:2014nha}
T.~Neumann and M.~Wiesemann, {\it {Finite top-mass effects in gluon-induced
  Higgs production with a jet-veto at NNLO}},  {\em JHEP} {\bf 11} (2014) 150,
  [\href{http://xxx.lanl.gov/abs/1408.6836}{{\tt arXiv:1408.6836}}].

\bibitem{deFlorian:2013jea}
D.~de~Florian and J.~Mazzitelli, {\it {Higgs Boson Pair Production at
  Next-to-Next-to-Leading Order in QCD}},  {\em Phys. Rev. Lett.} {\bf 111}
  (2013) 201801, [\href{http://xxx.lanl.gov/abs/1309.6594}{{\tt
  arXiv:1309.6594}}].

\bibitem{deFlorian:2016uhr}
D.~de~Florian, M.~Grazzini, C.~Hanga, S.~Kallweit, J.~M. Lindert,
  P.~Maierh{\"o}fer, J.~Mazzitelli, and D.~Rathlev, {\it {Differential Higgs
  Boson Pair Production at Next-to-Next-to-Leading Order in QCD}},  {\em JHEP}
  {\bf 09} (2016) 151, [\href{http://xxx.lanl.gov/abs/1606.09519}{{\tt
  arXiv:1606.09519}}].

\bibitem{Catani:2011qz}
S.~Catani, L.~Cieri, D.~de~Florian, G.~Ferrera, and M.~Grazzini, {\it {Diphoton
  production at hadron colliders: a fully-differential QCD calculation at
  NNLO}},  {\em Phys. Rev. Lett.} {\bf 108} (2012) 072001,
  [\href{http://xxx.lanl.gov/abs/1110.2375}{{\tt arXiv:1110.2375}}]. [Erratum:
  Phys. Rev. Lett.117,no.8,089901(2016)].

\bibitem{Campbell:2016yrh}
J.~M. Campbell, R.~K. Ellis, Y.~Li, and C.~Williams, {\it {Predictions for
  diphoton production at the LHC through NNLO in QCD}},  {\em JHEP} {\bf 07}
  (2016) 148, [\href{http://xxx.lanl.gov/abs/1603.02663}{{\tt
  arXiv:1603.02663}}].

\bibitem{Grazzini:2013bna}
M.~Grazzini, S.~Kallweit, D.~Rathlev, and A.~Torre, {\it {$Z\gamma$ production
  at hadron colliders in NNLO QCD}},  {\em Phys. Lett.} {\bf B731} (2014)
  204--207, [\href{http://xxx.lanl.gov/abs/1309.7000}{{\tt arXiv:1309.7000}}].

\bibitem{Grazzini:2015nwa}
M.~Grazzini, S.~Kallweit, and D.~Rathlev, {\it {$W\gamma$ and $Z\gamma$
  production at the LHC in NNLO QCD}},  {\em JHEP} {\bf 07} (2015) 085,
  [\href{http://xxx.lanl.gov/abs/1504.01330}{{\tt arXiv:1504.01330}}].

\bibitem{Campbell:2017aul}
J.~M. Campbell, T.~Neumann, and C.~Williams, {\it {$Z\gamma$ production at NNLO
  including anomalous couplings}},  {\em Submitted to: JHEP} (2017)
  [\href{http://xxx.lanl.gov/abs/1708.02925}{{\tt arXiv:1708.02925}}].

\bibitem{Cascioli:2014yka}
F.~Cascioli, T.~Gehrmann, M.~Grazzini, S.~Kallweit, P.~Maier{h\"o}fer, A.~von
  Manteuffel, S.~Pozzorini, D.~Rathlev, L.~Tancredi, and E.~Weihs, {\it {$ZZ$
  production at hadron colliders in NNLO QCD}},  {\em Phys. Lett.} {\bf B735}
  (2014) 311--313, [\href{http://xxx.lanl.gov/abs/1405.2219}{{\tt
  arXiv:1405.2219}}].

\bibitem{Grazzini:2015hta}
M.~Grazzini, S.~Kallweit, and D.~Rathlev, {\it {ZZ production at the LHC:
  fiducial cross sections and distributions in NNLO QCD}},  {\em Phys. Lett.}
  {\bf B750} (2015) 407--410, [\href{http://xxx.lanl.gov/abs/1507.06257}{{\tt
  arXiv:1507.06257}}].

\bibitem{Heinrich:2017bvg}
G.~Heinrich, S.~Jahn, S.~P. Jones, M.~Kerner, and J.~Pires, {\it {NNLO
  predictions for Z-boson pair production at the LHC}},
  \href{http://xxx.lanl.gov/abs/1710.06294}{{\tt arXiv:1710.06294}}.

\bibitem{Gehrmann:2014fva}
T.~Gehrmann, M.~Grazzini, S.~Kallweit, P.~Maierh{\"o}fer, A.~von Manteuffel,
  S.~Pozzorini, D.~Rathlev, and L.~Tancredi, {\it {$W^+W^-$ Production at
  Hadron Colliders in Next to Next to Leading Order QCD}},  {\em Phys. Rev.
  Lett.} {\bf 113} (2014), no.~21 212001,
  [\href{http://xxx.lanl.gov/abs/1408.5243}{{\tt arXiv:1408.5243}}].

\bibitem{Grazzini:2016ctr}
M.~Grazzini, S.~Kallweit, S.~Pozzorini, D.~Rathlev, and M.~Wiesemann, {\it
  {$W^{+}W^{−}$ production at the LHC: fiducial cross sections and
  distributions in NNLO QCD}},  {\em JHEP} {\bf 08} (2016) 140,
  [\href{http://xxx.lanl.gov/abs/1605.02716}{{\tt arXiv:1605.02716}}].

\bibitem{Grazzini:2016swo}
M.~Grazzini, S.~Kallweit, D.~Rathlev, and M.~Wiesemann, {\it {$W^{\pm}Z$
  production at hadron colliders in NNLO QCD}},  {\em Phys. Lett.} {\bf B761}
  (2016) 179--183, [\href{http://xxx.lanl.gov/abs/1604.08576}{{\tt
  arXiv:1604.08576}}].

\bibitem{Grazzini:2017ckn}
M.~Grazzini, S.~Kallweit, D.~Rathlev, and M.~Wiesemann, {\it {$W^\pm Z$
  production at the LHC: fiducial cross sections and distributions in NNLO
  QCD}},  {\em JHEP} {\bf 05} (2017) 139,
  [\href{http://xxx.lanl.gov/abs/1703.09065}{{\tt arXiv:1703.09065}}].

\bibitem{Czakon:2013goa}
M.~Czakon, P.~Fiedler, and A.~Mitov, {\it {Total Top-Quark Pair-Production
  Cross Section at Hadron Colliders Through $\mathcal{O}(\alpha_S^4)$}},  {\em
  Phys. Rev. Lett.} {\bf 110} (2013) 252004,
  [\href{http://xxx.lanl.gov/abs/1303.6254}{{\tt arXiv:1303.6254}}].

\bibitem{Czakon:2015owf}
M.~Czakon, D.~Heymes, and A.~Mitov, {\it {High-precision differential
  predictions for top-quark pairs at the LHC}},  {\em Phys. Rev. Lett.} {\bf
  116} (2016), no.~8 082003, [\href{http://xxx.lanl.gov/abs/1511.00549}{{\tt
  arXiv:1511.00549}}].

\bibitem{Brucherseifer:2014ama}
M.~Brucherseifer, F.~Caola, and K.~Melnikov, {\it {On the NNLO QCD corrections
  to single-top production at the LHC}},  {\em Phys. Lett.} {\bf B736} (2014)
  58--63, [\href{http://xxx.lanl.gov/abs/1404.7116}{{\tt arXiv:1404.7116}}].

\bibitem{Currie:2017eqf}
J.~Currie, A.~Gehrmann-De~Ridder, T.~Gehrmann, E.~W.~N. Glover, A.~Huss, and
  J.~Pires, {\it {Precise predictions for dijet production at the LHC}},  {\em
  Phys. Rev. Lett.} {\bf 119} (2017), no.~15 152001,
  [\href{http://xxx.lanl.gov/abs/1705.10271}{{\tt arXiv:1705.10271}}].

\bibitem{Cacciari:2015jma}
M.~Cacciari, F.~A. Dreyer, A.~Karlberg, G.~P. Salam, and G.~Zanderighi, {\it
  {Fully Differential Vector-Boson-Fusion Higgs Production at
  Next-to-Next-to-Leading Order}},  {\em Phys. Rev. Lett.} {\bf 115} (2015),
  no.~8 082002, [\href{http://xxx.lanl.gov/abs/1506.02660}{{\tt
  arXiv:1506.02660}}].

\bibitem{Boughezal:2015aha}
R.~Boughezal, C.~Focke, W.~Giele, X.~Liu, and F.~Petriello, {\it {Higgs boson
  production in association with a jet at NNLO using jettiness subtraction}},
  {\em Phys. Lett.} {\bf B748} (2015) 5--8,
  [\href{http://xxx.lanl.gov/abs/1505.03893}{{\tt arXiv:1505.03893}}].

\bibitem{Caola:2015wna}
F.~Caola, K.~Melnikov, and M.~Schulze, {\it {Fiducial cross sections for Higgs
  boson production in association with a jet at next-to-next-to-leading order
  in QCD}},  {\em Phys. Rev.} {\bf D92} (2015), no.~7 074032,
  [\href{http://xxx.lanl.gov/abs/1508.02684}{{\tt arXiv:1508.02684}}].

\bibitem{Chen:2016zka}
X.~Chen, J.~Cruz-Martinez, T.~Gehrmann, E.~W.~N. Glover, and M.~Jaquier, {\it
  {NNLO QCD corrections to Higgs boson production at large transverse
  momentum}},  {\em JHEP} {\bf 10} (2016) 066,
  [\href{http://xxx.lanl.gov/abs/1607.08817}{{\tt arXiv:1607.08817}}].

\bibitem{Campbell:2016lzl}
J.~M. Campbell, R.~K. Ellis, and C.~Williams, {\it {Direct Photon Production at
  Next-to--Next-to-Leading Order}},  {\em Phys. Rev. Lett.} {\bf 118} (2017),
  no.~22 222001, [\href{http://xxx.lanl.gov/abs/1612.04333}{{\tt
  arXiv:1612.04333}}].

\bibitem{Boughezal:2015ded}
R.~Boughezal, J.~M. Campbell, R.~K. Ellis, C.~Focke, W.~T. Giele, X.~Liu, and
  F.~Petriello, {\it {Z-boson production in association with a jet at
  next-to-next-to-leading order in perturbative QCD}},  {\em Phys. Rev. Lett.}
  {\bf 116} (2016), no.~15 152001,
  [\href{http://xxx.lanl.gov/abs/1512.01291}{{\tt arXiv:1512.01291}}].

\bibitem{Gehrmann-DeRidder:2016jns}
A.~Gehrmann-De~Ridder, T.~Gehrmann, E.~W.~N. Glover, A.~Huss, and T.~A. Morgan,
  {\it {NNLO QCD corrections for Drell-Yan $p_T^Z$ and $\phi^*$ observables at
  the LHC}},  {\em JHEP} {\bf 11} (2016) 094,
  [\href{http://xxx.lanl.gov/abs/1610.01843}{{\tt arXiv:1610.01843}}].

\bibitem{Boughezal:2015dva}
R.~Boughezal, C.~Focke, X.~Liu, and F.~Petriello, {\it {$W$-boson production in
  association with a jet at next-to-next-to-leading order in perturbative
  QCD}},  {\em Phys. Rev. Lett.} {\bf 115} (2015), no.~6 062002,
  [\href{http://xxx.lanl.gov/abs/1504.02131}{{\tt arXiv:1504.02131}}].

\bibitem{Gavin:2012sy}
R.~Gavin, Y.~Li, F.~Petriello, and S.~Quackenbush, {\it {W Physics at the LHC
  with FEWZ 2.1}},  {\em Comput. Phys. Commun.} {\bf 184} (2013) 208--214,
  [\href{http://xxx.lanl.gov/abs/1201.5896}{{\tt arXiv:1201.5896}}].

\bibitem{Anastasiou:2009kn}
C.~Anastasiou, S.~Bucherer, and Z.~Kunszt, {\it {HPro: A NLO Monte-Carlo for
  Higgs production via gluon fusion with finite heavy quark masses}},  {\em
  JHEP} {\bf 10} (2009) 068, [\href{http://xxx.lanl.gov/abs/0907.2362}{{\tt
  arXiv:0907.2362}}].

\bibitem{Boughezal:2016wmq}
R.~Boughezal, J.~M. Campbell, R.~K. Ellis, C.~Focke, W.~Giele, X.~Liu,
  F.~Petriello, and C.~Williams, {\it {Color singlet production at NNLO in
  MCFM}},  {\em Eur. Phys. J.} {\bf C77} (2017), no.~1 7,
  [\href{http://xxx.lanl.gov/abs/1605.08011}{{\tt arXiv:1605.08011}}].

\bibitem{Catani:1996jh}
S.~Catani and M.~Seymour, {\it {The Dipole formalism for the calculation of QCD
  jet cross-sections at next-to-leading order}},  {\em Phys. Lett.} {\bf B378}
  (1996) 287--301, [\href{http://xxx.lanl.gov/abs/hep-ph/9602277}{{\tt
  hep-ph/9602277}}].

\bibitem{Catani:1996vz}
S.~Catani and M.~Seymour, {\it {A General algorithm for calculating jet
  cross-sections in NLO QCD}},  {\em Nucl. Phys.} {\bf B485} (1997) 291--419,
  [\href{http://xxx.lanl.gov/abs/hep-ph/9605323}{{\tt hep-ph/9605323}}].

\bibitem{Cascioli:2011va}
F.~Cascioli, P.~Maierh{\"o}fer, and S.~Pozzorini, {\it {Scattering Amplitudes
  with Open Loops}},  {\em Phys. Rev. Lett.} {\bf 108} (2012) 111601,
  [\href{http://xxx.lanl.gov/abs/1111.5206}{{\tt arXiv:1111.5206}}].

\bibitem{Buccioni:2017yxi}
F.~Buccioni, S.~Pozzorini, and M.~Zoller, {\it {On-the-fly reduction of open
  loops}},  \href{http://xxx.lanl.gov/abs/1710.11452}{{\tt arXiv:1710.11452}}.

\bibitem{Gehrmann:2011ab}
T.~Gehrmann and L.~Tancredi, {\it {Two-loop QCD helicity amplitudes for $q\bar
  q \to W^\pm \gamma$ and $q\bar q \to Z^0 \gamma$}},  {\em JHEP} {\bf 02}
  (2012) 004, [\href{http://xxx.lanl.gov/abs/1112.1531}{{\tt
  arXiv:1112.1531}}].

\bibitem{Gehrmann:2015ora}
T.~Gehrmann, A.~von Manteuffel, and L.~Tancredi, {\it {The two-loop helicity
  amplitudes for $q \bar q' \to V_1 V_2 \to 4~\mathrm{leptons}$}},
  \href{http://xxx.lanl.gov/abs/1503.04812}{{\tt arXiv:1503.04812}}.

\bibitem{deFlorian:2013uza}
D.~de~Florian and J.~Mazzitelli, {\it {Two-loop virtual corrections to Higgs
  pair production}},  {\em Phys. Lett.} {\bf B724} (2013) 306--309,
  [\href{http://xxx.lanl.gov/abs/1305.5206}{{\tt arXiv:1305.5206}}].

\bibitem{Grazzini:2015wpa}
M.~Grazzini, S.~Kallweit, D.~Rathlev, and M.~Wiesemann, {\it
  {Transverse-momentum resummation for vector-boson pair production at
  NNLL+NNLO}},  {\em JHEP} {\bf 08} (2015) 154,
  [\href{http://xxx.lanl.gov/abs/1507.02565}{{\tt arXiv:1507.02565}}].

\bibitem{Kosower:1997zr}
D.~A. Kosower, {\it {Antenna factorization of gauge theory amplitudes}},  {\em
  Phys. Rev.} {\bf D57} (1998) 5410--5416,
  [\href{http://xxx.lanl.gov/abs/hep-ph/9710213}{{\tt hep-ph/9710213}}].

\bibitem{GehrmannDeRidder:2005cm}
A.~Gehrmann-De~Ridder, T.~Gehrmann, and E.~W.~N. Glover, {\it {Antenna
  subtraction at NNLO}},  {\em JHEP} {\bf 09} (2005) 056,
  [\href{http://xxx.lanl.gov/abs/hep-ph/0505111}{{\tt hep-ph/0505111}}].

\bibitem{Daleo:2006xa}
A.~Daleo, T.~Gehrmann, and D.~Maitre, {\it {Antenna subtraction with hadronic
  initial states}},  {\em JHEP} {\bf 04} (2007) 016,
  [\href{http://xxx.lanl.gov/abs/hep-ph/0612257}{{\tt hep-ph/0612257}}].

\bibitem{Currie:2013vh}
J.~Currie, E.~W.~N. Glover, and S.~Wells, {\it {Infrared Structure at NNLO
  Using Antenna Subtraction}},  {\em JHEP} {\bf 04} (2013) 066,
  [\href{http://xxx.lanl.gov/abs/1301.4693}{{\tt arXiv:1301.4693}}].

\bibitem{Somogyi:2005xz}
G.~Somogyi, Z.~Trocsanyi, and V.~Del~Duca, {\it {Matching of singly- and
  doubly-unresolved limits of tree-level QCD squared matrix elements}},  {\em
  JHEP} {\bf 06} (2005) 024,
  [\href{http://xxx.lanl.gov/abs/hep-ph/0502226}{{\tt hep-ph/0502226}}].

\bibitem{DelDuca:2015zqa}
V.~Del~Duca, C.~Duhr, G.~Somogyi, F.~Tramontano, and Z.~Trocsanyi, {\it {Higgs
  boson decay into b-quarks at NNLO accuracy}},  {\em JHEP} {\bf 04} (2015)
  036, [\href{http://xxx.lanl.gov/abs/1501.07226}{{\tt arXiv:1501.07226}}].

\bibitem{DelDuca:2016csb}
V.~Del~Duca, C.~Duhr, A.~Kardos, G.~Somogyi, and Z.~Trocsanyi, {\it {Three-Jet
  Production in Electron-Positron Collisions at Next-to-Next-to-Leading Order
  Accuracy}},  {\em Phys. Rev. Lett.} {\bf 117} (2016), no.~15 152004,
  [\href{http://xxx.lanl.gov/abs/1603.08927}{{\tt arXiv:1603.08927}}].

\bibitem{Czakon:2010td}
M.~Czakon, {\it {A novel subtraction scheme for double-real radiation at
  NNLO}},  {\em Phys. Lett.} {\bf B693} (2010) 259--268,
  [\href{http://xxx.lanl.gov/abs/1005.0274}{{\tt arXiv:1005.0274}}].

\bibitem{Czakon:2011ve}
M.~Czakon, {\it {Double-real radiation in hadronic top quark pair production as
  a proof of a certain concept}},  {\em Nucl. Phys.} {\bf B849} (2011)
  250--295, [\href{http://xxx.lanl.gov/abs/1101.0642}{{\tt arXiv:1101.0642}}].

\bibitem{Czakon:2014oma}
M.~Czakon and D.~Heymes, {\it {Four-dimensional formulation of the
  sector-improved residue subtraction scheme}},  {\em Nucl. Phys.} {\bf B890}
  (2014) 152--227, [\href{http://xxx.lanl.gov/abs/1408.2500}{{\tt
  arXiv:1408.2500}}].

\bibitem{Boughezal:2015eha}
R.~Boughezal, X.~Liu, and F.~Petriello, {\it {$N$-jettiness soft function at
  next-to-next-to-leading order}},  {\em Phys. Rev.} {\bf D91} (2015), no.~9
  094035, [\href{http://xxx.lanl.gov/abs/1504.02540}{{\tt arXiv:1504.02540}}].

\bibitem{Gaunt:2015pea}
J.~Gaunt, M.~Stahlhofen, F.~J. Tackmann, and J.~R. Walsh, {\it {N-jettiness
  Subtractions for NNLO QCD Calculations}},  {\em JHEP} {\bf 09} (2015) 058,
  [\href{http://xxx.lanl.gov/abs/1505.04794}{{\tt arXiv:1505.04794}}].

\bibitem{Caola:2017dug}
F.~Caola, K.~Melnikov, and R.~R{\"o}ntsch, {\it {Nested soft-collinear
  subtractions in NNLO QCD computations}},  {\em Eur. Phys. J.} {\bf C77}
  (2017), no.~4 248, [\href{http://xxx.lanl.gov/abs/1702.01352}{{\tt
  arXiv:1702.01352}}].

\bibitem{Collins:1984kg}
J.~C. Collins, D.~E. Soper, and G.~F. Sterman, {\it {Transverse Momentum
  Distribution in Drell-Yan Pair and W and Z Boson Production}},  {\em Nucl.
  Phys.} {\bf B250} (1985) 199.

\bibitem{Bozzi:2005wk}
G.~Bozzi, S.~Catani, D.~de~Florian, and M.~Grazzini, {\it {Transverse-momentum
  resummation and the spectrum of the Higgs boson at the LHC}},  {\em Nucl.
  Phys.} {\bf B737} (2006) 73--120,
  [\href{http://xxx.lanl.gov/abs/hep-ph/0508068}{{\tt hep-ph/0508068}}].

\bibitem{Bonciani:2015sha}
R.~Bonciani, S.~Catani, M.~Grazzini, H.~Sargsyan, and A.~Torre, {\it {The $q_T$
  subtraction method for top quark production at hadron colliders}},  {\em Eur.
  Phys. J.} {\bf C75} (2015), no.~12 581,
  [\href{http://xxx.lanl.gov/abs/1508.03585}{{\tt arXiv:1508.03585}}].

\bibitem{Frixione:1995ms}
S.~Frixione, Z.~Kunszt, and A.~Signer, {\it {Three jet cross-sections to
  next-to-leading order}},  {\em Nucl. Phys.} {\bf B467} (1996) 399--442,
  [\href{http://xxx.lanl.gov/abs/hep-ph/9512328}{{\tt hep-ph/9512328}}].

\bibitem{Frixione:1997np}
S.~Frixione, {\it {A General approach to jet cross-sections in QCD}},  {\em
  Nucl. Phys.} {\bf B507} (1997) 295--314,
  [\href{http://xxx.lanl.gov/abs/hep-ph/9706545}{{\tt hep-ph/9706545}}].

\bibitem{deFlorian:2001zd}
D.~de~Florian and M.~Grazzini, {\it {The Structure of large logarithmic
  corrections at small transverse momentum in hadronic collisions}},  {\em
  Nucl. Phys.} {\bf B616} (2001) 247--285,
  [\href{http://xxx.lanl.gov/abs/hep-ph/0108273}{{\tt hep-ph/0108273}}].

\bibitem{Catani:2013tia}
S.~Catani, L.~Cieri, D.~de~Florian, G.~Ferrera, and M.~Grazzini, {\it
  {Universality of transverse-momentum resummation and hard factors at the
  NNLO}},  {\em Nucl. Phys.} {\bf B881} (2014) 414--443,
  [\href{http://xxx.lanl.gov/abs/1311.1654}{{\tt arXiv:1311.1654}}].

\bibitem{Catani:2011kr}
S.~Catani and M.~Grazzini, {\it {Higgs Boson Production at Hadron Colliders:
  Hard-Collinear Coefficients at the NNLO}},  {\em Eur.Phys.J.} {\bf C72}
  (2012) 2013, [\href{http://xxx.lanl.gov/abs/1106.4652}{{\tt
  arXiv:1106.4652}}].

\bibitem{Catani:2012qa}
S.~Catani, L.~Cieri, D.~de~Florian, G.~Ferrera, and M.~Grazzini, {\it {Vector
  boson production at hadron colliders: hard-collinear coefficients at the
  NNLO}},  {\em Eur.Phys.J.} {\bf C72} (2012) 2195,
  [\href{http://xxx.lanl.gov/abs/1209.0158}{{\tt arXiv:1209.0158}}].

\bibitem{Kallweit:2014xda}
S.~Kallweit, J.~M. Lindert, P.~Maierh{\"o}fer, S.~Pozzorini, and
  M.~Sch{\"o}nherr, {\it {NLO electroweak automation and precise predictions
  for W+multijet production at the LHC}},  {\em JHEP} {\bf 04} (2015) 012,
  [\href{http://xxx.lanl.gov/abs/1412.5157}{{\tt arXiv:1412.5157}}].

\bibitem{Kallweit:2015dum}
S.~Kallweit, J.~M. Lindert, P.~Maierh{\"o}fer, S.~Pozzorini, and
  M.~Sch{\"o}nherr, {\it {NLO QCD+EW predictions for V + jets including
  off-shell vector-boson decays and multijet merging}},  {\em JHEP} {\bf 04}
  (2016) 021, [\href{http://xxx.lanl.gov/abs/1511.08692}{{\tt
  arXiv:1511.08692}}].

\bibitem{Catani:2002hc}
S.~Catani, S.~Dittmaier, M.~H. Seymour, and Z.~Trocsanyi, {\it {The Dipole
  formalism for next-to-leading order QCD calculations with massive partons}},
  {\em Nucl. Phys.} {\bf B627} (2002) 189--265,
  [\href{http://xxx.lanl.gov/abs/hep-ph/0201036}{{\tt hep-ph/0201036}}].

\bibitem{Denner:2014gla}
A.~Denner, S.~Dittmaier, and L.~Hofer, {\it {COLLIER - A fortran-library for
  one-loop integrals}},  {\em PoS} {\bf LL2014} (2014) 071,
  [\href{http://xxx.lanl.gov/abs/1407.0087}{{\tt arXiv:1407.0087}}].

\bibitem{Denner:2016kdg}
A.~Denner, S.~Dittmaier, and L.~Hofer, {\it {Collier: a fortran-based Complex
  One-Loop LIbrary in Extended Regularizations}},  {\em Comput. Phys. Commun.}
  {\bf 212} (2017) 220--238, [\href{http://xxx.lanl.gov/abs/1604.06792}{{\tt
  arXiv:1604.06792}}].

\bibitem{Ossola:2007ax}
G.~Ossola, C.~G. Papadopoulos, and R.~Pittau, {\it {CutTools: A Program
  implementing the OPP reduction method to compute one-loop amplitudes}},  {\em
  JHEP} {\bf 0803} (2008) 042, [\href{http://xxx.lanl.gov/abs/0711.3596}{{\tt
  arXiv:0711.3596}}].

\bibitem{vanHameren:2010cp}
A.~van Hameren, {\it {OneLOop: For the evaluation of one-loop scalar
  functions}},  {\em Comput.Phys.Commun.} {\bf 182} (2011) 2427--2438,
  [\href{http://xxx.lanl.gov/abs/1007.4716}{{\tt arXiv:1007.4716}}].

\bibitem{Wiesemann:2016tae}
M.~Wiesemann, {\it {Transverse-momentum resummation of colorless final states
  at the NNLL+NNLO}},  {\em PoS} {\bf RADCOR2015} (2016) 026,
  [\href{http://xxx.lanl.gov/abs/1602.03401}{{\tt arXiv:1602.03401}}].

\bibitem{Anastasiou:2002zn}
C.~Anastasiou, E.~W.~N. Glover, and M.~E. Tejeda-Yeomans, {\it {Two loop QED
  and QCD corrections to massless fermion boson scattering}},  {\em Nucl.
  Phys.} {\bf B629} (2002) 255--289,
  [\href{http://xxx.lanl.gov/abs/hep-ph/0201274}{{\tt hep-ph/0201274}}].

\bibitem{hepforge:VVamp}
The {\sc VVamp} project, by T.~Gehrmann, A.~von~Manteuffel, and L.~Tancredi, is
  publicly available at \url{http://vvamp.hepforge.org}.

\bibitem{Buckley:2014ana}
A.~Buckley, J.~Ferrando, S.~Lloyd, K.~Nordstr{\"o}m, B.~Page, M.~R{\"u}fenacht,
  M.~Sch{\"o}nherr, and G.~Watt, {\it {LHAPDF6: parton density access in the
  LHC precision era}},  {\em Eur. Phys. J.} {\bf C75} (2015) 132,
  [\href{http://xxx.lanl.gov/abs/1412.7420}{{\tt arXiv:1412.7420}}].

\bibitem{hepforge:OpenLoops}
The {\sc OpenLoops} one-loop generator, by F.~Cascioli, J.~Lindert,
  P.~Maierh{\"o}fer, and S.~Pozzorini, is publicly available at
  \url{http://openloops.hepforge.org}.

\bibitem{cln}
B.~Haible and R.~B. Kreckel, {\it {CLN: Class Library for Numbers}},
  \href{http://xxx.lanl.gov/abs/cs/0004015}{{\tt cs/0004015}}.

\bibitem{Bauer:2000cp}
C.~W. Bauer, A.~Frink, and R.~Kreckel, {\it {Introduction to the GiNaC
  framework for symbolic computation within the C++ programming language}},
  {\em J. Symb. Comput.} {\bf 33} (2000) 1,
  [\href{http://xxx.lanl.gov/abs/cs/0004015}{{\tt cs/0004015}}].

\bibitem{Dokshitzer:1997in}
Y.~L. Dokshitzer, G.~D. Leder, S.~Moretti, and B.~R. Webber, {\it {Better jet
  clustering algorithms}},  {\em JHEP} {\bf 08} (1997) 001,
  [\href{http://xxx.lanl.gov/abs/hep-ph/9707323}{{\tt hep-ph/9707323}}].

\bibitem{Wobisch:1998wt}
M.~Wobisch and T.~Wengler, {\it {Hadronization corrections to jet
  cross-sections in deep inelastic scattering}},  in {\em {Monte Carlo
  generators for HERA physics. Proceedings, Workshop, Hamburg, Germany,
  1998-1999}}, pp.~270--279, 1998.
\newblock \href{http://xxx.lanl.gov/abs/hep-ph/9907280}{{\tt hep-ph/9907280}}.

\bibitem{Catani:1993hr}
S.~Catani, Y.~L. Dokshitzer, M.~H. Seymour, and B.~R. Webber, {\it
  {Longitudinally invariant $K_t$ clustering algorithms for hadron hadron
  collisions}},  {\em Nucl. Phys.} {\bf B406} (1993) 187--224.

\bibitem{Cacciari:2008gp}
M.~Cacciari, G.~P. Salam, and G.~Soyez, {\it {The Anti-k(t) jet clustering
  algorithm}},  {\em JHEP} {\bf 0804} (2008) 063,
  [\href{http://xxx.lanl.gov/abs/0802.1189}{{\tt arXiv:0802.1189}}].

\bibitem{Frixione:1998jh}
S.~Frixione, {\it {Isolated photons in perturbative QCD}},  {\em Phys. Lett.}
  {\bf B429} (1998) 369--374,
  [\href{http://xxx.lanl.gov/abs/hep-ph/9801442}{{\tt hep-ph/9801442}}].

\bibitem{Skands:2003cj}
P.~Z. Skands et~al., {\it {SUSY Les Houches accord: Interfacing SUSY spectrum
  calculators, decay packages, and event generators}},  {\em JHEP} {\bf 07}
  (2004) 036, [\href{http://xxx.lanl.gov/abs/hep-ph/0311123}{{\tt
  hep-ph/0311123}}].

\bibitem{Olive:2016xmw}
{\bf Particle Data Group} Collaboration, C.~Patrignani et~al., {\it {Review of
  Particle Physics}},  {\em Chin. Phys.} {\bf C40} (2016), no.~10 100001.

\bibitem{Aad:2013izg}
{\bf ATLAS} Collaboration, G.~Aad et~al., {\it {Measurements of $W \gamma$ and
  $Z \gamma$ production in $pp$ collisions at $\sqrt{s}$=7  TeV with the
  ATLAS detector at the LHC}},  {\em Phys. Rev.} {\bf D87} (2013), no.~11
  112003, [\href{http://xxx.lanl.gov/abs/1302.1283}{{\tt arXiv:1302.1283}}].
  [Erratum: Phys. Rev.D91,no.11,119901(2015)].

\bibitem{Aad:2016ett}
{\bf ATLAS} Collaboration, G.~Aad et~al., {\it {Measurements of $W^\pm Z$
  production cross sections in $pp$ collisions at $\sqrt{s} = 8$ TeV with the
  ATLAS detector and limits on anomalous gauge boson self-couplings}},  {\em
  Phys. Rev.} {\bf D93} (2016), no.~9 092004,
  [\href{http://xxx.lanl.gov/abs/1603.02151}{{\tt arXiv:1603.02151}}].

\bibitem{Denner:2005fg}
A.~Denner, S.~Dittmaier, M.~Roth, and L.~H. Wieders, {\it {Electroweak
  corrections to charged-current $e^+ e^-\to$ 4 fermion processes: Technical
  details and further results}},  {\em Nucl. Phys.} {\bf B724} (2005) 247--294,
  [\href{http://xxx.lanl.gov/abs/hep-ph/0505042}{{\tt hep-ph/0505042}}].
  [Erratum: Nucl. Phys.B854,504(2012)].

\bibitem{Ball:2014uwa}
{\bf NNPDF} Collaboration, R.~D. Ball et~al., {\it {Parton distributions for
  the LHC Run II}},  {\em JHEP} {\bf 1504} (2015) 040,
  [\href{http://xxx.lanl.gov/abs/1410.8849}{{\tt arXiv:1410.8849}}].

\bibitem{Harlander:2012pb}
R.~V. Harlander, S.~Liebler, and H.~Mantler, {\it {SusHi: A program for the
  calculation of Higgs production in gluon fusion and bottom-quark annihilation
  in the Standard Model and the MSSM}},  {\em Comput. Phys. Commun.} {\bf 184}
  (2013) 1605--1617, [\href{http://xxx.lanl.gov/abs/1212.3249}{{\tt
  arXiv:1212.3249}}].

\bibitem{Frixione:1997ks}
S.~Frixione and G.~Ridolfi, {\it {Jet photoproduction at HERA}},  {\em Nucl.
  Phys.} {\bf B507} (1997) 315--333,
  [\href{http://xxx.lanl.gov/abs/hep-ph/9707345}{{\tt hep-ph/9707345}}].

\bibitem{Alioli:2016fum}
S.~Alioli et~al., {\it {Precision studies of observables in $p p \rightarrow W
  \rightarrow l\nu _l$ and $ pp \rightarrow \gamma ,Z \rightarrow l^+ l^-$
  processes at the LHC}},  {\em Eur. Phys. J.} {\bf C77} (2017), no.~5 280,
  [\href{http://xxx.lanl.gov/abs/1606.02330}{{\tt arXiv:1606.02330}}].

\bibitem{Mikaelian:1979nr}
K.~O. Mikaelian, M.~A. Samuel, and D.~Sahdev, {\it {The Magnetic Moment of Weak
  Bosons Produced in p p and p anti-p Collisions}},  {\em Phys. Rev. Lett.}
  {\bf 43} (1979) 746.

\bibitem{Caola:2015psa}
F.~Caola, K.~Melnikov, R.~R{\"o}ntsch, and L.~Tancredi, {\it {QCD corrections
  to ZZ production in gluon fusion at the LHC}},  {\em Phys. Rev.} {\bf D92}
  (2015), no.~9 094028, [\href{http://xxx.lanl.gov/abs/1509.06734}{{\tt
  arXiv:1509.06734}}].

\bibitem{Caola:2015rqy}
F.~Caola, K.~Melnikov, R.~R{\"o}ntsch, and L.~Tancredi, {\it {QCD corrections
  to $W^+W^-$ production through gluon fusion}},  {\em Phys. Lett.} {\bf B754}
  (2016) 275--280, [\href{http://xxx.lanl.gov/abs/1511.08617}{{\tt
  arXiv:1511.08617}}].

\bibitem{Baur:1994ia}
U.~Baur, T.~Han, and J.~Ohnemus, {\it {Amplitude zeros in $W^\pm Z$
  production}},  {\em Phys. Rev. Lett.} {\bf 72} (1994) 3941--3944,
  [\href{http://xxx.lanl.gov/abs/hep-ph/9403248}{{\tt hep-ph/9403248}}].

\bibitem{Ridder:2016nkl}
A.~Gehrmann-De~Ridder, T.~Gehrmann, E.~W.~N. Glover, A.~Huss, and T.~A. Morgan,
  {\it {The NNLO QCD corrections to Z boson production at large transverse
  momentum}},  {\em JHEP} {\bf 07} (2016) 133,
  [\href{http://xxx.lanl.gov/abs/1605.04295}{{\tt arXiv:1605.04295}}].

\bibitem{Aad:2016sau}
{\bf ATLAS} Collaboration, G.~Aad et~al., {\it {Measurements of $Z\gamma$ and
  $Z\gamma\gamma$ production in $pp$ collisions at $\sqrt{s}=$ 8 TeV with the
  ATLAS detector}},  {\em Phys. Rev.} {\bf D93} (2016), no.~11 112002,
  [\href{http://xxx.lanl.gov/abs/1604.05232}{{\tt arXiv:1604.05232}}].

\bibitem{CMS:2016xzm}
{\bf CMS} Collaboration, {\it {Measurement of the production cross section for
  $pp \rightarrow Z\gamma \rightarrow \nu \bar\nu\gamma$ at $\sqrt{s} = 13 TeV$
  at CMS}},  \href{http://xxx.lanl.gov/abs/CMS-PAS-SMP-16-004}{{\tt
  CMS-PAS-SMP-16-004}}.

\bibitem{Sirunyan:2017lvq}
{\bf CMS} Collaboration, A.~M. Sirunyan et~al., {\it {Measurements of the pp
  $\to W\gamma\gamma$ and pp $\to Z\gamma\gamma$ cross sections and limits on
  anomalous quartic gauge couplings at $ \sqrt{s}=8 $ TeV}},  {\em JHEP} {\bf
  10} (2017) 072, [\href{http://xxx.lanl.gov/abs/1704.00366}{{\tt
  arXiv:1704.00366}}].

\bibitem{Aad:2015rka}
{\bf ATLAS} Collaboration, G.~Aad et~al., {\it {Measurements of four-lepton
  production in $pp$ collisions at $\sqrt{s}=$ 8 TeV with the ATLAS detector}},
   {\em Phys. Lett.} {\bf B753} (2016) 552--572,
  [\href{http://xxx.lanl.gov/abs/1509.07844}{{\tt arXiv:1509.07844}}].

\bibitem{Aaboud:2016urj}
{\bf ATLAS} Collaboration, M.~Aaboud et~al., {\it {Measurement of the $ZZ$
  production cross section in proton-proton collisions at $\sqrt s =$ 8 TeV
  using the $ZZ\to\ell^{-}\ell^{+}\ell^{\prime -}\ell^{\prime +}$ and
  $ZZ\to\ell^{-}\ell^{+}\nu\bar{\nu}$ channels with the ATLAS detector}},  {\em
  JHEP} {\bf 01} (2017) 099, [\href{http://xxx.lanl.gov/abs/1610.07585}{{\tt
  arXiv:1610.07585}}].

\bibitem{Aaboud:2017rwm}
{\bf ATLAS} Collaboration, M.~Aaboud et~al., {\it {$ZZ \to
  \ell^{+}\ell^{-}\ell^{\prime +}\ell^{\prime -}$ cross-section measurements
  and search for anomalous triple gauge couplings in 13 TeV $pp$ collisions
  with the ATLAS detector}},  \href{http://xxx.lanl.gov/abs/1709.07703}{{\tt
  arXiv:1709.07703}}.

\bibitem{CMS:2016kxu}
{\bf CMS} Collaboration, {\it {Measurement of the differential cross section
  for $pp \rightarrow ZZ \rightarrow 4 \ell$ produced in association with jets
  in $pp$ collisions at $\sqrt s = 8 TeV$}},
  \href{http://xxx.lanl.gov/abs/CMS-PAS-SMP-15-012}{{\tt CMS-PAS-SMP-15-012}}.

\bibitem{CMS:2015fnj}
{\bf CMS} Collaboration, {\it {Measurement of the ZZ production cross section
  in $\ell\ell\ell'\ell'$ decays in pp collisions at $\sqrt{s} = 13$ TeV}},
  \href{http://xxx.lanl.gov/abs/CMS-PAS-SMP-15-005}{{\tt CMS-PAS-SMP-15-005}}.

\bibitem{Khachatryan:2016txa}
{\bf CMS} Collaboration, V.~Khachatryan et~al., {\it {Measurement of the ZZ
  production cross section and Z $\to \ell^+\ell^-\ell'^+\ell'^-$ branching
  fraction in pp collisions at $\sqrt s$=13 TeV}},  {\em Phys. Lett.} {\bf
  B763} (2016) 280--303, [\href{http://xxx.lanl.gov/abs/1607.08834}{{\tt
  arXiv:1607.08834}}]. [Erratum: Phys. Lett.B772,884(2017)].

\bibitem{Sirunyan:2017zjc}
{\bf CMS} Collaboration, A.~M. Sirunyan et~al., {\it {Measurements of the
  pp$\to$ZZ production cross section and the Z$\to 4\ell$ branching fraction,
  and constraints on anomalous triple gauge couplings at $\sqrt{s} =$ 13 TeV}},
   \href{http://xxx.lanl.gov/abs/1709.08601}{{\tt arXiv:1709.08601}}.

\bibitem{Aad:2016wpd}
{\bf ATLAS} Collaboration, G.~Aad et~al., {\it {Measurement of total and
  differential $W^+W^-$ production cross sections in proton-proton collisions
  at $\sqrt{s}=$ 8 TeV with the ATLAS detector and limits on anomalous
  triple-gauge-boson couplings}},  {\em JHEP} {\bf 09} (2016) 029,
  [\href{http://xxx.lanl.gov/abs/1603.01702}{{\tt arXiv:1603.01702}}].

\bibitem{Aaboud:2016mrt}
{\bf ATLAS} Collaboration, M.~Aaboud et~al., {\it {Measurement of $W^+W^-$
  production in association with one jet in proton--proton collisions at
  $\sqrt{s} =8$ TeV with the ATLAS detector}},  {\em Phys. Lett.} {\bf B763}
  (2016) 114--133, [\href{http://xxx.lanl.gov/abs/1608.03086}{{\tt
  arXiv:1608.03086}}].

\bibitem{Aaboud:2017qkn}
{\bf ATLAS} Collaboration, M.~Aaboud et~al., {\it {Measurement of the $W^+W^-$
  production cross section in $pp$ collisions at a centre-of-mass energy of
  $\sqrt{s}$ = 13 TeV with the ATLAS experiment}},  {\em Phys. Lett.} {\bf
  B773} (2017) 354--374, [\href{http://xxx.lanl.gov/abs/1702.04519}{{\tt
  arXiv:1702.04519}}].

\bibitem{Khachatryan:2015sga}
{\bf CMS} Collaboration, V.~Khachatryan et~al., {\it {Measurement of the
  ${{\mathrm{W} }^{+} }\mathrm{W}^{-} $ cross section in pp collisions at
  $\sqrt{s} =$ 8 TeV and limits on anomalous gauge couplings}},  {\em Eur.
  Phys. J.} {\bf C76} (2016), no.~7 401,
  [\href{http://xxx.lanl.gov/abs/1507.03268}{{\tt arXiv:1507.03268}}].

\bibitem{CMS:2016vww}
{\bf CMS} Collaboration, {\it {Measurement of the WW cross section pp
  collisions at $\sqrt{s}=13 TeV$}},
  \href{http://xxx.lanl.gov/abs/CMS-PAS-SMP-16-006}{{\tt CMS-PAS-SMP-16-006}}.

\bibitem{Aaboud:2016yus}
{\bf ATLAS} Collaboration, M.~Aaboud et~al., {\it {Measurement of the
  $W^{\pm}Z$ boson pair-production cross section in $pp$ collisions at
  $\sqrt{s}=13$ TeV with the ATLAS Detector}},  {\em Phys. Lett.} {\bf B762}
  (2016) 1--22, [\href{http://xxx.lanl.gov/abs/1606.04017}{{\tt
  arXiv:1606.04017}}].

\bibitem{ATLAS:2016qzn}
{\bf ATLAS} Collaboration, {\it {Measurement of $W^{\pm}Z$ boson
  pair-production in $pp$ collisions at $\sqrt{s}=13$ TeV with the ATLAS
  Detector and confidence intervals for anomalous triple gauge boson
  couplings}},  \href{http://xxx.lanl.gov/abs/ATLAS-CONF-2016-043}{{\tt
  ATLAS-CONF-2016-043}}.

\bibitem{Khachatryan:2016poo}
{\bf CMS} Collaboration, V.~Khachatryan et~al., {\it {Measurement of the WZ
  production cross section in pp collisions at $\sqrt{s} = 7$ and 8
  $\,\text{TeV}$ and search for anomalous triple gauge couplings at $\sqrt{s} =
  8\,\text{TeV} $}},  {\em Eur. Phys. J.} {\bf C77} (2017), no.~4 236,
  [\href{http://xxx.lanl.gov/abs/1609.05721}{{\tt arXiv:1609.05721}}].

\bibitem{Khachatryan:2016tgp}
{\bf CMS} Collaboration, V.~Khachatryan et~al., {\it {Measurement of the WZ
  production cross section in pp collisions at $\sqrt(s) =$ 13 TeV}},  {\em
  Phys. Lett.} {\bf B766} (2017) 268--290,
  [\href{http://xxx.lanl.gov/abs/1607.06943}{{\tt arXiv:1607.06943}}].

\bibitem{deFlorian:2016spz}
{\bf LHC Higgs Cross Section Working Group} Collaboration, D.~de~Florian
  et~al., {\it {Handbook of LHC Higgs Cross Sections: 4. Deciphering the Nature
  of the Higgs Sector}},  \href{http://xxx.lanl.gov/abs/1610.07922}{{\tt
  arXiv:1610.07922}}.

\bibitem{Chatrchyan:2013fya}
{\bf CMS} Collaboration, S.~Chatrchyan et~al., {\it {Measurement of the
  $W\gamma$ and $Z\gamma$ inclusive cross sections in $pp$ collisions at $\sqrt
  s=7$  TeV and limits on anomalous triple gauge boson couplings}},  {\em
  Phys. Rev.} {\bf D89} (2014), no.~9 092005,
  [\href{http://xxx.lanl.gov/abs/1308.6832}{{\tt arXiv:1308.6832}}].

\end{thebibliography}\endgroup

\end{document}